\documentclass[letterpaper,12pt]{article}

\usepackage{tabularx} 
\usepackage{amsmath,amsfonts,mathrsfs}
\usepackage{dsfont}  
\usepackage{graphicx} 
\usepackage[margin=1in,letterpaper]{geometry} 
\usepackage{cite} 
\usepackage[final]{hyperref} 
\usepackage{slashed}
\usepackage{mathtools}
\usepackage{subcaption}
\usepackage{import}
\usepackage{multicol}
\usepackage[font=small,
            width=0.8\textwidth]{caption}
\hypersetup{
	colorlinks=true,       
	linkcolor=black,       
	citecolor=blue,        
	filecolor=magenta,     
	urlcolor=black         
}
\usepackage{blindtext}
\DeclareUnicodeCharacter{2009}{\\,}

\NewDocumentCommand{\evalat}{sO{\big}mm}{%
  \IfBooleanTF{#1}
   {\mleft. #3 \mright|_{#4}}
   {#3#2|_{#4}}%
}

\usepackage{comment}
\usepackage[dvipsnames]{xcolor}
\newcommand{\ed}{\end{document}}

\usepackage[normalem]{ulem}

\newcommand{\A}{\mathcal{A}}

\renewcommand{\O}{\mathbf{O}}

\newcommand{\BBd}{B_{\mu\nu}}

\newcommand{\WWd}{W_{\mu\nu}}

\newcommand{\cl}{\%~\text{C.L.}}

\renewcommand{\to}{\rightarrow}

\newcommand{\de}{\partial}

\newcommand{\beq}{\begin{equation}}
\newcommand{\eeq}{\end{equation}}
\newcommand{\bea}{\begin{eqnarray}}
\newcommand{\eea}{\end{eqnarray}}
\renewcommand{\[}{\begin{equation}}
\renewcommand{\]}{\end{equation}}

\newcommand{\LL}{\mathscr{L}}

\newcounter{diagram}

\newcommand{\agg}{a\gamma\gamma}
\newcommand{\ga}{\gamma}

\newcommand{\etmiss}{E_T^{miss}}

\newcommand{\mgg}{m_{\gamma\gamma}}
\newcommand{\drgg}{\Delta R_{\gamma\gamma}}

\newcommand{\ptmiss}{\vec{p}_T^{\;\text{miss}}}
\newcommand{\ptl}{p_T^\ell}
\newcommand{\ptg}{p_T^\gamma}
\newcommand{\ptgone}{p_T^{\gamma_1}}
\newcommand{\ptgi}{p_T^{\gamma_i}}

\newcommand{\ptll}{p_{T}^{\ell\ell}}
\newcommand{\drll}{\Delta R_{\ell\ell}}
\newcommand{\ptz}{p_{T}^{Z}}

\newcommand{\etag}{\eta_{\gamma}}
\newcommand{\etal}{\eta_{\ell}}

\newcommand{\mll}{m_{\ell\ell}}

\newcommand{\linv}{\ell+\mr{inv}}
\newcommand{\mtlinv}{m_T^{\linv}}
\newcommand{\mtlinvg}{m_T^{\linv,\gamma}}

\newcommand{\mtginv}{m_T^{\gamma,\mr{inv}}}
\newcommand{\mtllinv}{m_T^{\ell\ell, \mr{inv}}}

\newcommand{\dpginv}{\Delta\phi_{\gamma, \mr{inv}}}
\newcommand{\dplinvg}{\Delta\phi_{\linv,\gamma}}
\newcommand{\dpllinv}{\Delta\phi_{\ell\ell,\mr{inv}}}
\newcommand{\dpll}{\Delta\phi_{\ell\ell}}

\newcommand{\wpm}{W^+W^-}

\newcommand{\awpg}{aW^+\gamma}
\newcommand{\awmg}{aW^-\gamma}
\newcommand{\awg}{aW\gamma}
\newcommand{\azg}{aZ\gamma}

\newcommand{\azw}{aZW}
\newcommand{\aww}{a\wpm}
\newcommand{\azz}{aZZ}

\newcommand{\GeV}{\mathrm{GeV}}
\newcommand{\mc}{\mathcal}
\newcommand{\mr}{\mathrm}

\begin{document}

	\title{
  Constraints on axion-like particles via associated diboson production in hadronic collisions}
\author{
  B. Jäger\thanks{\href{jaeger@itp.uni-tuebingen.de}{jaeger@itp.uni-tuebingen.de}},\hspace{0.3cm}
  O. Semin\thanks{\href{ozan.semin@uni-tuebingen.de}{ozan.semin@uni-tuebingen.de}}}
\date{
  {\small \href{https://uni-tuebingen.de/fakultaeten/mathematisch-naturwissenschaftliche-fakultaet/fachbereiche/physik/institute/institut-fuer-theoretische-physik/arbeitsgruppen/}{\textit{Institute for Theoretical Physics, Auf der Morgenstelle 14, University of Tübingen, Tübingen 72076, Germany}}}\\\vspace{1cm}    
	\today}
\maketitle
\vspace{-0.5cm}

\begin{abstract}
We investigate the sensitivity of current and future hadron-collider experiments to axion-like particles (ALPs) through associated diboson production, focusing on a linear effective field theory framework with bosonic ALP couplings. We analyze the dominant production mechanisms and relevant backgrounds, considering the impact of jet misidentification rates on the diboson background. We present our results using conservative jet-misidentification rates, and derive four dimensional constraints on the ALP couplings to gluons, weak bosons, and photons. Our findings highlight the potential of the high-luminosity phase of the CERN Large Hadron Collider to probe the ALP parameter space in the sub-GeV mass range, as well as the codependencies of the various ALP couplings.
\end{abstract}

\section{Introduction}\label{sec:intro}

Axion-like particles (ALPs) have become promising candidates for a possible extension of the Standard Model (SM) of elementary particles, as they may help address several unresolved problems in particle physics and cosmology~\cite{Zwicky1933Die,Rubin:1980zd,Massey:2010hh}. Axions were first proposed by Peccei and Quinn in the context of QCD to solve the strong CP problem~\cite{Peccei:1977hh,Weinberg:1977ma}. In contrast to these QCD axions, whose properties are strictly tied to the Peccei-Quinn symmetry breaking scale, ALPs represent a broader class of pseudo-Nambu-Goldstone bosons. While giving rise to similar phenomenological signatures  as the axions, the mass and coupling strengths of ALPs are independent parameters. ALPs now appear in many theoretical frameworks, such as string theory compactifications~\cite{Arvanitaki:2009fg} and models of dark matter (DM)~\cite{Saikumar:2024ahz,Jungman:1995df,Gehrlein:2019iwl,FileviezPerez:2019cyn}. 
Their feeble couplings to SM fields and their light masses make them hard to detect, but they promise valuable insights into new physics at both low- and high-energy experiments.

Traditionally, searches for ALPs at colliders have focused on ALP masses from $\mc{O}$(0.1~GeV) up to $\mc{O}$(100~GeV)~\cite{Knapen:2021elo, Ghebretinsaea:2022djg,Bauer:2018uxu,Sharma:2025vsh,Tentori:2024xju}. In this mass range ALPs feature quick decays within the detector giving rise to a variety of possible final states. However, the sub-GeV range, particularly for ALPs with masses below the MeV~scale, is less studied. In this range, ALPs tend to be long-lived, escaping the detector and leaving signatures marked by missing energy or, sometimes, displaced vertices~\cite{Ferber:2022rsf,Chenarani:2025uay,Bonilla:2022pxu}. 
In contrast, when ALPs do decay within the detector, the resulting photons can be hard to separate from SM backgrounds, especially at hadron colliders where QCD-induced jet-production processes are prevalent. 

Our work aims at exploring the sensitivity of current and future hadron colliders to ALPs in the sub-GeV mass range, focusing on the associated production of ALPs with dibosons. Similar processes involving pairs of $Z$ and $W$ bosons in context of the LHC have already been considered in the literature~\cite{Chenarani:2025uay}. However, such studies typically assume fixed mass and coupling values to reduce the number of free parameters. We use a linear effective field theory (EFT) framework containing bosonic ALP couplings~\cite{Brivio:2017ije}. In order to keep our analysis as general as possible, we do not assume any predetermined constraints, and consider all possible ALP couplings at once. This provides a very broad way to describe ALP interactions with SM gauge bosons. 

Section~\ref{sec:alp_model} introduces this linear ALP model, detailing the relevant operators and their implications. We explain the structure of the effective Lagrangian, the significance of electroweak symmetry breaking in generating couplings to photons, fermions and weak gauge bosons, and the relationship between different Wilson coefficients. We explore the effect of the Higgs coupling, which can create effective fermion couplings even without tree-level interactions, after appropriate field redefinitions. 

In Sec.~\ref{sec:pheno}, we perform a comprehensive analysis of associated ALP production at the LHC and its high-luminosity upgrade (HL-LHC), focusing on diboson associated production. We systematically explore the dominant production mechanisms, including gluon-gluon fusion and electroweak processes, and evaluate their sensitivity to the relevant couplings. A key component of our study is the detailed treatment of SM backgrounds, particularly those arising from detector effects such as jet misidentification, which often mimic photon signatures in hadronic environments. 
By employing multivariate analysis techniques, specifically Boosted Decision Trees (BDTs), we aim to maximize signal discrimination against the overwhelming QCD and electroweak backgrounds. 

We then provide updated constraints on relevant ALP couplings, derived from algorithmically selected signal-background discriminating kinematic distributions. 
In our conclusions we interpret our results in the light of current and future collider capabilities. We also discuss what our findings mean for future experimental efforts, emphasizing the need for better background modeling and the potential for complementary searches at lepton colliders. 

\section{ALP interactions in the linear bosonic model}\label{sec:alp_model}

The most general model accounting for pseudo-scalar ALPs coupling to the bosons of the SM exhibits a Lagrangian of the form 
\beq 
\mathscr{L}^\text{SM+ALP}=\mathscr{L}^\text{SM}+\mathscr{L}^\text{bosonic}_a+\frac{1}{2}(\partial_\mu a)(\partial^\mu a)-\frac{m_a^2}{2}a^2,
\label{eq:lag}
\eeq
where $\mathscr{L}^\text{SM}$ denotes the Lagrangian of the SM,  $\mathscr{L}^\text{bosonic}_a$  accounts for all  ALP-boson interactions, and the remaining two terms denote kinetic and mass term of the ALP field $a$, respectively. 
For the ALP-boson interaction we use the linear bosonic ALP model~\cite{Brivio:2017ije}, such that
\beq
\delta\LL_a^\text{bosonic}\,=\,c_{W}\A_{W}+c_{B}\A_{B}+c_{G}\A_{G}+c_{a\Phi}\O_{a\Phi}\,
\label{deltaLbosonic-lin-initial}
\eeq
with
\begin{eqnarray}
  \A_{B} &=&-\BBd \tilde{B}^{\mu\nu}\dfrac{a}{f_a}\,,\label{ABtilde}\\
  \A_{W} &=&-\WWd^a\tilde{W}^{a\mu\nu}\dfrac{a}{f_a}\,,\label{AWtilde}\\
  \A_{G} &=&-G^a_{\mu\nu}\tilde{G}^{a\mu\nu}\dfrac{a}{f_a}\,,\label{AGtilde}\\
\O_{a\Phi}&=& i (\Phi^\dag\overleftrightarrow{D}_\mu\Phi)\frac{\de^\mu a}{f_a}\,,\label{OaPhi}
\end{eqnarray}
where the $B_{\mu\nu}$, $W_{\mu\nu}$, and $G_{\mu\nu}$ correspond to the gauge field strength tensors of the symmetry groups $U(1)_Y$, $SU(2)_L$ and $SU(3)_c$, respectively, and the dual of  each tensor $X_{\mu\nu}$ is defined as $\tilde{X}^{\mu\nu}\equiv \frac12 \epsilon^{\mu\nu\rho\sigma}X_{\rho\sigma}$. The Higgs field is denoted by $\Phi$, and $D_\mu$ is the covariant derivative, defined by $D_\mu = \partial_\mu - i g' Y B_\mu - i g \frac{\sigma^a}{2} W^a_\mu$, where $g'$ and $g$ are the gauge couplings of $U(1)_Y$ and $SU(2)_L$, respectively, $Y$ is the hypercharge and $\sigma^a$ are the Pauli matrices. The ALP mass is given by $m_a$, and the ALP scale $f_a$ corresponds to the energy scale below which the EFT approach of the model remains valid. Note that $f_a$ is usually considered as a free parameter. The $c_{W}$, $c_{B}$, $c_{G}$, and $c_{a\Phi}$ in Eq.~\eqref{deltaLbosonic-lin-initial} are dimensionless Wilson coefficients, which can be related to the strength of the respective ALP-boson couplings.  

In this work, we aim at putting \textit{codependent} constraints on these coefficients with minimum assumptions. Since interaction terms of the form $c_i \A_{i}$ always contain the \textit{ratio} of the a Wilson coefficient $c_i$ with the EFT scale $f_a$, we will present constraints in terms of the ratios $c_i /f_a$. 

After electroweak symmetry breaking, linear combinations of the Wilson coefficients produce the relevant couplings for interactions of the ALP with a gluon $g$, a photon $\gamma$, a $Z$~boson or a $W$~boson, 

\begin{eqnarray}
\label{eq:g-aVV}
  &&g_{a\gamma\gamma}=\frac{4}{f_a}(c_{B}c_\theta^2+c_{W}s_\theta^2)\,,\quad
  g_{a\gamma Z}=\frac{8}{f_a}c_\theta s_\theta(c_{W}-c_{B})\,,\,\nonumber\\
  &&g_{a Z Z}=\frac{4}{f_a}(c_{B}s_\theta^2+c_{W}c_\theta^2)\,,\quad
  g_{aWW}=\frac{4}{f_a}c_{W}\,,
\end{eqnarray}

where $c_\theta$ and $s_\theta$ are the cosine and sine of the weak mixing angle $\theta$, respectively. 

Furthermore, in the bosonic model, one can still obtain small, yet non-negligible fermionic couplings from the Higgs operators by appropriate field redefinitions or equations of motion (EOM). The operator that couples the axion to the Higgs doublet contains the neutral component of the Higgs isospin current, which projects onto the third $SU(2)_L$ generator. At leading order, the $SU(2)_L$ gauge-field equation of motion relates this Higgs current directly to the corresponding fermion isospin current. Using this EOM relation, which is valid within the usual EFT equivalence rules, the Higgs current can be traded for a flavor-universal, chirality-conserving fermionic interaction. As a result, the original bosonic operator and the induced fermionic operator describe the same physical effects and should not be included simultaneously in a non-redundant operator basis. We thus make use of  
\begin{equation}\label{eq:coup_higgs_alp}
  \O_{a\Phi} \to -\frac{\partial_\mu a}{2f_a}\sum\limits_{f}\bar{\psi}_f\gamma^\mu\gamma_5\sigma_3\psi_f,
\end{equation}
where the $\psi_f$ are $SU(2)_L$ weak-isospin doublets, i.e.\ the quark and lepton doublets of a specific flavor $f$.  

This relation implies that the $\sigma_3$ matrix acting on the weak-isospin doublet space assigns opposite signs to the upper and lower components of each doublet. Thus, after electroweak symmetry breaking, up-type and down-type fermions acquire couplings of equal magnitude but opposite sign, i.e.\ 
\begin{equation}
  c_{add}=-c_{auu}=c_{a\Phi},
\end{equation}
where $d$ and $u$ correspond to down-type and up-type fermions, respectively.

Since we use the purely bosonic linear ALP model, there are no other tree level fermionic ALP couplings. This will allow us to constrain Higgs-ALP Wilson coefficients, without explicit reference to processes involving the Higgs particle. 

The linear bosonic ALP EFT employed in this work provides a minimal and systematically controlled framework for describing interactions between a pseudo-scalar ALP and the SM gauge sector. The operator set introduced in Eq.~\eqref{deltaLbosonic-lin-initial} constitutes the complete basis of gauge-invariant, dimension-five ALP-boson interactions in a linear realisation of electroweak symmetry. Any additional structures at this order can be eliminated through integration by parts, total derivative identities, or application of the SM EOM, thereby removing redundant operators without affecting physical observables. In particular, the Higgs-current operator $\O_{a\Phi}$ may be traded for an equivalent fermionic interaction via equations of motion, as demonstrated in Eq.~\eqref{eq:coup_higgs_alp}, ensuring that the operator basis employed throughout this work is non-redundant. The induced fermionic couplings are entirely determined by the bosonic operator from which they originate and introduce no additional free parameters.

The linear EFT framework adopted here assumes that electroweak symmetry breaking is realised through a fundamental Higgs doublet. This assumption excludes non-linear realisations of the electroweak sector, such as those arising in composite Higgs or strong dynamics scenarios, where an extended operator basis and modified coupling correlations would emerge at leading order. The present analysis is therefore valid within the domain of applicability of the linear model, where the SM symmetry-breaking mechanism and Higgs interactions retain their canonical structure.

A characteristic feature of the linear EFT is the pattern of physical ALP couplings that emerges after electroweak symmetry breaking. Interactions with photons, $Z$~bosons, and $W$~bosons arise from only two independent gauge operators, $\A_B$ and $\A_W$, leading to definite correlations among production and decay channels. These correlations, intrinsic to the linear EFT, are central to the combined constraints derived in subsequent sections and enable simultaneous probes of multiple Wilson coefficients through complementary processes. 
\section{Phenomenological analysis}\label{sec:pheno}

For the ALP model considered in this work, several phenomenological studies have been performed in the context of searches at colliders, including mono-photon searches~\cite{Mimasu:2014nea, Brivio:2017ije, ATLAS:2020uiq,Bao:2025tqs}, mono-$Z/W^\pm$ signatures~\cite{Brivio:2017ije, Bauer:2017ris,Bao:2025tqs}, as well as VBF and VBS processes~\cite{ Bonilla:2022pxu, Florez:2021zoo}. These studies either employ the general bosonic ALP framework used here, or consider specific ALP models that can be obtained from our generalized model by imposing certain assumptions on the Wilson coefficients. Most of these studies focus on ALPs  in the mass range $m_a\geq10~\mathrm{GeV}$, which have shorter decay lengths than lighter ALPs. 
Complementing these efforts, we focus on the less explored regime of light, long-lived ALPs and systematically study their production in association with two electroweak bosons. 

In this section, we present a detailed discussion of ALP production in association with two bosons in proton-proton collisions, 
\begin{equation}
  p+p\to a + V+V^\prime\,,
\end{equation}
where the $V,V^\prime$ denote a photon or an off-shell massive gauge boson $(Z,W^\pm)$ which in turn decays leptonically. The individual final states will be discussed separately below. 

For all of them, to derive codependent constraints for the $\{\frac{c_W}{f_a},\frac{c_B}{f_a},\frac{c_G}{f_a},\frac{c_{a\Phi}}{f_a}\}$ parameter space of the bosonic ALP model, we consider center-of-mass energies of $\sqrt{s}=14$~TeV at the LHC and the HL-LHC.  
 While the LHC operated at $\sqrt{s}=13.6~\text{TeV}$ until now, the difference in cross sections and kinematic distributions between 13.6 and 14~TeV is very small for the processes studied here, and adopting the HL-LHC design value simplifies the comparison between the two collider modes.

Throughout, we consider integrated luminosities of $450~\text{fb}^{-1}$ for the LHC and $3000~\text{fb}^{-1}$ for the HL-LHC~\cite{ZurbanoFernandez:2020cco}.

The LHC value represents the approximate expected dataset by the conclusion of Run~3, combining the recorded Run~2 data with the projected yields of the current run. The latter corresponds to the ultimate target for the HL era, providing a significant increase in statistics. 
Throughout, we use an ALP mass of $m_a=1~\text{MeV}$. In this mass region, ALPs are long-lived and decay outside the detector, thus giving rise to missing energy signatures. 

We consider three benchmark points  BP1, BP2, and BP3 for the ALP-boson couplings: 
\begin{align}
  \text{BP1:} \quad & \frac{c_{a\Phi}}{f_a} = \frac{c_W}{f_a} = \frac{c_G}{f_a} = 1~\text{TeV}^{-1}, \quad \frac{c_B}{f_a} = 0, \label{eq:BP1} \\
  \text{BP2:} \quad & \frac{c_{a\Phi}}{f_a} = \frac{c_W}{f_a} = \frac{c_G}{f_a} = \frac{c_B}{f_a} = 1~\text{TeV}^{-1}, \label{eq:BP2} \\
  \text{BP3:} \quad & \frac{c_{a\Phi}}{f_a} = \frac{c_G}{f_a} = \frac{c_B}{f_a} = 1~\text{TeV}^{-1}, \quad \frac{c_W}{f_a} = 2~\text{TeV}^{-1}. \label{eq:BP3}
\end{align}

In our analysis, hard scattering events are generated with \texttt{MadGraph5\_aMC\@NLO}~\cite{Alwall:2011uj} with the linear ALP \texttt{UFO} file provided in the \texttt{FeynRules} database~\cite{Alloul:2013bka,Brivio:2017ije}.  Parton showering is simulated with \texttt{Pythia8}~\cite{Bierlich:2022pfr}, and detector effects are modeled using \texttt{Delphes3}~\cite{deFavereau:2013fsa} with the \texttt{delphes\_card\_CMS.tcl} card for the CMS detector configuration provided by \texttt{Delphes3}, appropriately modified to account for jet misidentification.

For the parton distribution functions (PDFs) of the proton we use the \texttt{NNPDF23} set~\cite{Ball:2012cx,Buckley:2014ana} and the corresponding value of the strong coupling, $\alpha_s(m_Z)=0.119$. For all processes, we choose the renormalization and factorization scales as the average transverse mass of the hard-process final-state particles,
\begin{equation}
  \mu_R=\mu_F=\frac{1}{N_{\mathrm{hard}}}\sum_{i\in\mathrm{hard}} m_{T,i}\,,\qquad 
  m_{T,i}=\sqrt{m_i^2+p_{T,i}^2}\,,
\end{equation}
where the sum runs over the $N_{\mathrm{hard}}$ particles produced in the hard process , $m_i$ denotes the mass and $p_{T,i}$ the magnitude of the transverse momentum of particle $i$, i.e.\ $p_{T,i} = |\vec{p}_{T,i}|$. In processes involving weak gauge bosons, such as $pp\to a+V+V^\prime$ with subsequent decays of the $V,V^\prime$ bosons, the reconstructed momenta of the gauge bosons are used rather than those of their decay products. At the generator level, we impose minimum transverse momentum requirements of $p_T^{\mathrm{min}}=10~\text{GeV}$ for photons and $p_T^{\mathrm{min}}=20~\text{GeV}$ for jets on the hard-process final-state particles.
This results in an effective lower cutoff value for $\mu_R$ and $\mu_F$ even in process involving only massless particles.
Values for the masses $m_V$ and widths $\Gamma_V$ of the massive SM bosons have been extracted from the particle data group~\cite{ParticleDataGroup:2024cfk} as quoted in Tab.~\ref{tab:sm_parameters}.  

\begin{table}[t]
  \centering
  \begin{tabular}{c|c}
    parameter & value \\
    \hline
    $m_Z$ & $91.1876~\text{GeV}$ \\
    $\Gamma_Z$ & $2.4955~\text{GeV}$\\\hline
    $m_W$ & $80.379~\text{GeV}$ \\
    $\Gamma_W$ & $2.14~\text{GeV}$\\\hline
    $m_H$ & $125.20~\text{GeV}$ \\
    $\Gamma_H$ & $3.7~\text{MeV}$\\\hline
    $G_F$ & $1.1663787\times10^{-5}~\text{GeV}^{-2}$ \\
  \end{tabular}
  \caption{Values of SM parameters used in the simulation~\cite{ParticleDataGroup:2024cfk}.}
  \label{tab:sm_parameters}
  \end{table}
As EW input parameters, $m_Z, m_W$ and the Fermi constant $G_F$ have been used. The corresponding values for the electromagnetic coupling, $\alpha_\text{em}$, and the sine of the weak mixing angle, $\sin\theta_W$, are computed internally using the above parameters. 

Processes with diboson final states at hadron colliders are plagued by large backgrounds, mostly due to processes involving jets with huge production rates. Such jets are occasionally misidentified, mostly because of neutral meson decays within  the jets, where the resulting photons have a negligible angular separation and are misidentified as  isolated photons. These have to be modeled using data-driven methods. 
Other SM backgrounds, for instance from misidentified leptons, are typically many orders of magnitude less prevalent, and thus will be ignored. 
The transfer rate for jets being misidentified as photons at the LHC, which we conservatively assume to remain unchanged for the HL-LHC projections in the absence of significant detector upgrades targeting jet-photon discrimination, 
has been analyzed in Ref.~\cite{ATLAS:2016ukn,Koksal:2023qch}. As a conservative proxy, throughout this paper we are assuming a constant rate of $f_{j\to\gamma}=0.1\%$. 

The optimal signal-background discrimination is identified via binary decision trees (BDTs) using the \texttt{Toolkit for Multivariate Data Analysis} (\texttt{TMVA}) tools provided by \texttt{ROOT}~\cite{TMVA:2007ngy,Brun:1997pa}. 

Using the binned likelihood ratio approach, statistical 95\%~confidence level (\text{C.L.}) bounds can be derived using these BDT distributions. Specifically, for every computed BDT distribution, we calculate the optimized cut that maximizes statistical significance of the signal, and use the numbers of signal and background events above the cut to construct ratios of probabilities. The likelihood function is defined as 
\begin{equation}
  \mathcal{L}(\mu)=\prod\limits_{i=1}^{N_{bins}}\frac{(\mu s_i+b_i)^{n_i}}{n_i!}e^{-(\mu s_i+b_i)},
\end{equation}
where $s_i$ and $b_i$ are the expected numbers of signal and background events in the $i$-th bin, respectively, $n_i$ is the observed number of events in that bin, and $\mu$ is the signal strength parameter. To derive expected exclusion limits in the absence of actual data, we employ the Asimov dataset approach, where the observed event counts are set to the background-only expectation, $n_i = b_i$. This procedure yields median expected limits under the assumption that the true underlying model contains no signal contribution. The test statistic is defined as
\begin{equation}
  q_\mu = -2\ln\frac{\mathcal{L}(\mu)}{\mathcal{L}(\hat{\mu})},
\end{equation}
where $\hat{\mu}$ is the value that maximizes the likelihood. With these, the confidence level for the signal, $\text{CL}_s$,  is computed as
\begin{equation}
  \text{CL}_s = \frac{P_{s+b}(q_\mu\geq q_\mu^{obs}|\mu)}{1-P_{b}(q_\mu\leq q_\mu^{obs}|0)},
\end{equation}
where $P_{s+b}$ and $P_b$ are the probability distributions of the test statistic under the signal-plus-background and background-only hypotheses, respectively, and the notation $P(\text{condition}|\mu)$ denotes the probability that the condition is satisfied, given a signal strength $\mu$. The 95\cl~upper limit on the signal strength $\mu$ is determined by finding the value of $\mu$ for which $\text{CL}_s = 0.05$
Let us now turn to a discussion of the individual final states of our analysis.

In the following, when we refer to a background process by its production label (e.g.\ $ZZ$), we implicitly assume the specific decay channel that is considered in the relevant signal process (e.g.\ $ZZ\to 4\ell$ in the case of the $4\ell$ decay mode of the  $\azz$ signal process).

\subsection{The $\agg$ final state}
\label{sec:agg}

\begin{figure}[!t]
  \centering
  \begin{subfigure}{0.45\textwidth}
    \includegraphics[width=\textwidth]{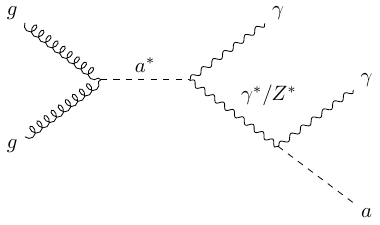}
    \subcaption{}
    \label{fig:gg-agg_s}
  \end{subfigure}
  \hfill
  \begin{subfigure}{0.45\textwidth}
    \includegraphics[width=\textwidth]{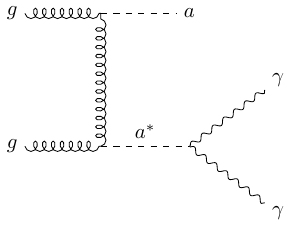}
   \subcaption{}
   \label{fig:gg-agg_u}
 \end{subfigure}

  \begin{subfigure}{0.45\textwidth}
    \includegraphics[width=\textwidth]{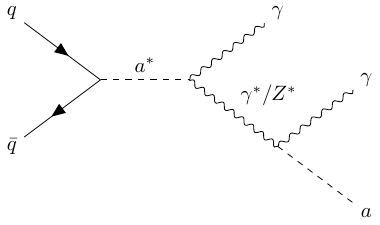}
   \subcaption{}
   \label{fig:qq-agg_s}
  \end{subfigure}
  \hfill
  \begin{subfigure}{0.45\textwidth}
    \includegraphics[width=\textwidth]{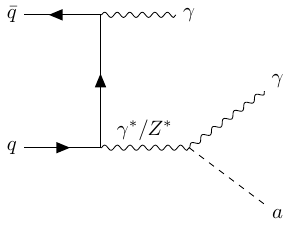}
    \subcaption{}
    \label{fig:qq-agg_u}
  \end{subfigure}
  \caption{Representative Feynman diagrams for the $\agg$ production process.  \label{fig:agg-diagrams}
  }
\end{figure}

In hadronic collisions the production of a $\agg$ final state at tree level can proceed either via the exchange of a virtual ALP in gluon-gluon or quark-anti-quark annihilation processes, or via QED interactions of SM particles that in turn give rise to an $\agg$ final state. Representative diagrams for each of these classes are shown in Figs.~\ref{fig:agg-diagrams}\subref{fig:gg-agg_s}--\subref{fig:qq-agg_s} and Fig.~\ref{fig:agg-diagrams}\subref{fig:qq-agg_u}, respectively.

Note that these two classes of production mode differ by the number of ALP couplings involved. For our analysis we take both into account.  

In each case we consider diagrams involving the exchange of virtual $Z$~bosons or photons. 
However,  $Z$-exchange contributions  are numerically strongly suppressed because of the large mass of the $Z$~boson. Moreover, the ALP coupling to an EW boson occurring in this kind of diagrams is predominantly photonic, resulting in a dependence of the relevant cross section on the ALP-photon coupling $g_{\agg}$ of Eq.~\eqref{eq:g-aVV}. Because of the dependence of $g_{\agg}$ on both $c_B$ and $c_W$, it is not possible to decouple these two Wilson coefficients from each other in the $\agg$ production process. 

As apparent from the diagrams of Fig.~\ref{fig:agg-diagrams}, subprocesses with gluon-gluon and quark-anti-quark initial states contribute to the $\agg$ production process, with the gluon channels providing the dominant contribution to the signal cross section.
The $\agg$ production mode is thus very sensitive to the ALP-gluon coupling. 
The dominance of the gluon-induced channels is due to a combination of factors: the larger gluon luminosity at the LHC compared to the quark luminosity, and the derivative nature of the ALP-gluon interaction operator of Eq.~\eqref{AGtilde}, which enhances contributions at higher partonic energies. Therefore, even if we assumed all the Wilson coefficients to be non-zero, gluonic contributions would dominate over quark-induced ones at high energies. The analysis of the $\agg$ process thus allows us to constrain the operator coefficients in the $\{c_G/f_a,c_{a\Phi}/f_a,g_{a\gamma\gamma}\}$ space.  
In the limit $c_G \to 0$, only quark-initiated contributions remain, whereas for $c_{a\Phi} \to 0$, gluon-initiated channels continue to dominate.

The statistically dominant backgrounds to $pp\to \agg$ are the $jj, j\gamma, \gamma\gamma$, and $\gamma\gamma Z(Z\to\nu\bar{\nu})$ production processes, 
where the jets are assumed to be misidentified as photons. As mentioned before, lepton misidentification is ignored due to the negligible contribution of the corresponding channels.  
In the histograms below, contributions from diagrams with either one or three new physics couplings are shown separately to illustrate the relevance of each class. We classify contributions by the number $N_a$ of ALPs in each diagram. Diagrams with $N_a=1$ contain a single ALP coupling vertex and one external ALP, whereas diagrams with $N_a=2$ include two ALPs,  resulting in three ALP coupling vertices and one additional ALP propagator. Interference effects between these types of contributions are negligible. Even though the ALP couplings are assumed to be much smaller than the SM couplings, $N_a=2$ contributions dominate in most scenarios with a non-vanishing ALP-gluon coupling, due to the derivative gluonic interactions being enhanced in the energy range probed at the LHC. As the ALP-gluon coupling goes to zero, $N_a=1$ diagrams become more relevant. 

For our analysis, we only consider events with exactly two photons after detector simulation fulfilling minimum requirements on transverse momentum, $\ptgi$, and pseudorapidity, $\eta_{\ga_i}$ for $i=1,2$,  
\beq 
\label{eq:agg-cuts1}
\ptgi>50~\GeV\,,\quad 
|\eta_{\ga_i}|<2.5\,. 
\eeq
Additionally, the two photons have to be well separated in the pseudorapidity-azimuthal angle plane, 
\beq 
\label{eq:agg-cuts2}
\drgg =  \sqrt{(\eta_{\ga_1}-\eta_{\ga_2})^2 + (\phi_{\ga_1}-\phi_{\ga_2})^2} > 0.4\,, 
\eeq
and invariant mass, 
\beq 
\label{eq:agg-cuts3}
\mgg>20~\GeV\,. 
\eeq

\begin{figure}[!t]
  \centering
  \begin{subfigure}{0.45\textwidth}
    \includegraphics[width=\textwidth]{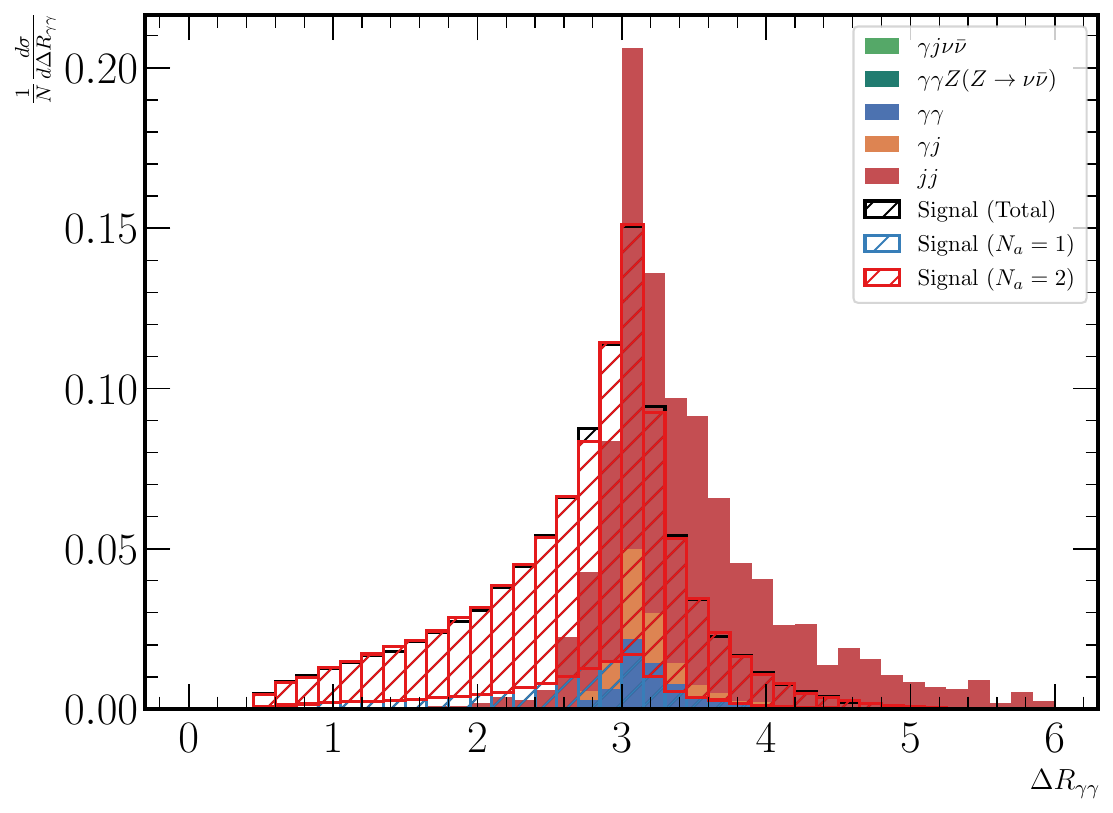}
    \subcaption{}
    \label{fig:dr}
  \end{subfigure}
  \hfill
  \begin{subfigure}{0.45\textwidth}
    \includegraphics[width=\textwidth]{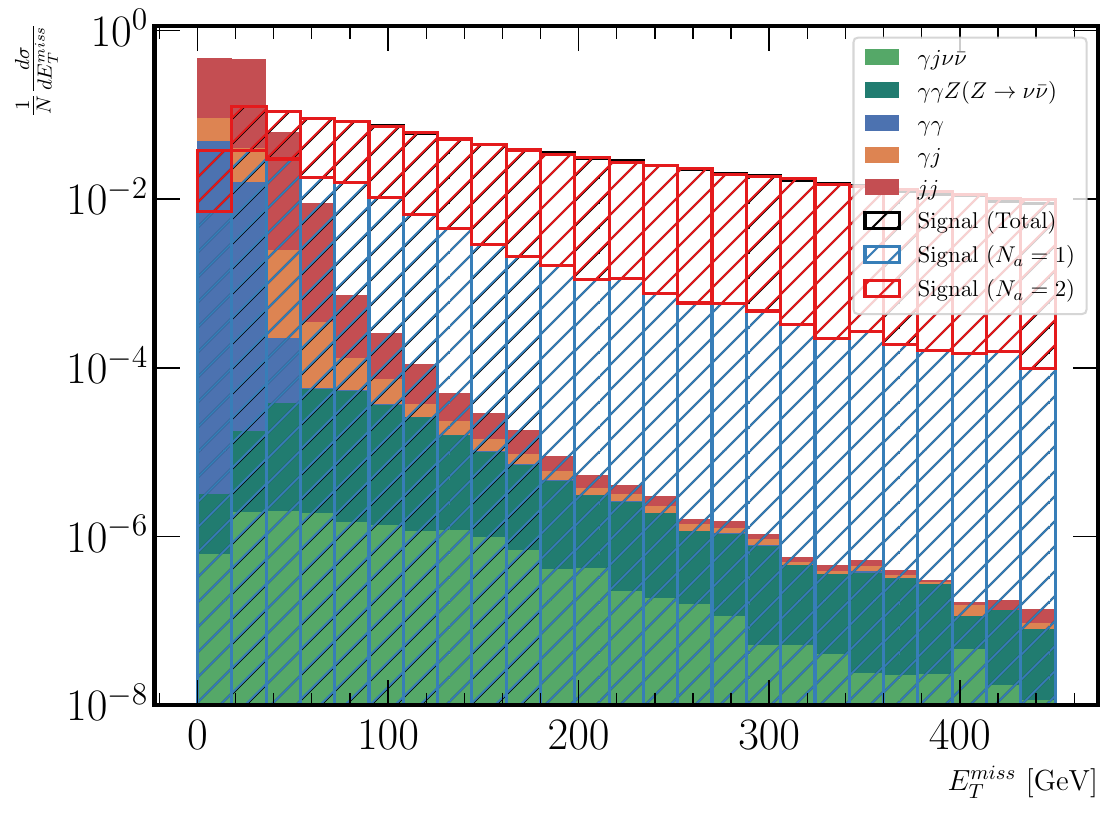}
    \subcaption{}
    \label{fig:met}
  \end{subfigure}
  \\[5mm]
  \begin{subfigure}{0.45\textwidth}
    \includegraphics[width=\textwidth]{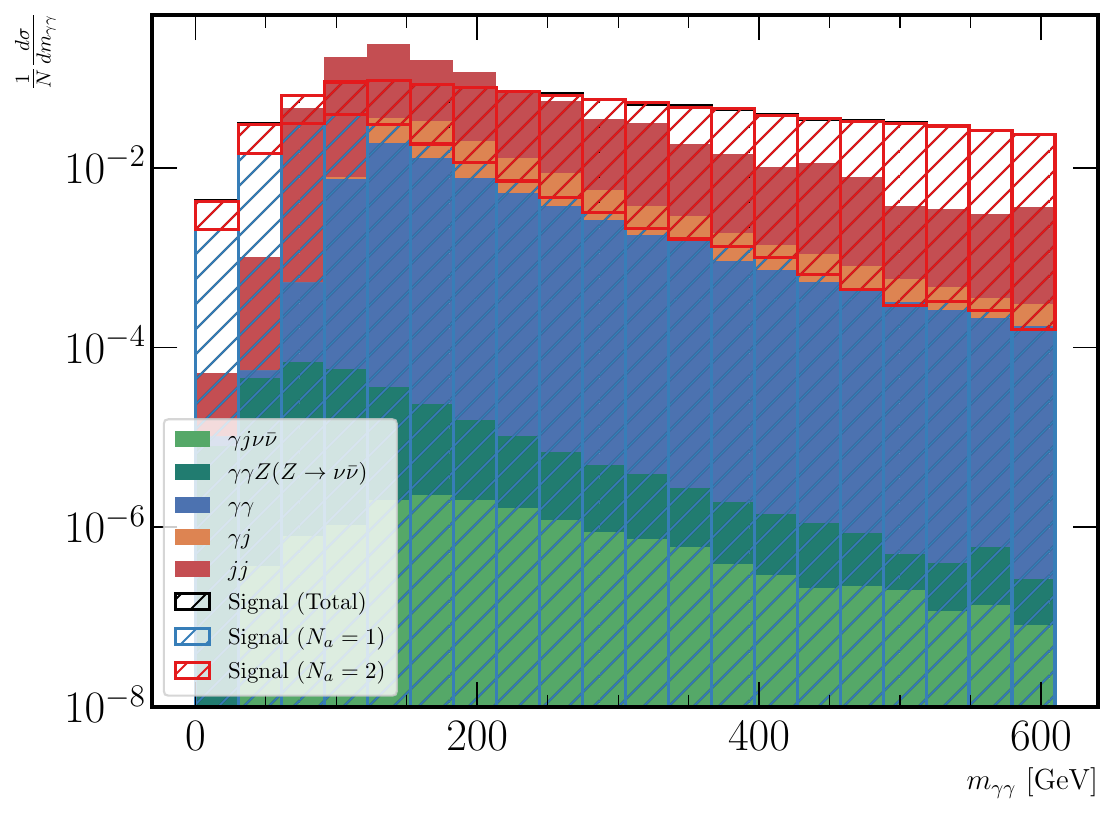}
    \subcaption{}
    \label{fig:mgg}
  \end{subfigure}
  \hfill 
  \begin{subfigure}{0.45\textwidth}
    \includegraphics[width=\textwidth]{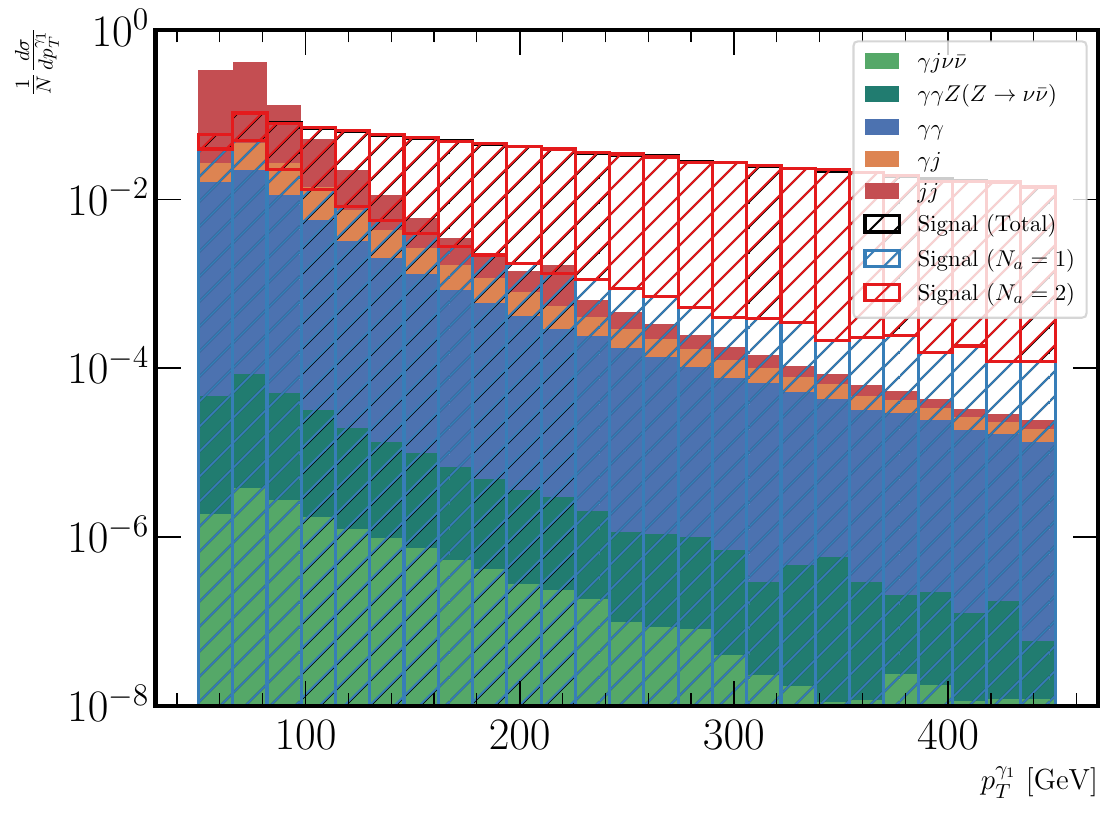}
    \subcaption{}
    \label{fig:pt}
  \end{subfigure}
  \caption{Normalized distributions for the simulated signal and background contributions to the $a\gamma\gamma$ channel at the LHC with $\sqrt{s}=14$~TeV using the benchmark point BP1 specified in Eq.~\eqref{eq:BP1}.  
  In each case the hatched histograms represent the signal scenarios with $N_a=1$~(blue), $N_a=2$~(red) and their sum~(black), whereas the filled histograms represent the $\ga j \nu\bar{\nu}$~(light green), $\gamma\gamma Z(Z\to\nu\bar{\nu})$~(dark green), $\ga\ga$~(blue), $\ga j$~(orange), and $jj$~(red) background contributions.}
\label{fig:agg-sig-back}
\end{figure}

In Fig.~\ref{fig:agg-sig-back}  

several kinematic distributions are presented for the $\agg$ signal and the most relevant background processes. 
In Fig.~\ref{fig:dr} we show the normalized distribution of the angular separation between the two photons. The signal exhibits a broader distribution extending to smaller values of $\drgg$, indicating that the photons tend to be more collimated in signal than in background contributions.
Figure~\ref{fig:met} displays the missing transverse energy, $\etmiss$, computed as the magnitude of the negative vector sum of the transverse momenta of all reconstructed objects (i.e.\ photons, jets, charged leptons) in an event. This observable captures contributions from all undetected particles, including the ALP. The missing transverse energy distribution exhibits a pronounced tail for the signal extending to high values, reflecting the escaping ALP, while the backgrounds peak at lower values of $\etmiss$. This variable provides the strongest discrimination between signal and background.
The diphoton invariant mass distribution is shown in Fig.~\ref{fig:mgg}. The signal distribution peaks at lower values and shows a slightly flatter tail than the backgrounds, which exhibit a broader spectrum. Out of the shown observables, this one provides the least discriminating power.
In Fig.~\ref{fig:pt}, we show the transverse momentum of the leading photon, $\ptgone$. For this distribution, the signal develops a flatter tail at high transverse momentum values than the backgrounds, providing additional discriminating power.

\begin{figure}[!t]
  \centering
    \includegraphics[width=0.65\textwidth]{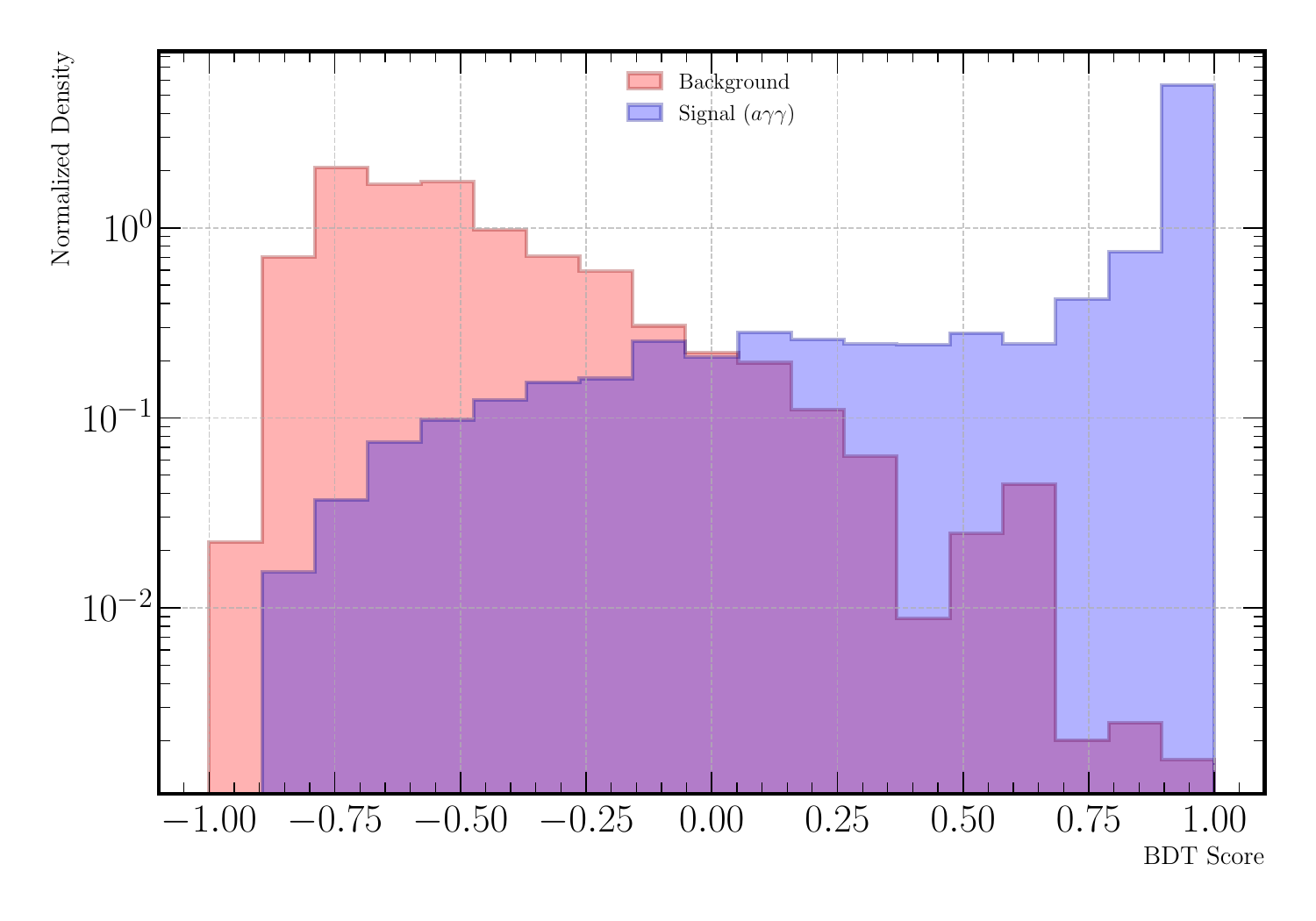}
\caption{BDT score distributions of the $\agg$~signal versus the sum of the $jj$, $j\gamma$, $\gamma\gamma$, $\gamma j \nu\bar{\nu}$ and $\gamma\gamma Z(Z\to\nu\bar{\nu})$ background processes for the benchmark point BP1 specified  in Eq.~\eqref{eq:BP1}, with both histograms normalized to unit area. } 
\label{fig:agg_a}
\end{figure}

To extract the maximum amount of information, we choose the variables $\etmiss$, $\ptgone$, $\mgg$, $\drgg$, $\eta_{\gamma_1}$ 
to train binary decision trees. Considering individual variables, as shown in  Fig.~\ref{fig:agg-sig-back}, the missing transverse energy distribution provides the highest discriminating power, followed by the transverse  momentum of the leading photon. 
For the ALP-model BP1 benchmark point already used above, the resulting BDT graph is shown in Fig.~\ref{fig:agg_a}.  
%
%%%%%%%%%%%%%
%
%
%%%%%%
%
The separating power obtained by the BDT algorithm is listed for the various input variables in Tab.~\ref{tab:bdt_var_separation}. 
%
%%%%%
%
\begin{table}[t!]
  \centering
  \begin{tabular}{c||c|c|c|c|c}
    variable & $\etmiss$ & $\ptgone$ & $\Delta R_{\gamma\gamma}$ & $m_{\gamma\gamma}$ & $\eta_{\gamma_1}$ \\
    \hline
    separation & 0.495 & 0.195 & 0.165 & 0.098 & 0.048 \\
  \end{tabular}
  \caption{BDT input variables and their separation power (TMVA ranking).}
  \label{tab:bdt_var_separation}
\end{table}
%
%%%%
%
Here, the separation of each variable is defined as the difference in the mean values of the signal and background distributions for that variable, normalized by the sum of their root-mean-square (RMS) values, showing the discriminatory power of each variable.

\begin{figure}[!t]
  \centering
  \begin{subfigure}{0.45\textwidth}
    \includegraphics[width=\textwidth]{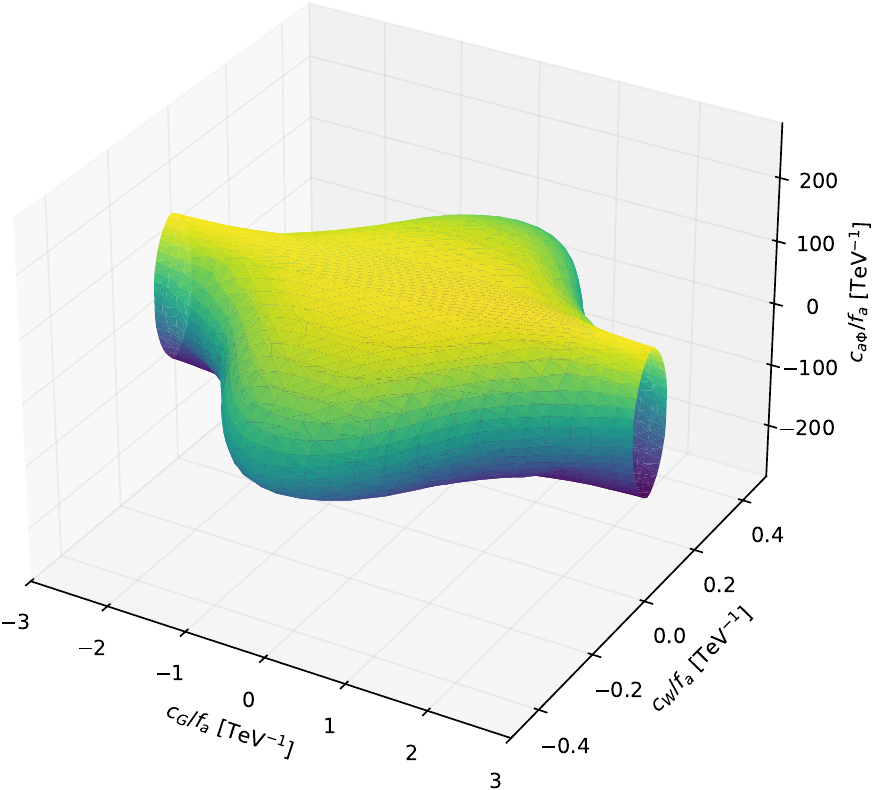}
    \caption{}
    \label{fig:agg_b}
  \end{subfigure}
  \hfill
  \begin{subfigure}{0.45\textwidth}
    \includegraphics[width=\textwidth]{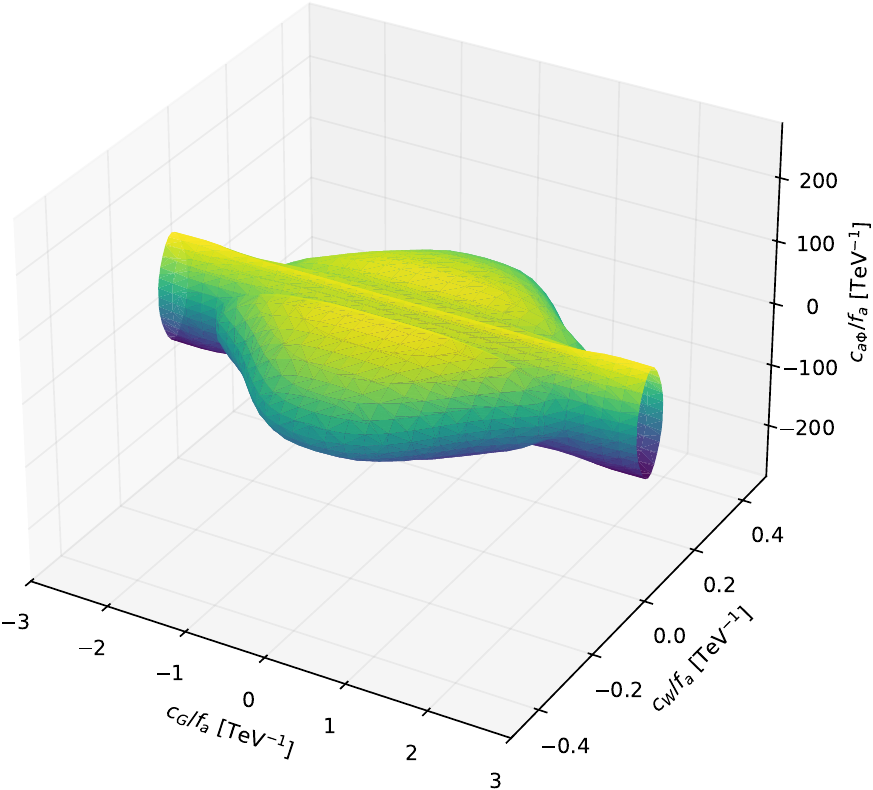}
    \caption{}
    \label{fig:agg_c}
  \end{subfigure}
\caption{Results for 95\cl~contours of the \texttt{TMVA} analysis of the BDT score distributions at the LHC with an integrated luminosity of $\int L=450~\text{fb}^{-1}$ (left) and at the HL-LHC with  $\int L=3000~\text{fb}^{-1}$ for the benchmark point BP1 of Eq.~\eqref{eq:BP1}.
}
\label{fig:agg-limits}
\end{figure}

The resulting exclusion limits are shown in Fig.~\ref{fig:agg-limits}. 
%
%%%%%%%%%%%%%
%
%
%%%%%%
%
As expected from the fact that every diagram containing gluons also contains ALP-photon couplings, $c_G/f_a$ becomes unconstrained for vanishing $g_{\agg}$. 
This behavior is limited by the ALP-Higgs coupling, since it produces diagrams with $N_a=1$ depending solely on $c_{a\Phi}$. The resulting contour is mirror-symmetric in each direction, and has an ellipsoid shape in its center. For larger values of the ALP-gluon coupling the boundaries converge to a three-dimensional slab extending along the $c_G$~direction. This behavior persists for both LHC and HL-LHC luminosities with significantly stronger constraints in the latter case.

%%%%%%%%%%%%%%%%%%%%%%%%%%
\subsection{The $\awg$ final state}
\begin{figure}[!t]
  \centering
  \begin{subfigure}{0.45\textwidth}
    \includegraphics[width=\textwidth]{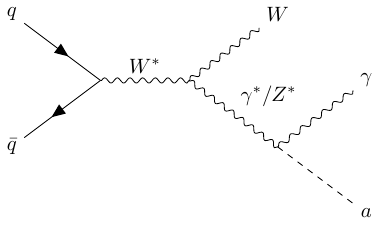}
    \caption{}
    \label{fig:2_tree_s}
  \end{subfigure}
  \hspace{2cm}
  \begin{subfigure}{0.20\textwidth}
    \includegraphics[width=\textwidth]{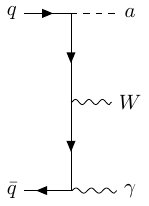}
    \caption{}
    \label{fig:2_tree_u}
  \end{subfigure}
  \caption{Some of the relevant Feynman diagrams for $aW\gamma$ production. 
  }
  \label{fig:Wphoton-diagrams}
\end{figure}

We now consider the case where an ALP is produced in association with a $W^\pm$ boson and a photon. Although the $W^\pm$ boson prefers to decay into a quark-antiquark pair, we consider only the leptonic decay channels with the $W^\pm$~boson decaying into a charged lepton $\ell^\pm$ and a neutrino of either of the first two generations, i.e.\ $\ell^\pm\in\{e^\pm,\mu^\pm\}$. Thus, the final state consists of one isolated lepton, one isolated photon and missing energy from the undetected ALP and neutrino, resulting in an $\ell^\pm \gamma + E_T^{miss}$ signature. Representative Feynman diagrams are shown in Fig.~\ref{fig:Wphoton-diagrams}. 
%
%%%%%
%
%
%%%%
%
We note that, at tree-level, only diagrams with one ALP coupling occur. 
Different from the diphoton case, both the $c_W$ and $c_B$ couplings contribute independently to the electroweak vertices, which allows us to decouple them from each other. We then combine the resulting constraints with our results from the $\agg$ channel.  
Despite the $\awg$ process including the $W^\pm$~boson as a final state particle, which couples only via $c_W$ to the ALP, because of contributions from off-shell intermediate $Z$ bosons, the process exhibits a dependence on $c_B$. Furthermore, due to the final state being electromagnetically charged, there are no diagrams containing gluonic partons. Therefore, the resulting constraints are independent of $c_G$. 

The dominant backgrounds for the $\ell^\pm \gamma + E_T^{miss}$ signature are constituted by $W^\pm\gamma$ production within the SM %(irreducible) 
and $W^\pm+\text{jets}$ events where a jet is misidentified as a photon. %(reducible). 
Other potential backgrounds include $t\bar{t}\gamma$ production, where the top quark decays produce real $W^\pm$~bosons and photons, and $Z\gamma$ events where one lepton from the $Z\to\ell^+\ell^-$ decay is lost or fails reconstruction criteria. The $t\bar{t}\gamma$ contribution is effectively suppressed by imposing a veto on $b$-tagged jets and restricting the total jet multiplicity. Similarly, contributions from $Z\gamma$ and QCD multijet events are negligible after requiring a high-quality, isolated charged lepton and substantial missing transverse energy. Consequently, we focus on the $W^\pm\gamma$ and $W^\pm+\text{jets}$ channels. The latter is conservatively estimated using the jet-to-photon misidentification rate $f_{j\to\gamma}=0.1\%$ introduced above. 

In addition to the variables employed in the analysis of the $\agg$ final state, i.e.\ $E_T^{miss}$, $\ptg$, and $\eta_\gamma$, we now also use the transverse momentum and pseudorapidity of the charged lepton, $\ptl$ and $\eta_{\ell}$, and quantities related to the momentum of the $W^\pm$~boson. 
In the SM backgrounds, the missing transverse momentum $\ptmiss$ mostly stems from the neutrino resulting from $W^\pm$~boson decays, while in the $\awg$ signal it receives an additional contribution from the invisible ALP. Consequently, we do not attempt to  reconstruct the true momentum of the $W^\pm$~boson event-by-event, but define a proxy for the system consisting of the charged lepton and invisible particles, referred to as $\linv$, with a transverse mass 
\beq 
\label{eq:mtlinv}
\mtlinv = \sqrt{2\, \ptl \, E_T^{miss} \, (1 - \cos\Delta\phi_{\ell,\text{miss}})}\,,
\eeq 
using the angular separation $\Delta\phi_{\ell,\text{miss}}$ between the  lepton and the invisible particle system. 
The transverse momentum of the $(\linv)$ system is computed from the sum $\vec{p}_T^{\;\linv} = \vec{p}_T^{\;\ell} + \ptmiss$.
From $\vec{p}_T^{\;\linv}$ we then infer 
the azimuthal angle separation between the $(\linv)$ system and the photon, $\dplinvg$,  
and the transverse mass of the $(\linv)+$photon system, which is defined as 
\beq
m_{T}^{\linv,\gamma}= \sqrt{2\, p_T^{\;\linv} \, p_T^\gamma \, \bigl(1 - \cos\dplinvg\bigr)}\,.
\eeq

For our $\awg$ analysis we add contributions from the $\awpg$ and $\awmg$ final states and apply preselection cuts requiring exactly one isolated lepton and exactly one isolated photon, satisfying 
\bea
\ptl>25~\text{GeV},  \quad |\etal|<2.5,\quad 
\ptg>25~\text{GeV}, \quad |\etag|<2.5\,.
\eea
Additionally, we impose the following selection criteria:
\beq
\label{eq:awg-cuts}
  E_T^{miss} > 30~\text{GeV}, \quad
  \mtlinv > 30~\text{GeV}, \quad 
  \Delta R_{\ell,\gamma} > 0.4, 
\eeq 
with $\Delta R_{\ell,\gamma}$ denoting the separation of the charged lepton from the photon in the pseudorapidity-azimuthal angle plane. 

\begin{figure}[!t]
  \centering
  
  \begin{subfigure}{0.45\textwidth}
    \includegraphics[width=\textwidth]{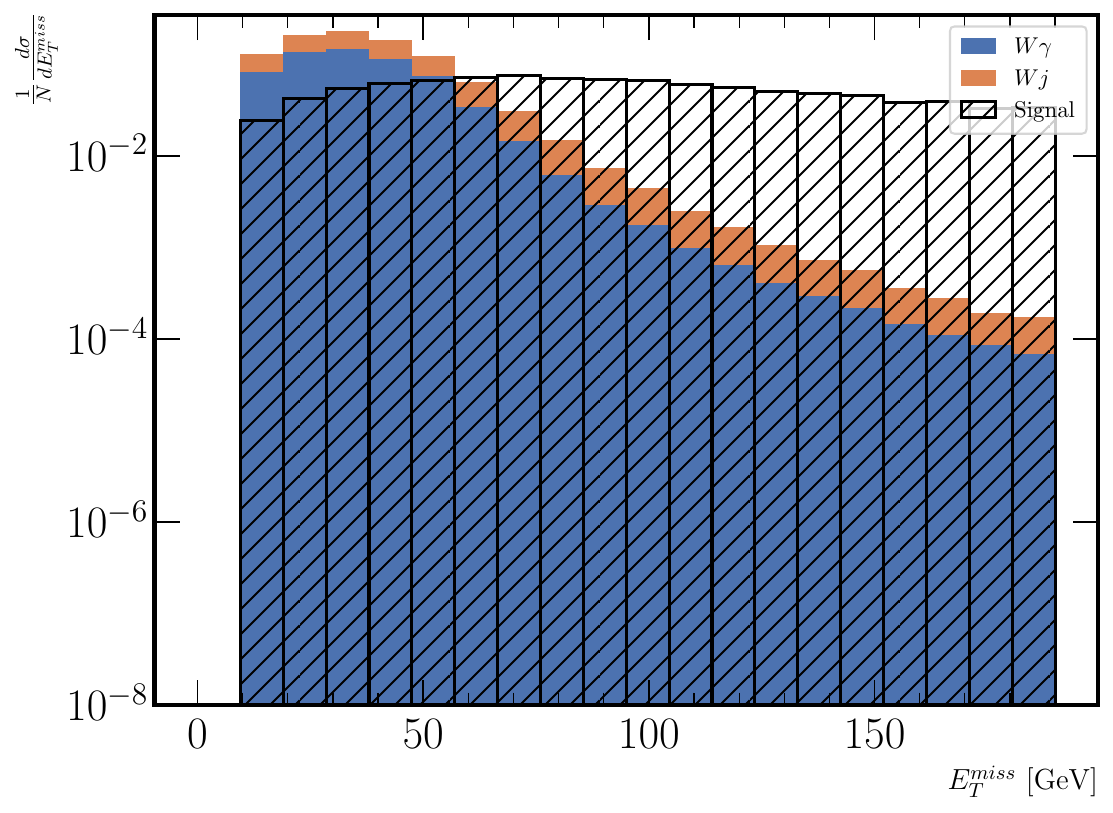}
    \caption{}
    \label{fig:E_T_awg}
  \end{subfigure}
  \hfill
  \begin{subfigure}{0.45\textwidth}
    \includegraphics[width=\textwidth]{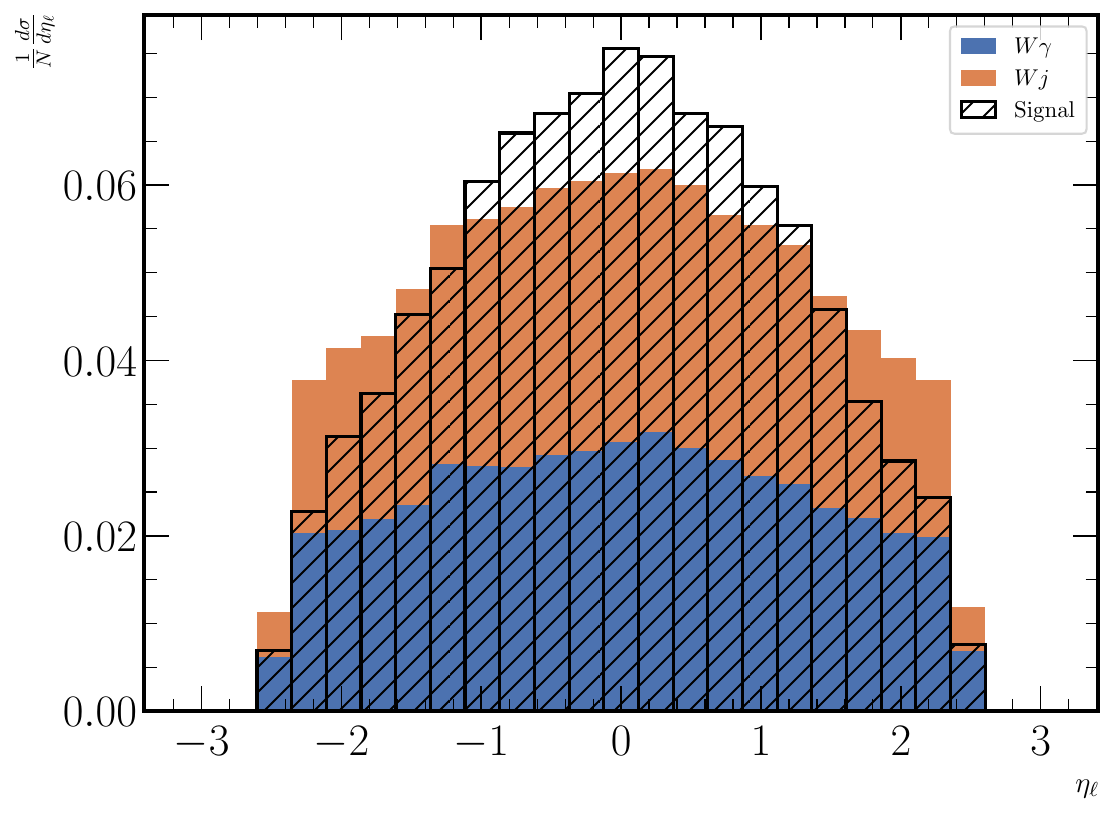}
    \caption{}
    \label{fig:eta_ell_awg}
  \end{subfigure}
  \\[5mm]
  \begin{subfigure}{0.45\textwidth}
    \includegraphics[width=\textwidth]{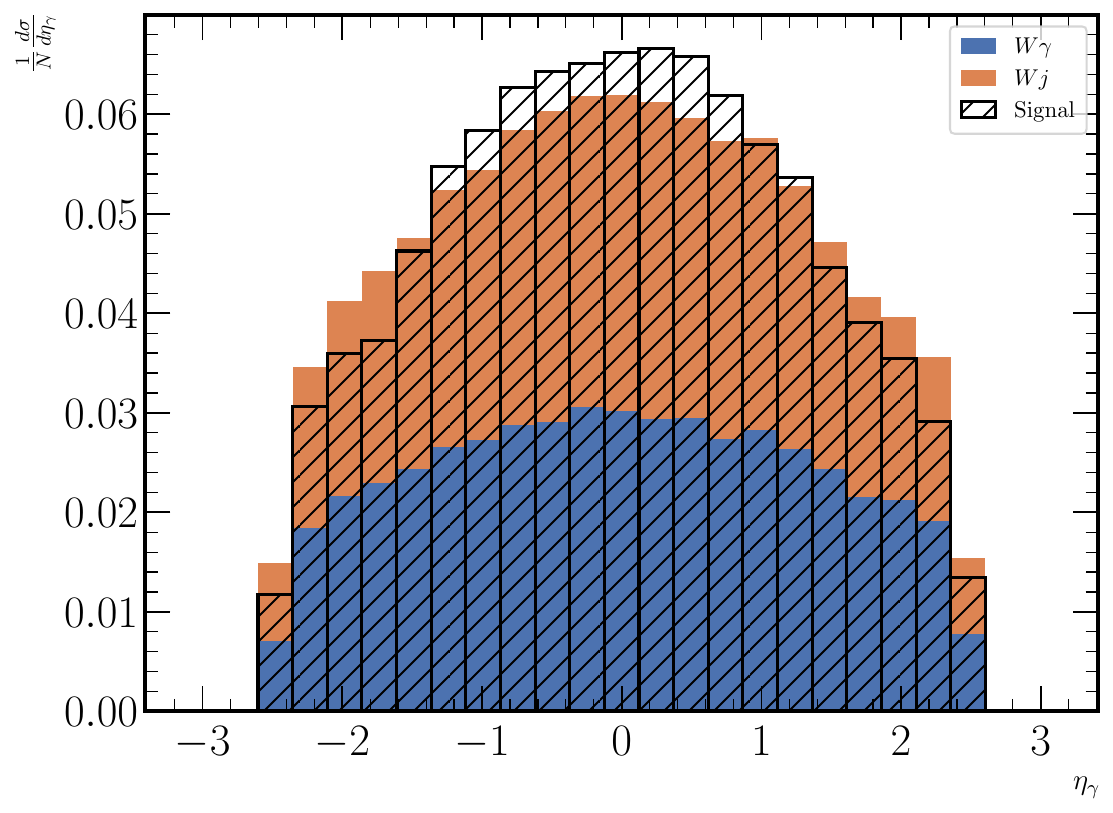}
    \caption{}
    \label{fig:eta_gamma_awg}
  \end{subfigure}
  \hfill 
  \begin{subfigure}{0.45\textwidth}
    \includegraphics[width=\textwidth]{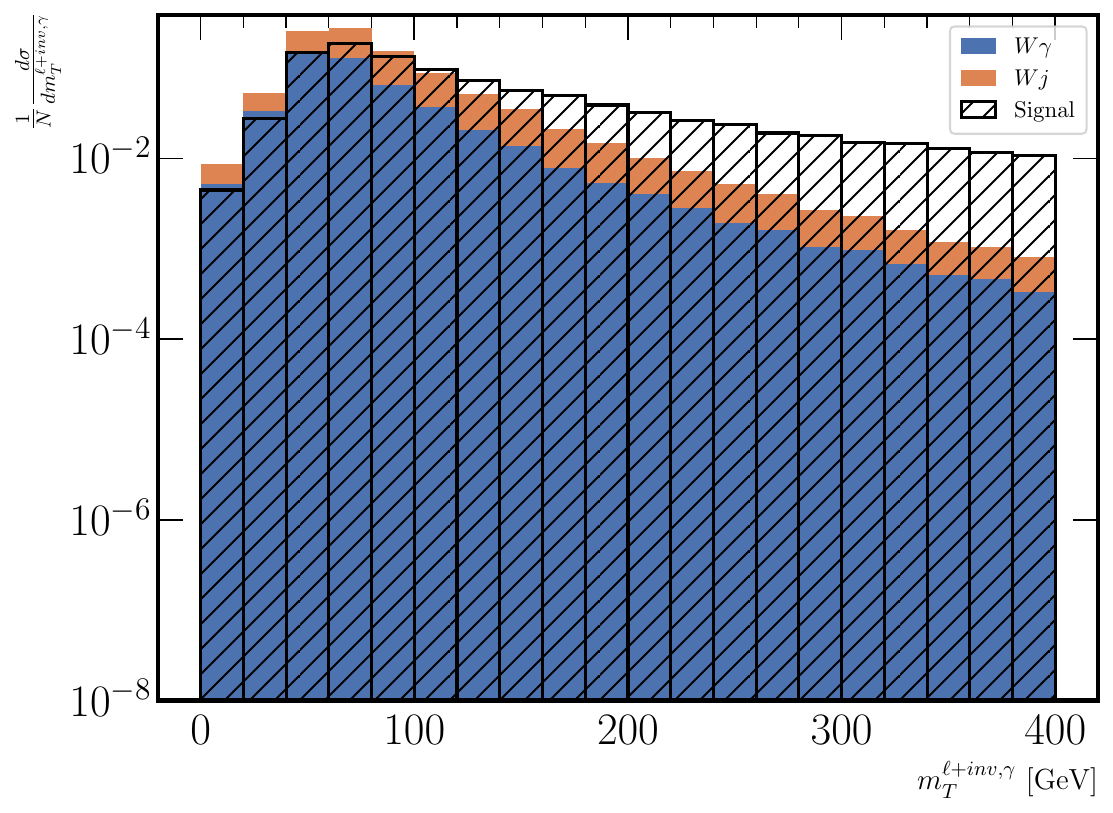}
    \caption{}
    \label{fig:m_W_gamma_awg}
  \end{subfigure}
  \\[5mm]
  \begin{subfigure}{0.45\textwidth}
    \includegraphics[width=\textwidth]{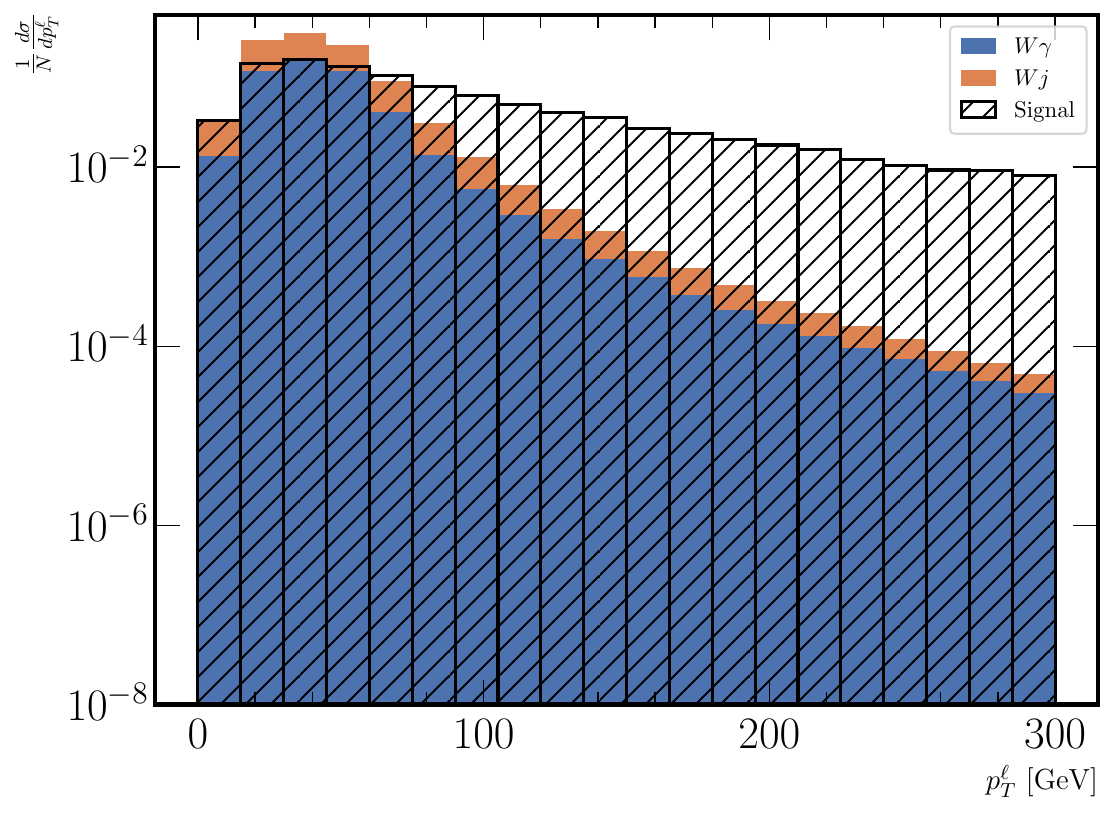}
    \caption{}
    \label{fig:p_T_ell_awg}
  \end{subfigure}
  \hfill 
  \begin{subfigure}{0.45\textwidth}
    \includegraphics[width=\textwidth]{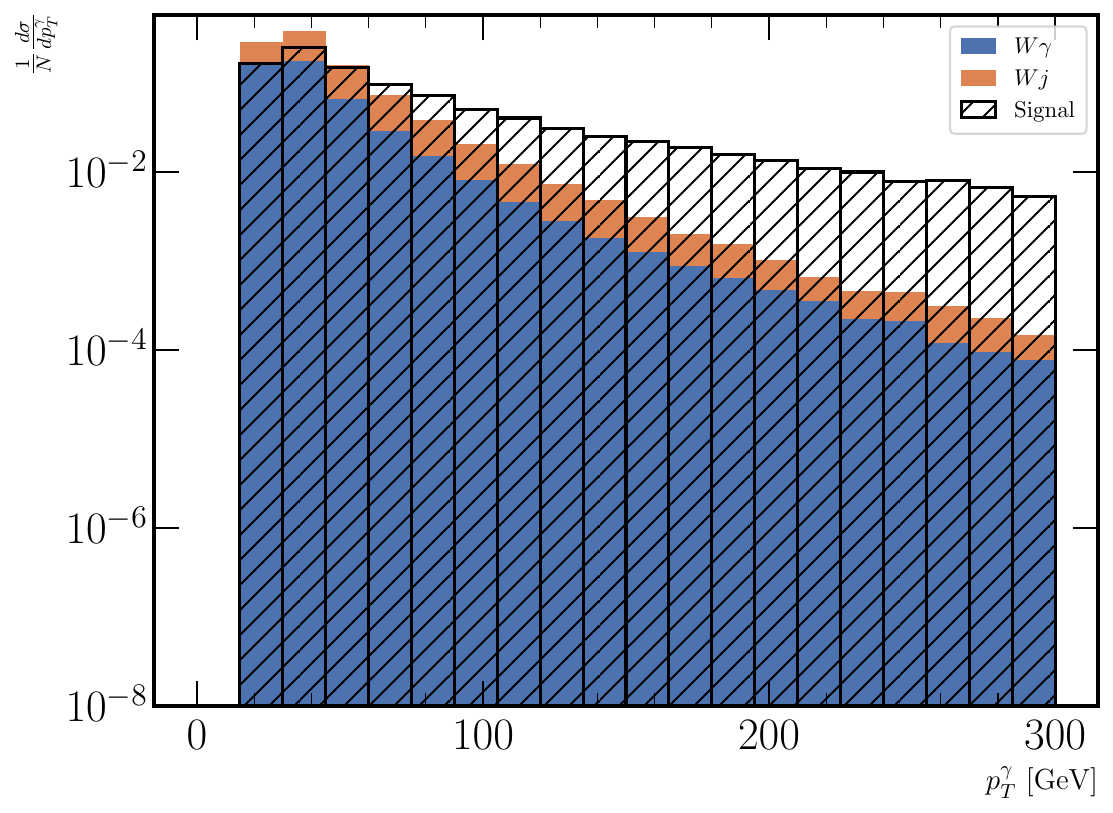}
    \caption{}
    \label{fig:p_T_gamma_awg}
  \end{subfigure}
  \caption{Normalized distributions for the simulated signal and background contributions to the $aW\gamma$ channel {at the LHC with $\sqrt{s}=14$~TeV} using the benchmark point BP2 specified in Eq.~\eqref{eq:BP2}. The hatched histograms represent the signal, whereas the filled histograms represent the $W\gamma$~(blue) and $Wj$~(orange) background contributions.}
  \label{fig:agg-sig-back_awg}
\end{figure}

Individual distributions are shown in Fig.~\ref{fig:agg-sig-back_awg}. 
In Fig.~\ref{fig:E_T_awg} we show the normalized distribution of the missing transverse energy which receives contributions from the undetected ALP and the neutrino stemming from the leptonic decay of the $W^\pm$~boson. Consequently, and as expected from the previous discussion, the signal distribution spans a broader range towards higher $E_T^{miss}$ values compared to the backgrounds.
The pseudorapidity of the charged lepton is shown in Fig.~\ref{fig:eta_ell_awg}. Signal and background events populate the central region, with only a mild shape difference, such that this observable provides limited separation power. The signal distribution is more centered, while the background is slightly more spread out. Similarly, Fig.~\ref{fig:eta_gamma_awg} displays the pseudorapidity of the photon, which is dominantly produced in the central detector region for both signal and backgrounds, and thus has less discriminating power than $\eta_\ell$.
Figure~\ref{fig:m_W_gamma_awg} depicts the transverse mass distribution $\mtlinvg$, which exhibits a peak around $80~\GeV$ and decreases slightly more slowly for the signal than the backgrounds towards larger values of $\mtlinvg$.
Finally, Figs.~\ref{fig:p_T_ell_awg} and \ref{fig:p_T_gamma_awg} show the transverse momentum distributions of the charged lepton and the photon, respectively. While the low-$p_T$ regions are dominated by the SM backgrounds, the signal develops slightly wider tails towards large transverse momenta, consistent with the higher momentum flow induced by the additional invisible particle in the final state, similar to the $\etmiss$ distribution.

\begin{figure}[!tp]
  \centering

      \includegraphics[width=0.65\textwidth]{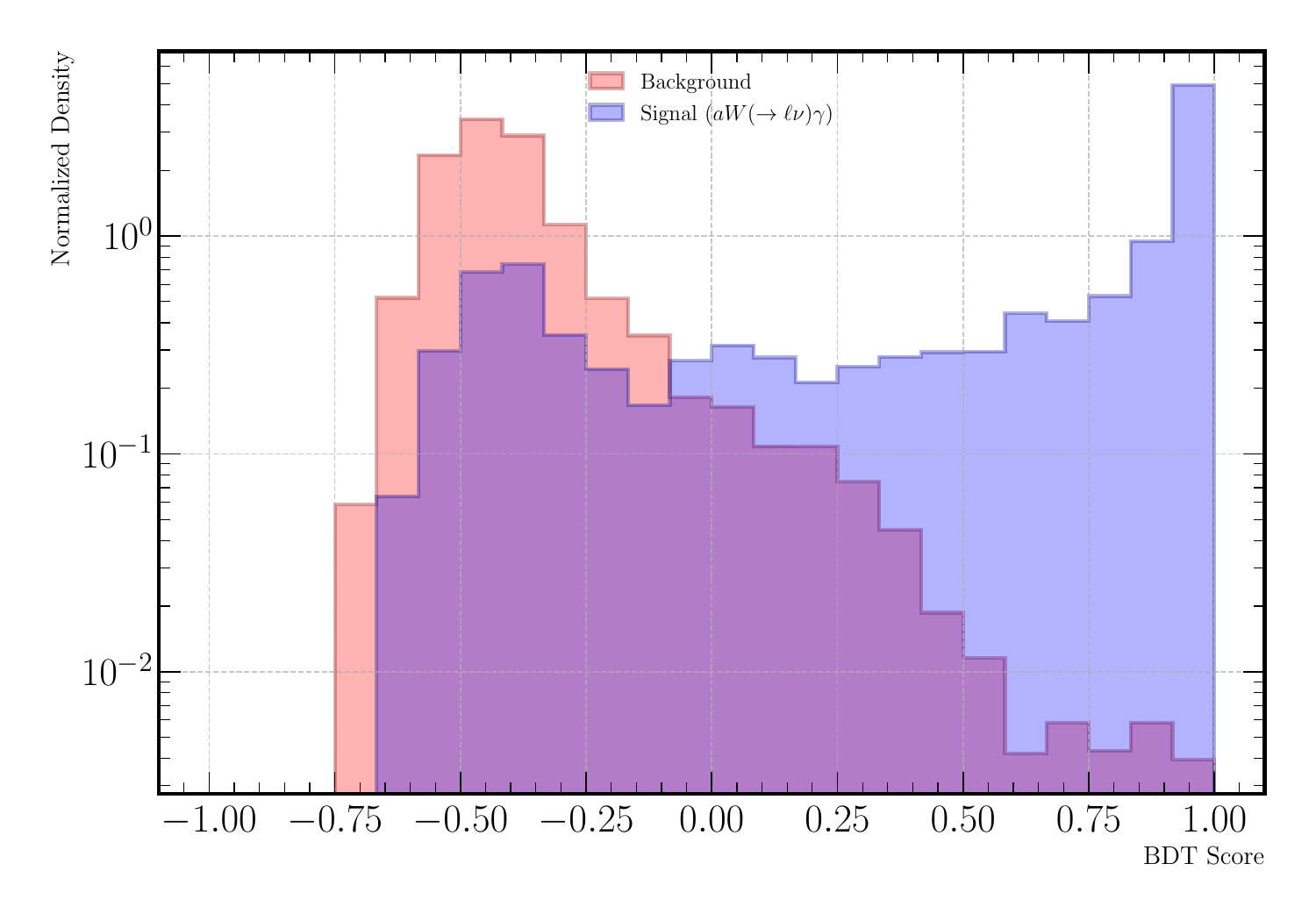}
    \caption{
      Normalized BDT score distributions of the $aW\gamma$ signal versus the sum of the $W\gamma$ and $Wj$ background processes for the benchmark point BP2 of Eq.~\eqref{eq:BP2}. }
\label{fig:awg_a}
\end{figure}

The separation power of the various variables is listed in Tab.~\ref{tab:bdt_var_separation_awg}, and 
the BDT score distribution for the $aW\gamma$ channel is shown in Fig.~\ref{fig:awg_a}. The signal and background distributions exhibit clear separation across the entire BDT score range. The background distribution peaks at low BDT scores and shows a rapid decline towards higher values, while the signal distribution extends broadly across the score range and rises sharply as the score approaches unity. This behavior indicates strong discriminating power of our chosen observables. 
%
%%%%%%
%
%
%%%%%%%
%
\begin{table}[t!]
  \centering
  \begin{tabular}{c||c|c|c|c|c|c|c}
    variable & $E_T^{miss}$ & $\ptl$ & $\mtlinvg$ & $\ptg$ & $\dplinvg$ & $\eta_{\ell}$ & $\eta_{\gamma}$ \\
    \hline
    separation & 0.4574 & 0.2947 & 0.1344 & 0.0858 & 0.0170 & 0.0133 & 0.0103 \\
  \end{tabular}
  \caption{BDT input variables and their separation power for the $aW\gamma$ channel.}
  \label{tab:bdt_var_separation_awg}
\end{table} 
%
%%%%
%
%
%%%%%%%%%%
%
%
%%%%%%%%%
%

\begin{figure}[!tp]
  \centering
    \begin{subfigure}{0.45\textwidth}
      \includegraphics[width=\textwidth]{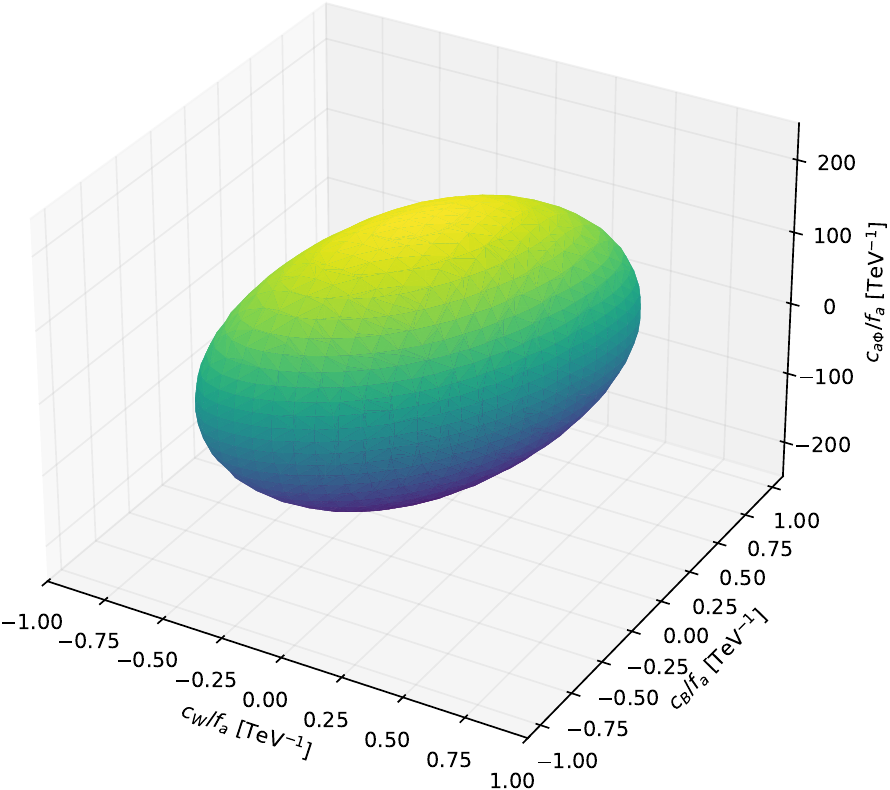}
      \caption{}
      \label{fig:awg_b}
    \end{subfigure}
    \hfill
    \begin{subfigure}{0.45\textwidth}
      \includegraphics[width=\textwidth]{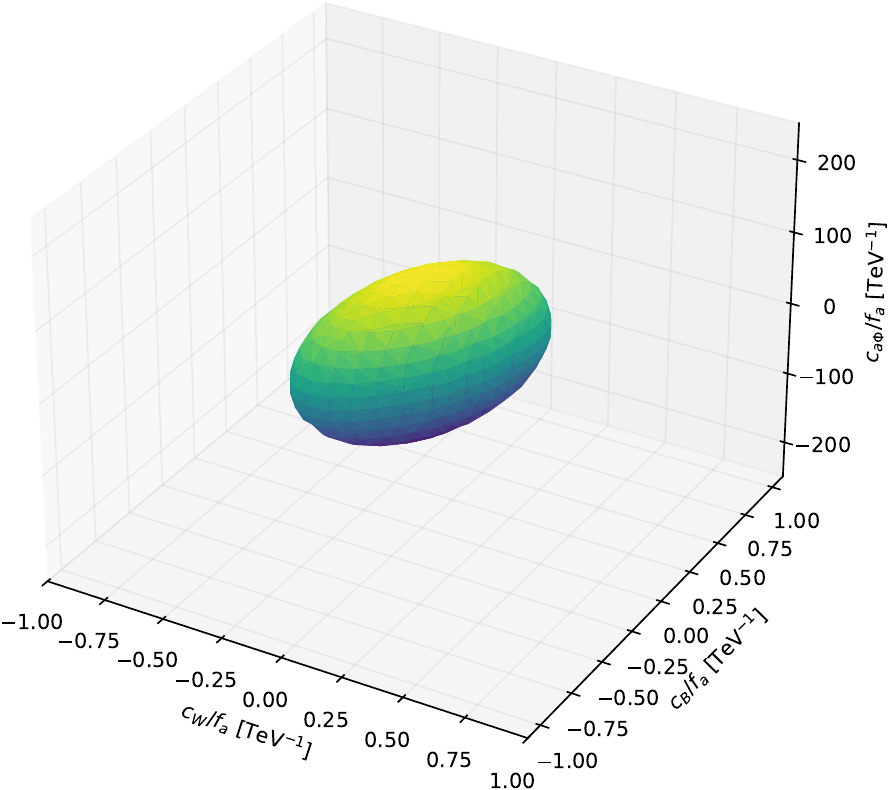}
      \caption{}
      \label{fig:awg_c}
    \end{subfigure}
    \caption{
Results for 95\cl~contours of the \texttt{TMVA} analysis of the BDT score distributions for the $aW\gamma$ final state at the LHC with an integrated luminosity of $\int L=450~\text{fb}^{-1}$ (left) and at the HL-LHC with  $\int L=3000~\text{fb}^{-1}$.    
    } 
\label{fig:awg}
\end{figure}

Using the binned likelihood ratio method explained previously, the statistical 95\cl~bounds can be derived using the obtained BDT distribution, as presented in Fig.~\ref{fig:awg}. 
%
%%%%%%%%%%
%
%
%%%%%
%
The resulting contours are both centered at the origin, and have a slightly skewed ellipsoid shape, very different from the results obtained in the diphoton case. 

%%%%%%%%%%%%%%%%%%%%%%%%%%
%
\subsection{The $\azg$ final state} 
The channel where an ALP is produced in association with a $Z$ boson and a photon is particularly interesting as it probes the linear combination of EW couplings orthogonal to those encountered in the diphoton case discussed in Sec.~\ref{sec:agg}. Specifically, the effective $g_{aZ\gamma}$ coupling is proportional to $c_W - c_B$, whereas the $g_{a\gamma\gamma}$ coupling depends on $c_W s^2_\theta + c_B c^2_\theta$.  Consequently, combining the results of the $\azg$ channel with those obtained for the $\agg$ final state allows for a more comprehensive exploration of the $(c_W, c_B)$ parameter space. Furthermore, since the $\azg$ final state is electrically neutral, diagrams involving initial state gluons via the effective $c_G$ coupling contribute significantly, analogous to the $a\gamma\gamma$ process but different from the charged $aW\gamma$ channel.

To ensure a clean signal extraction, we focus on the charged lepton and the neutrino decay modes of the $Z$ boson, $Z \to \ell^+\ell^-$ ($\ell = e, \mu$) and $Z \to \nu\bar{\nu}$, respectively\footnote{For simplicity, in the following we will refer to the charged lepton decay modes of the $Z$~boson simply as ``leptonic decay modes''. }. Although the hadronic branching fraction is higher, the overwhelming QCD background makes the hadronic channel less sensitive for this specific topology. The neutrino decay mode is particularly advantageous due to its branching ratio of about 20\% being significantly larger than the one for charged lepton decays which amounts to ca.~6.7\%. Furthermore, since the ALP escapes detection, the additional missing energy from the neutrinos enhances the missing energy signature, providing a strong handle against backgrounds with little to no genuine missing energy. 
The signal signature for the leptonic case is defined by two opposite-sign, same-flavor leptons, a high-transverse momentum isolated photon, and significant missing transverse energy attributed to the escaping ALP. To identify the $Z$ boson in this channel, we reconstruct the dilepton pair and require its invariant mass $\mll$ to lie within the window 
\beq
\label{eq:mll}
|\mll - m_Z| < 15~\text{GeV}\,,  
\eeq 
For neutrino decays, direct mass reconstruction is impossible; instead, the final state is characterized by a photon of large transverse momentum  and large missing transverse energy, with no additional visible objects.

The background processes relevant for the leptonic and neutrino decay channels are summarized in Tab.~\ref{tab:zga_backgrounds}. 
%
%%%%%
 %
\begin{table}[t!]
  \centering
  \begin{tabular}{c|c}
    \hline
    \textbf{leptonic decay mode} & \textbf{neutrino decay mode} \\
    \hline
    $\gamma t\bar{t}$, $\gamma \wpm$, $\gamma Z$ & $W\gamma$, $Z\gamma$, $ZZ\gamma$ \\
    $j t\bar{t}$, $j \wpm$, $j Z$ & $Wj$, $Zj$ \\
    \hline
  \end{tabular}
  \caption{Dominant background processes for the $Z(\to \ell\ell)\gamma a$ and $Z(\to \nu\nu)\gamma a$ channels.}
  \label{tab:zga_backgrounds}
\end{table}
%
%%%%%
%
For the leptonic decay mode, the irreducible backgrounds arise from SM processes producing a photon in association with a $Z$~boson, a $W^\pm$~boson, or a $t\bar{t}$ pair, where the leptons originate from the electroweak boson or top quark decays. The reducible backgrounds stem from processes where a jet is misidentified as a photon, including $Zj$, $W^\pm j$, and $t\bar{t} j$ production. For the neutrino decay mode, the main irreducible backgrounds are $W\gamma$, $Z\gamma$, and $ZZ\gamma$ production, where the missing transverse energy arises from neutrinos in $W^\pm$ or $Z$ decays. The reducible $Wj$ and $Zj$ backgrounds contribute when a jet fakes a photon signature. As in Sec.~\ref{sec:agg}, we model the jet-to-photon misidentification using the constant transfer rate $f_{j\to\gamma}=0.1\%$. 
Contributions from the $t\bar{t}\gamma$ final state are effectively suppressed by vetoing events with high jet multiplicity and $b$-tagged jets, while the $WZ$ background becomes negligible after requiring exactly two charged  leptons and a photon. 

For the multivariate analysis, we construct separate BDTs for the leptonic and neutrino decay channels to exploit their distinct kinematic features. For the $Z(\to\ell\ell)\gamma$ channel, having full control on the charged leptons' momenta allows us to use variables such as the invariant mass $m_{\ell\ell}$ and the transverse momentum of the dilepton system $p_{T,\ell\ell}$. 
In contrast, the $Z(\to\nu\bar{\nu})\gamma$ channel relies heavily on missing transverse energy correlations, specifically the separation and the transverse mass of the photon and the missing energy,
\beq
\mtginv=\sqrt{2\, p_T^\gamma \, E_T^{miss} \, \bigl(1 - \cos\dpginv\bigr)},
\eeq 
where $\dpginv$ denotes the azimuthal-angle separation of the two objects. 
 
For the leptonic channel, we apply preselection cuts requiring exactly two same-flavor, opposite-sign leptons, each satisfying 
\beq
\ptl > 25~\text{GeV} \quad \text{and} \quad |\eta_\ell| < 2.5\,, 
\eeq
with a dilepton invariant mass satisfying Eq.~\eqref{eq:mll} and exactly one photon satisfying 
\beq
\label{eq:azg-photon}
\ptg > 25~\text{GeV} \quad \text{and} \quad |\eta_\gamma| < 2.5\,.
\eeq
For the neutrino channel, we again require the presence of a photon fulfilling the cuts of Eq.~\eqref{eq:azg-photon}, along with significant missing transverse energy, 
\beq
E_T^{miss} > 100~\text{GeV}\,, 
\eeq
for the suppression of QCD backgrounds. 

\begin{figure}[!tp]
  \centering  
  \begin{subfigure}{0.45\textwidth}
    \includegraphics[width=\textwidth]{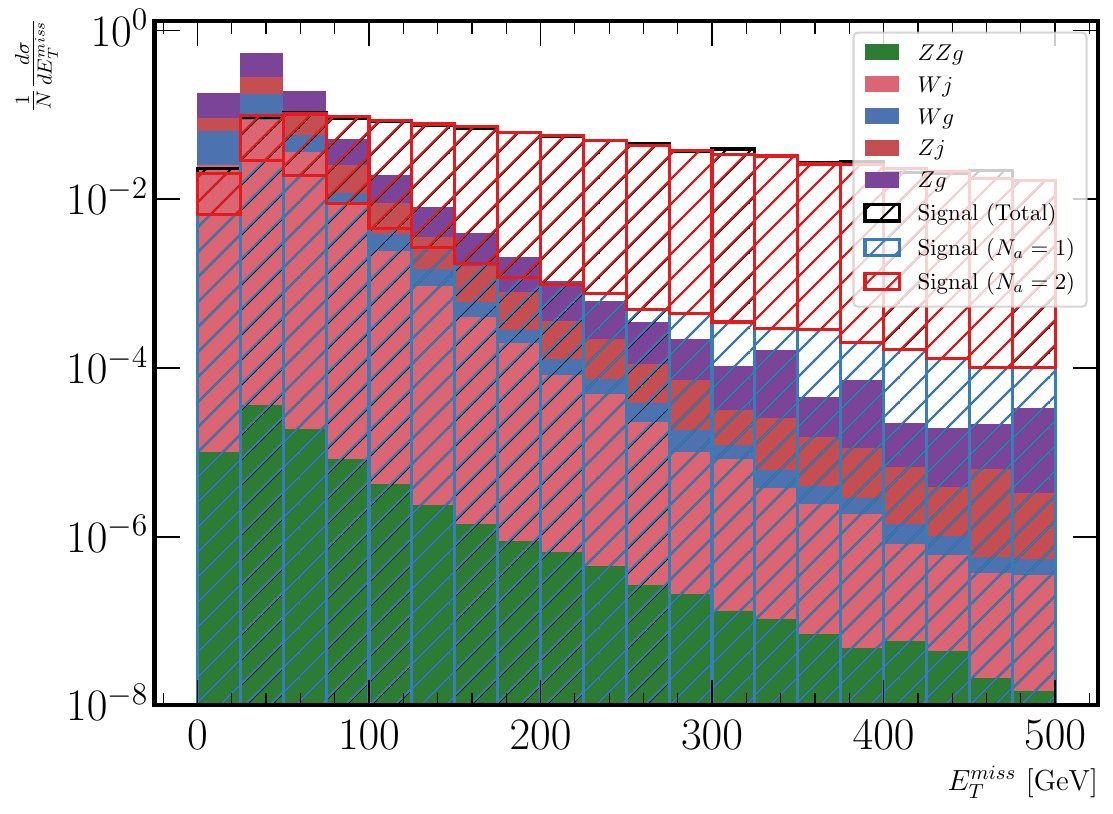}
    \caption{}
    \label{fig:E_T_azg}
  \end{subfigure}
  \hfill
  \begin{subfigure}{0.45\textwidth}
    \includegraphics[width=\textwidth]{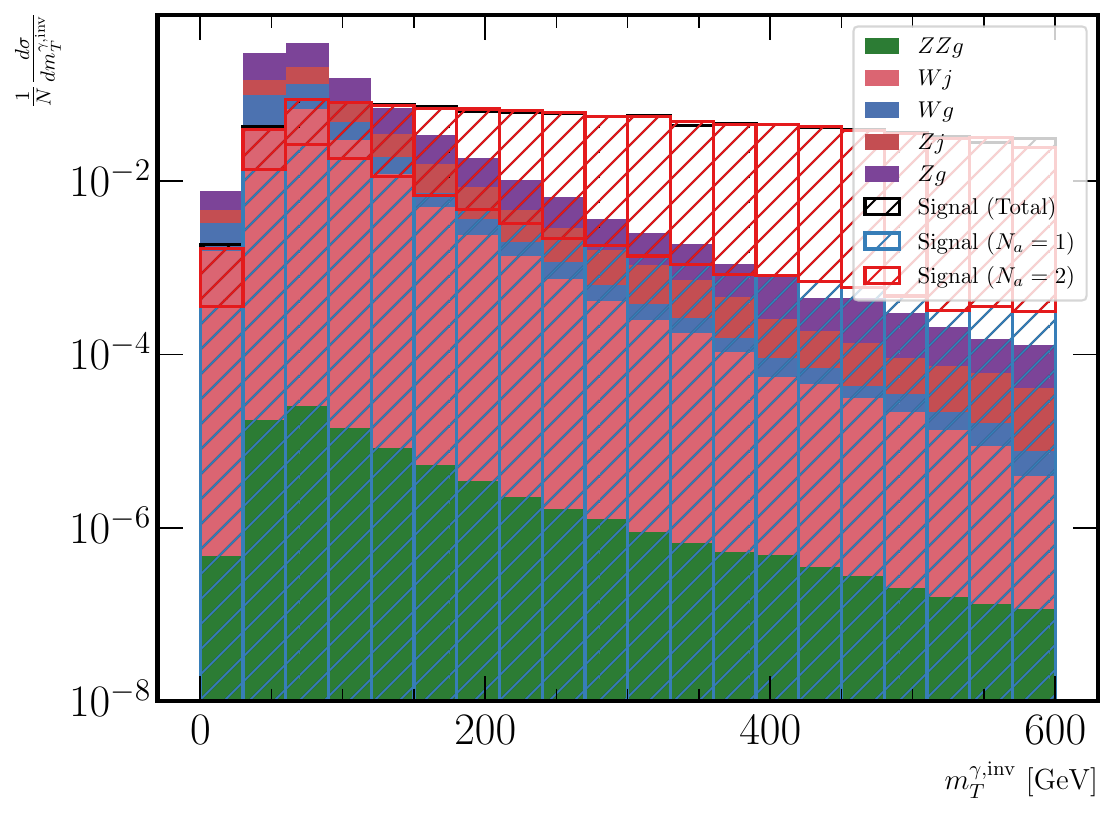}
    \caption{}
    \label{fig:eta_miss_azg}
  \end{subfigure}
  \\[5mm]
  \begin{subfigure}{0.45\textwidth}
    \includegraphics[width=\textwidth]{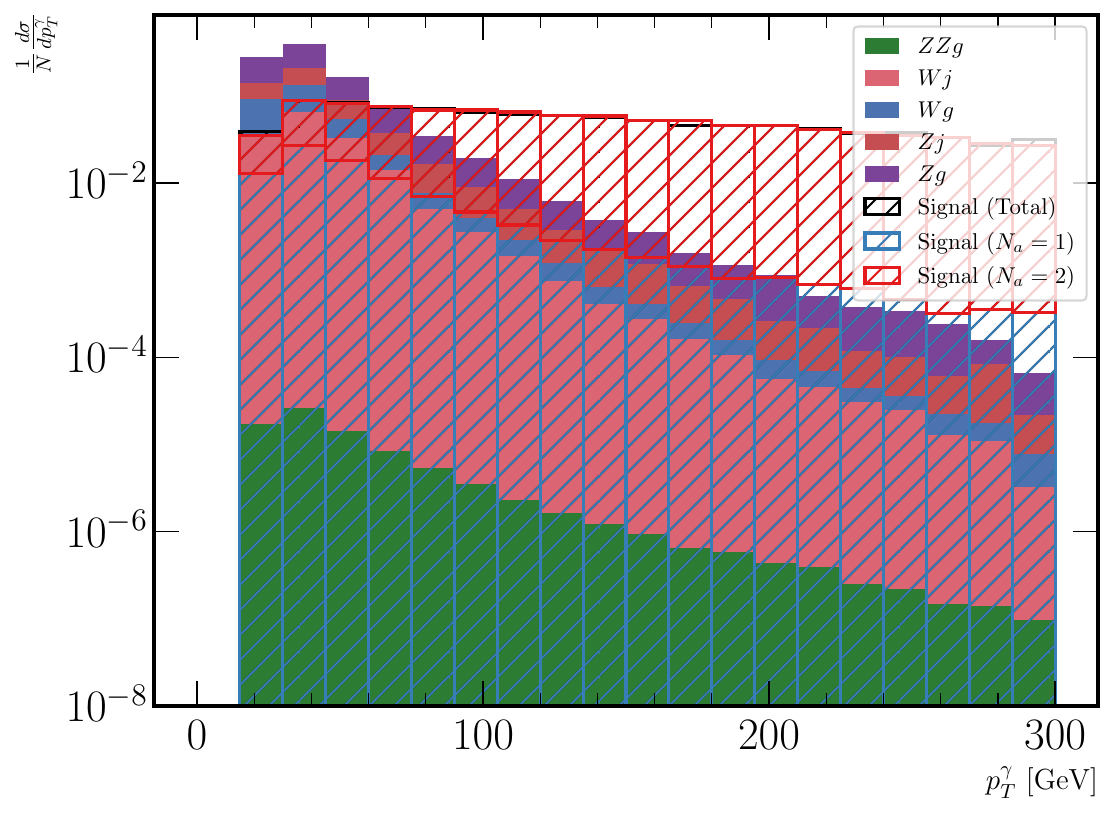}
    \caption{}
    \label{fig:p_T_gamma_azg}
  \end{subfigure}
  \hfill 
  \begin{subfigure}{0.45\textwidth}
    \includegraphics[width=\textwidth]{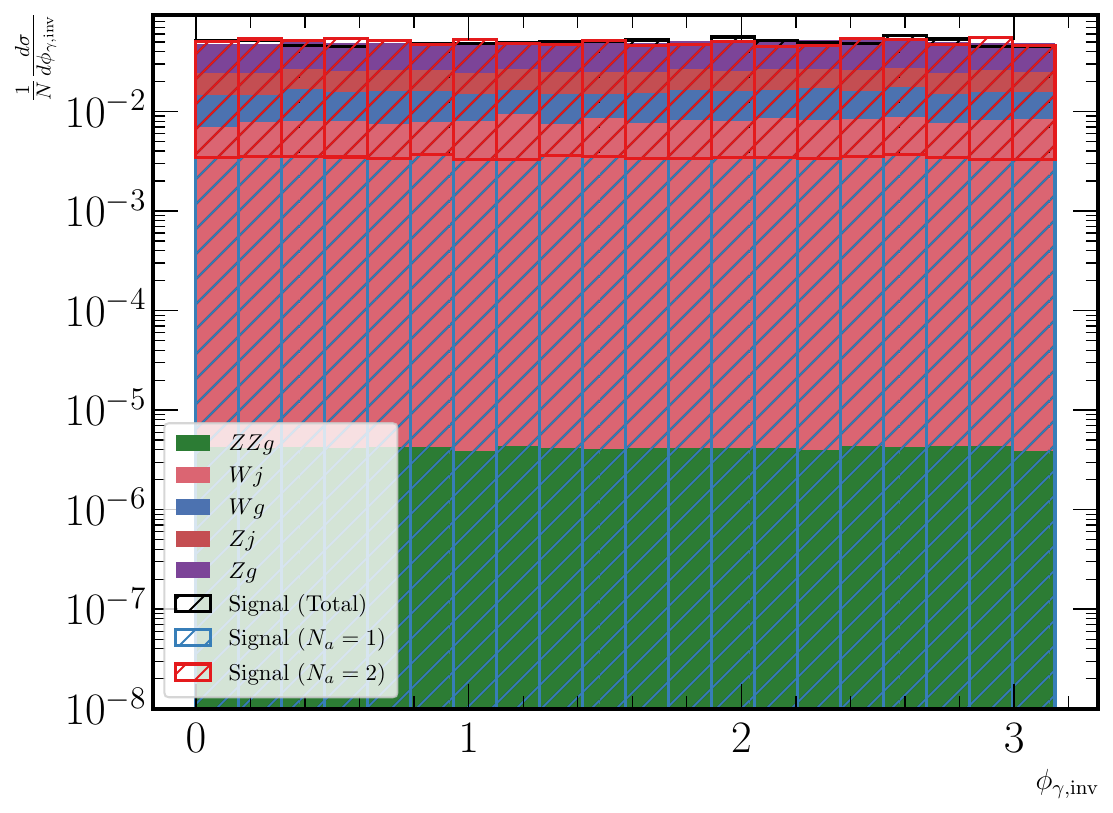}
    \caption{}
    \label{fig:phi_miss_azg}
  \end{subfigure}
  \caption{Normalized distributions for the simulated signal and background contributions to the $aZ\gamma(Z\to\nu\bar{\nu})$ channel at the LHC with $\sqrt{s}=14$~TeV using the benchmark point BP3 specified in Eq.~\eqref{eq:BP3}. In each case the hatched histograms represent the signal scenarios with $N_a=1$~(blue), $N_a=2$~(red) and their sum~(black), whereas the filled histograms represent the $ZZ\gamma$~(green), $Wj$~(orange), $W\gamma$~(blue), $Zj$~(red), and $Z\gamma$~(purple) background contributions.}
\label{fig:agg-sig-back_azg}
\end{figure}

Selected distributions for the $Z(\to\nu\bar{\nu})\gamma a$ channel are shown in Fig.~\ref{fig:agg-sig-back_azg}.  
%
%%%%%%%
%
%
%%%%%%%%%%%
%
The missing transverse energy distribution, depicted in Fig.~\ref{fig:E_T_azg}, exhibits a pronounced tail for the signal extending to high values, reflecting the escaping ALP and neutrinos, while the backgrounds peak at lower values of $E_T^{miss}$, similar to the $\agg$ and $\awg$ cases. The transverse mass of the system comprised of the photon and missing energy, shown in Fig.~\ref{fig:eta_miss_azg}, exhibits a similar behavior, with the signal distribution extending to higher values compared to the backgrounds, which cluster at lower masses, making it the best discriminating variable. The transverse momentum of the photon, depicted in Fig.~\ref{fig:p_T_gamma_azg}, also shows a broader distribution for the signal, while the backgrounds are more concentrated at lower $\ptg$ values. The azimuthal angle separation between the photon and the missing energy vector, illustrated in Fig.~\ref{fig:phi_miss_azg}, however, has no discriminating power at all, as both signal and background distributions are relatively flat across the entire range. Throughout, the $N_a=1$ and $N_a=2$ signal contributions that are shown separately in each case exhibit similar shapes, yet the one with $N_a=2$ dominates the overall signal yield due to its larger cross section. 
This behavior is observed not only in the $Z\to\nu\bar{\nu}$ channel, but also in the $Z\to\ell\ell$ channel. 

\begin{figure}[!tp]
  \centering
  \includegraphics[width=0.65\textwidth]{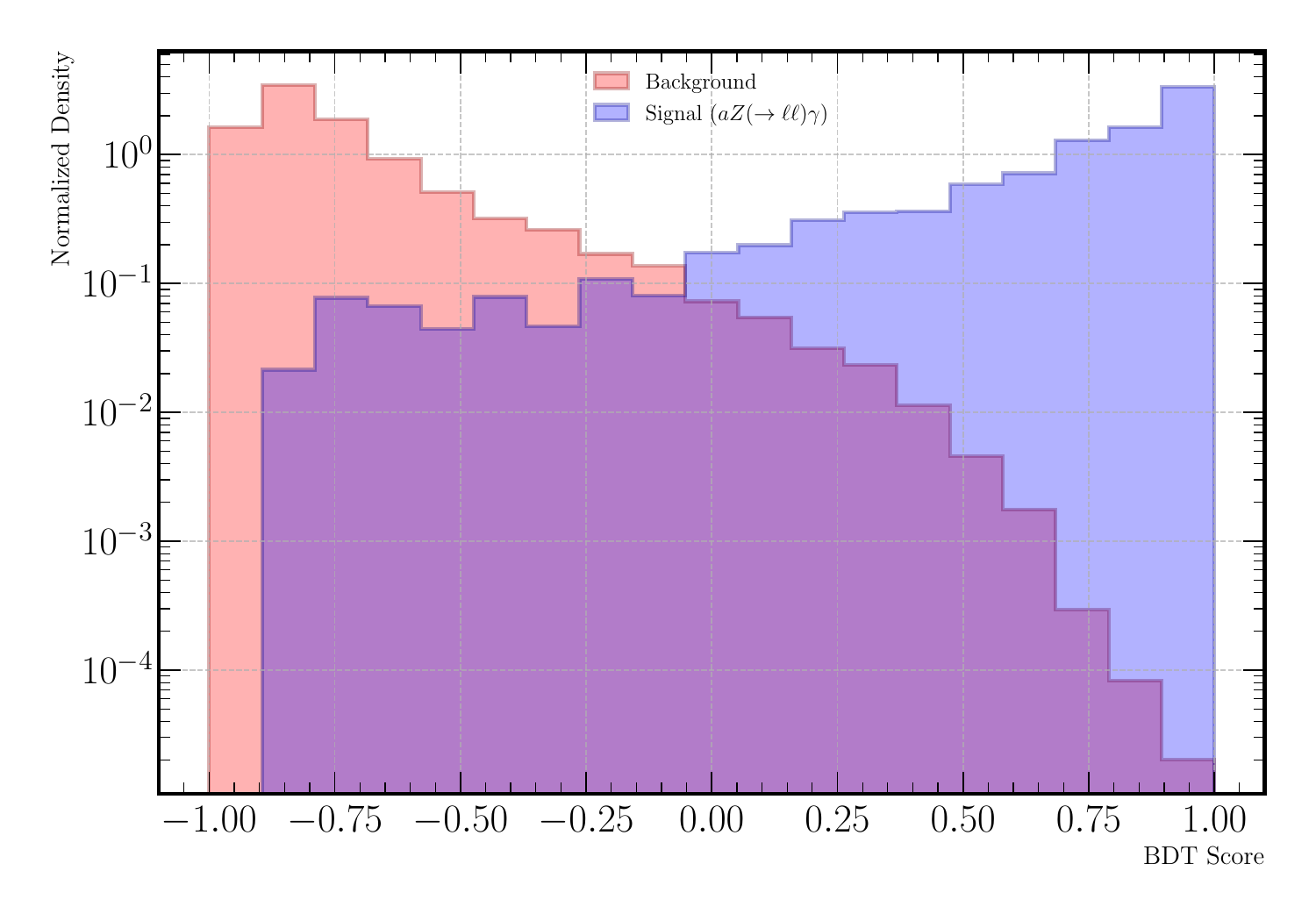}
  \caption{Normalized BDT score distribution for the $\azg$ signal in the ($Z\to\ell\ell$) decay channel versus the sum of the relevant background processes given in Tab.~\ref{tab:zga_backgrounds} for the benchmark point BP3 of Eq.~\eqref{eq:BP3}. 
  }
  \label{fig:zga_bdt}
\end{figure}

The ranking of the input variables based on their separation power for the charged-lepton and the neutrino  channels is summarized in Tab.~\ref{tab:bdt_var_separation_azg}. 
\begin{table}[t]
  \centering
  \begin{tabular}{c|c||c|c}
    \multicolumn{2}{c||}{\textbf{$Z(\to\nu\bar{\nu})\gamma$}} & \multicolumn{2}{c}{\textbf{$Z(\to\ell\ell)\gamma$}} \\
    \hline
    variable & separation & variable & separation \\
    \hline
    $\ptg$ & 0.3031 & $E_T^{miss}$ & 0.2860 \\
    $\mtginv$ & 0.1797 & $m_{\ell\ell}$ & 0.1772 \\
    $E_T^{miss}$ & 0.1481 & $\ptll$ & 0.1151 \\
    $\Delta R_{\gamma, \mr{inv}}$ & 0.1224 & $\ptg$ & 0.1031 \\
    $\eta_{\gamma}$ & 0.1015 & $\eta_{\gamma}$ & 0.0663 \\
    $\eta_{\mr{inv}}$& 0.0925 & $\Delta R_{\ell\ell, \gamma}$ & 0.0532 \\
    $\phi_{\mr{inv}}$ & 0.0458 & - & - \\
  \end{tabular}
  \caption{BDT input variables and their separation power for the $Z(\to\nu\bar{\nu})\gamma a$ and $Z(\to\ell\ell)\gamma a$ channels.
  \label{tab:bdt_var_separation_azg}}
\end{table}
As expected, leptonic $Z$-boson decays provide more kinematic handles to discriminate signal from background, as reflected in the generally higher separation values. This is due to the fact that the missing energy in the leptonic case arises mainly from the undetected ALP, making it easier to distinguish signal from background processes, as the latter is concentrated at low $E_T^{miss}$ values. In contrast, for the neutrino channel, a reduced discriminating power is expected, since both the ALP and the neutrinos from the $Z$-boson decay contribute to the missing transverse energy, making it inherently difficult to distinguish the signal from SM processes with invisible final states. The BDT distribution of this leptonic channel is shown in Fig.~\ref{fig:zga_bdt}.
%
%%%%%
%
%%%%
%
As expected from the higher separation power of the input variables in the leptonic channel, the corresponding BDT score distribution of Fig.~\ref{fig:zga_bdt} shows a clear separation between signal and background.

\begin{figure}[!tp]
  \centering
      \begin{subfigure}{0.45\textwidth}
        \includegraphics[width=\textwidth]{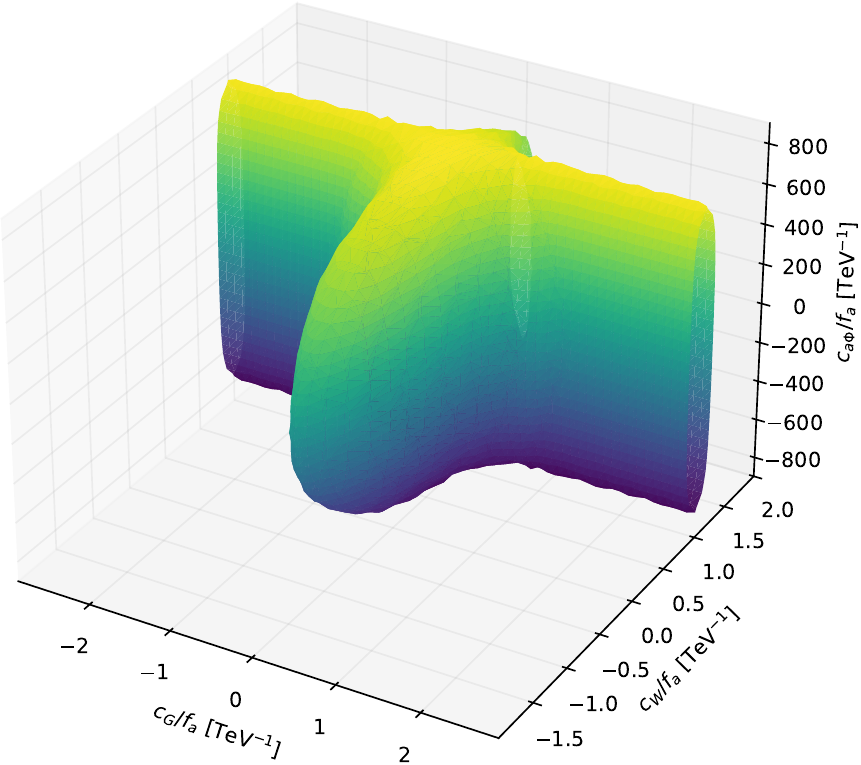}
        \caption{}
        \label{fig:zga_a}
        \end{subfigure}
        \hfill
        \begin{subfigure}{0.45\textwidth}
      \includegraphics[width=\textwidth]{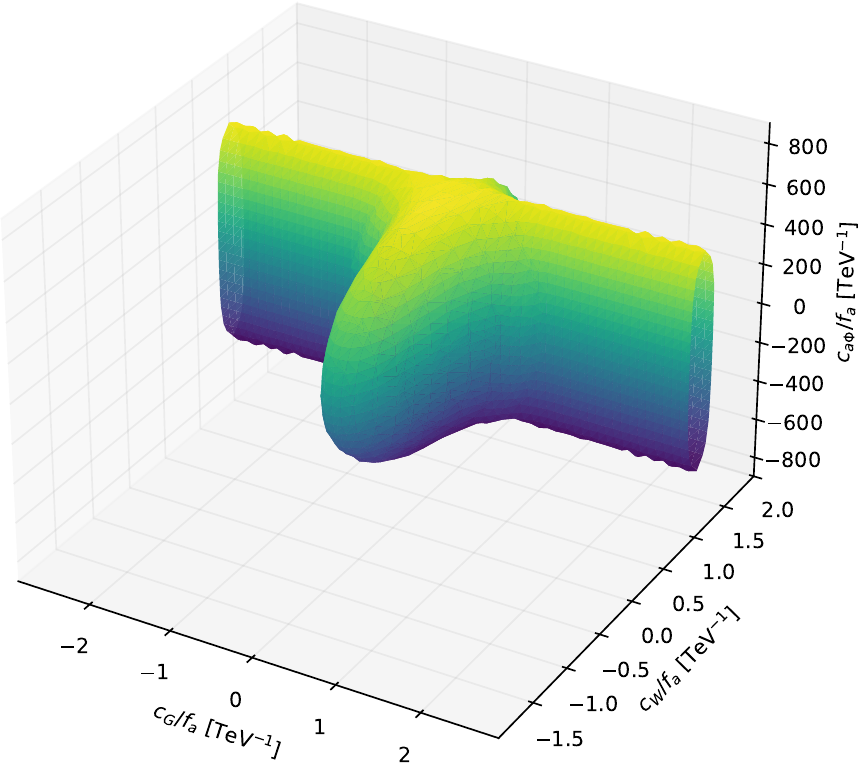}
      \caption{}
      \label{fig:zga_b}
        \end{subfigure}
        
        \par\bigskip
        \begin{subfigure}{0.45\textwidth}
      \includegraphics[width=\textwidth]{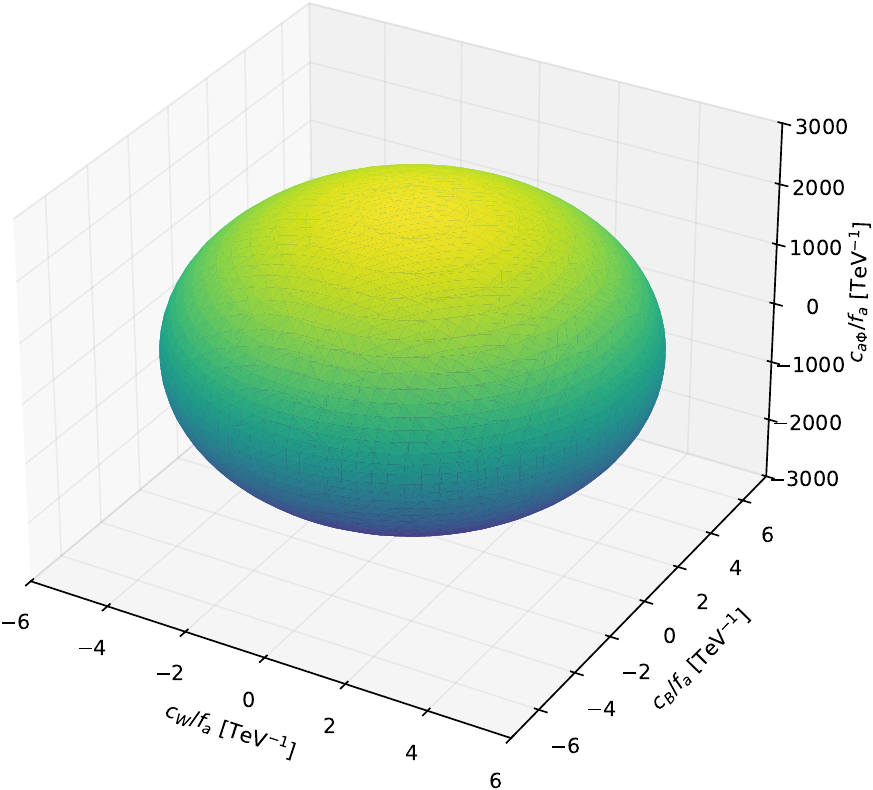}
      \caption{}
      \label{fig:zga_c}
        \end{subfigure}
        \hfill
        \begin{subfigure}{0.45\textwidth}
      \includegraphics[width=\textwidth]{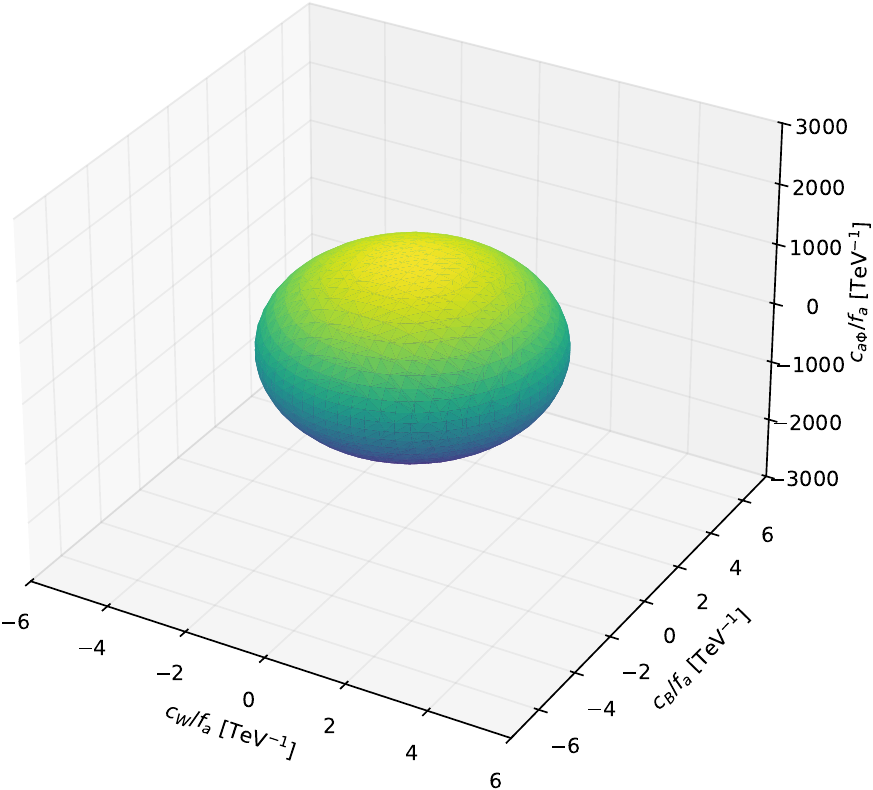}
      \caption{}
      \label{fig:zga_d}
        \end{subfigure}
        \caption{Projected 95\cl~exclusion contours for the $Z(Z\to\nu\bar{\nu})\gamma a$ channel in the $(c_W/f_a, c_G/f_a, c_{a\Phi}/f_a)$ and $(c_W/f_a, c_B/f_a, c_{a\Phi}/f_a)$ spaces for the upper and lower panels,  respectively. The assumed integrated luminosity is $\int L=450~\text{fb}^{-1}$ for panels (a) and (c) and $\int L=3000~\text{fb}^{-1}$ for panels (b) and (d). For the upper panels, $c_B/f_a$ is set to $1~\text{TeV}^{-1}$, whereas for the lower panels, $c_G/f_a$ is set to $0.1~\text{TeV}^{-1}$.}
\label{fig:zga}
\end{figure}

The resulting exclusion limits for the neutrino decay channel are presented in Fig.~\ref{fig:zga}, 
%%%%%%%%%
%
%%%%%%%%%
%
where we consider benchmark scenarios with $c_B/f_a=1~\text{TeV}^{-1}$ and $c_G/f_a=0.1~\text{TeV}^{-1}$. As discussed above, the analysis is primarily driven by the transverse mass of the photon$+$missing energy system and the magnitude of the missing transverse energy. This is a direct consequence of the signal topology: Since both the ALP and the neutrinos from the $Z$-boson decay escape detection, the process is fundamentally characterized by a large invisible component, making the kinematic correlation between the single visible photon and the missing momentum vector the most powerful discriminator. 

\begin{figure}[!tp]
  \centering
  \begin{subfigure}{0.45\textwidth}
    \includegraphics[width=\textwidth]{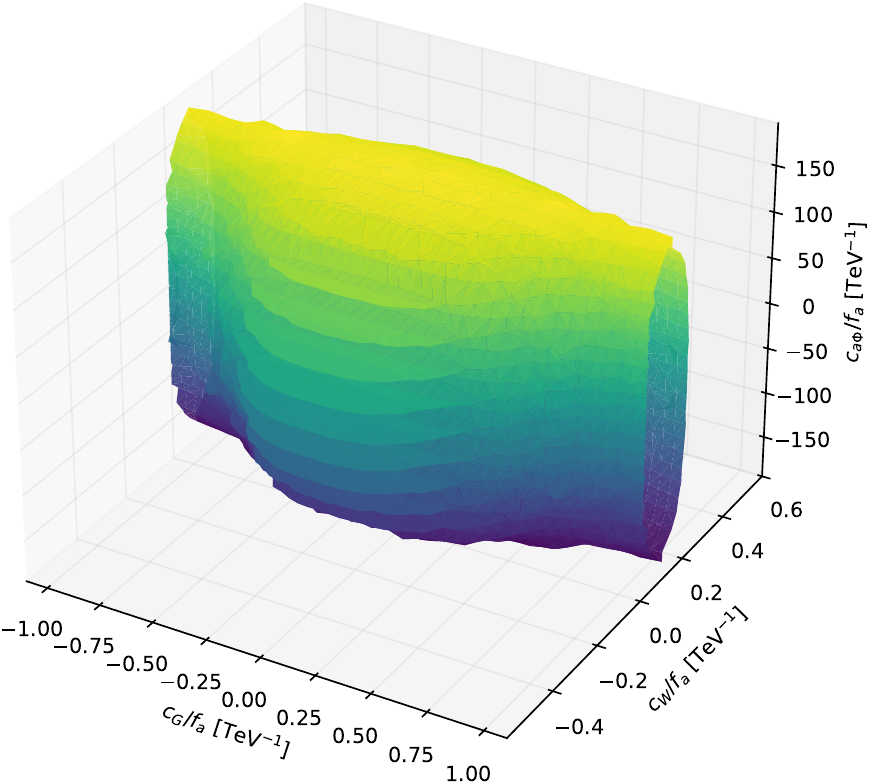}
    \caption{}
    \label{fig:zga_lep_a}
  \end{subfigure}
  \hfill
  \begin{subfigure}{0.45\textwidth}
    \includegraphics[width=\textwidth]{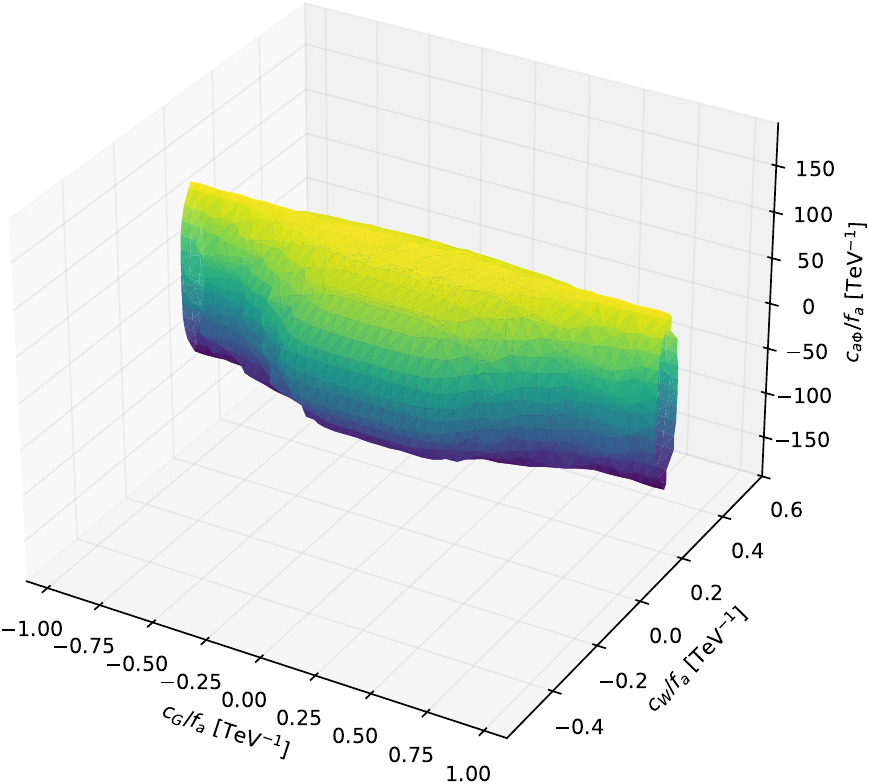}
    \caption{}
    \label{fig:zga_lep_b}
  \end{subfigure}
  \caption{Projected 95\cl~exclusion contours for the $Z(\to\ell\ell)\gamma a$ channel in the $(c_G/f_a, c_W/f_a, c_{a\Phi}/f_a)$ plane with $c_B/f_a$ set to $0.1~\text{TeV}^{-1}$. The assumed integrated luminosity is $\int L=450~\text{fb}^{-1}$ for (a) and $\int L=3000~\text{fb}^{-1}$ for (b).}
  \label{fig:zga_lep}
\end{figure}

While the neutrino channel benefits from a larger branching ratio, the leptonic channel offers a cleaner experimental signature allowing for a full reconstruction of the $Z$ boson. Although its branching fraction is smaller, the ability to constrain the kinematics of the $Z$~boson provides a complementary handle on the signal, particularly in the high-transverse momentum regime where backgrounds are lower. The constraints derived from the analysis of the leptonic decay mode for the $c_B/f_a=0.1~\text{TeV}^{-1}$ benchmark scenario are shown in Fig.~\ref{fig:zga_lep}.
The constraints in the $(c_G/f_a, c_W/f_a)$ plane exhibit a more distinct alignment than the $\agg$ and $\awg$ channels, highlighting the complementarity of the $\azg$ process. For the $c_B=1$ case, the contours are centered along the $c_W=1$ line, forming an ellipsoid that becomes significantly narrower with increased luminosity, with a slight bulge towards the coordinate origin.
%
%%%%%%%%%%%%%%%%%%%%%%%%%%
%
\subsection{The $\azw$ final state}
Next, we consider the associated production of an ALP with a $Z$ and a $W^+$ or $W^-$ boson. Only diagrams including one ALP ($N_a=1$) are present in this case.   Similarly to the $aW\gamma$ channel, the final-state $ZW^\pm$ pair carries a net electric charge of $\pm 1$. Consequently, there are no tree-level diagrams involving initial state gluons that can produce this final state without additional charged particles. This implies that the process is insensitive to the gluonic coupling $c_G$ at leading order, and the constraints derived are primarily on the electroweak ALP couplings $c_W$ and $c_B$.

We focus on the fully leptonic decay channel where the $Z$~boson decays into a pair of charged leptons, and the $W^\pm$~boson into a lepton-neutrino pair, with leptons from the first two generations.
The trilepton signal is thus characterized by a pair of same-flavor, opposite-sign leptons with an invariant mass consistent with the mass of the $Z$~boson, an additional isolated lepton, and significant missing transverse energy due to the neutrino and the ALP. 
The dominant SM backgrounds for this signature are the $ZW^\pm$, the $W^+W^-W^\pm$ and $W^+W^-Z$, and the $t\bar{t}W$ production processes, all of which can give rise to trilepton final states. 
For our analysis we consider the sum of contributions from $\azw^+$ and $\azw^-$ final states, referred to as $\azw$ from now on.  Similarly, we add positively and negatively charged background contributions and in the following drop charge superscripts accordingly. 

We require the presence of exactly three charged leptons, among which at least two are of the same flavor and opposite sign. On these three leptons we apply the following preselection cuts: 
\beq
\label{eq:azw-cuts}
\ptl>25~\mr{GeV}\,,\quad
|\etal| < 2.5\,. 
\eeq
The two same-flavor, opposite-sign leptons with invariant mass closest to $m_Z$ are identified as candidates for the $Z$-boson decay and are additionally required to exhibit an invariant mass $\mll$ close to $m_Z$, 
\beq
|\mll - m_Z| < 15~\mr{GeV}\,.
\eeq  
To train the BDT for our multivariate analysis, we utilize the missing transverse energy, the transverse momentum of the $Z$ boson, $\ptz$, computed from the leptons identified as its decay products;   
the transverse mass of the $\linv$ system, $m_T^{\linv}$,  constructed from the third lepton and the missing energy vector, according to Eq.~\eqref{eq:mtlinv}; 
and the transverse momentum of the $\linv$ system, $p_T^{\linv}$.  

\begin{figure}[!tp]
  \centering
  
  \begin{subfigure}{0.45\textwidth}
    \includegraphics[width=\textwidth]{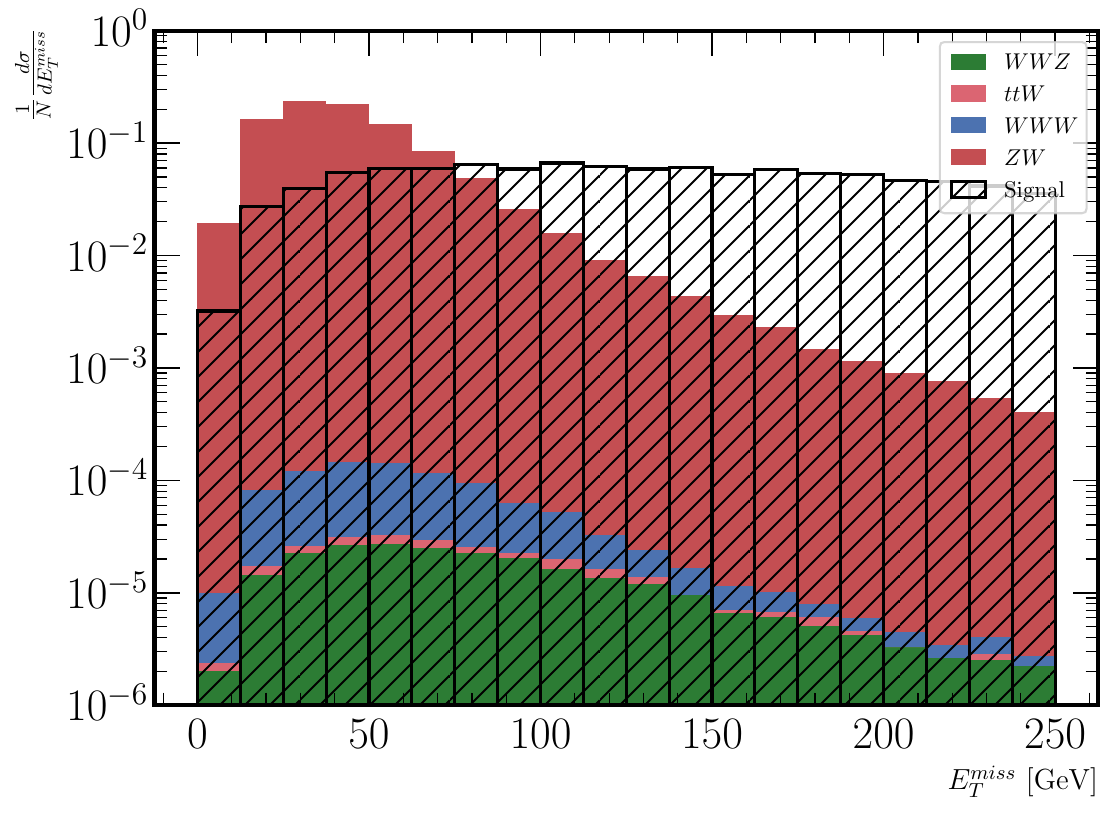}
    \caption{}
    \label{fig:dr_azw}
  \end{subfigure}
%  \hfill
  \begin{subfigure}{0.45\textwidth}
    \includegraphics[width=\textwidth]{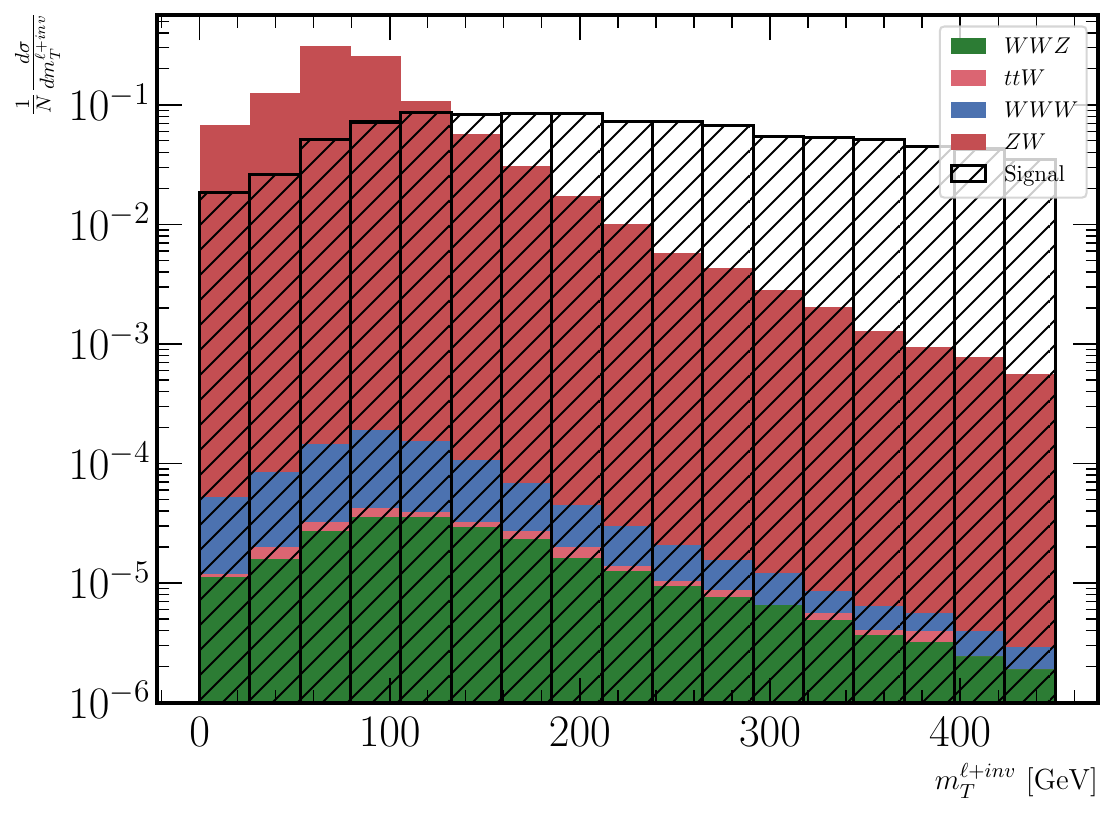}
    \caption{}
    \label{fig:met_azw}
  \end{subfigure}
  \\[5mm]
  \begin{subfigure}{0.45\textwidth}
    \includegraphics[width=\textwidth]{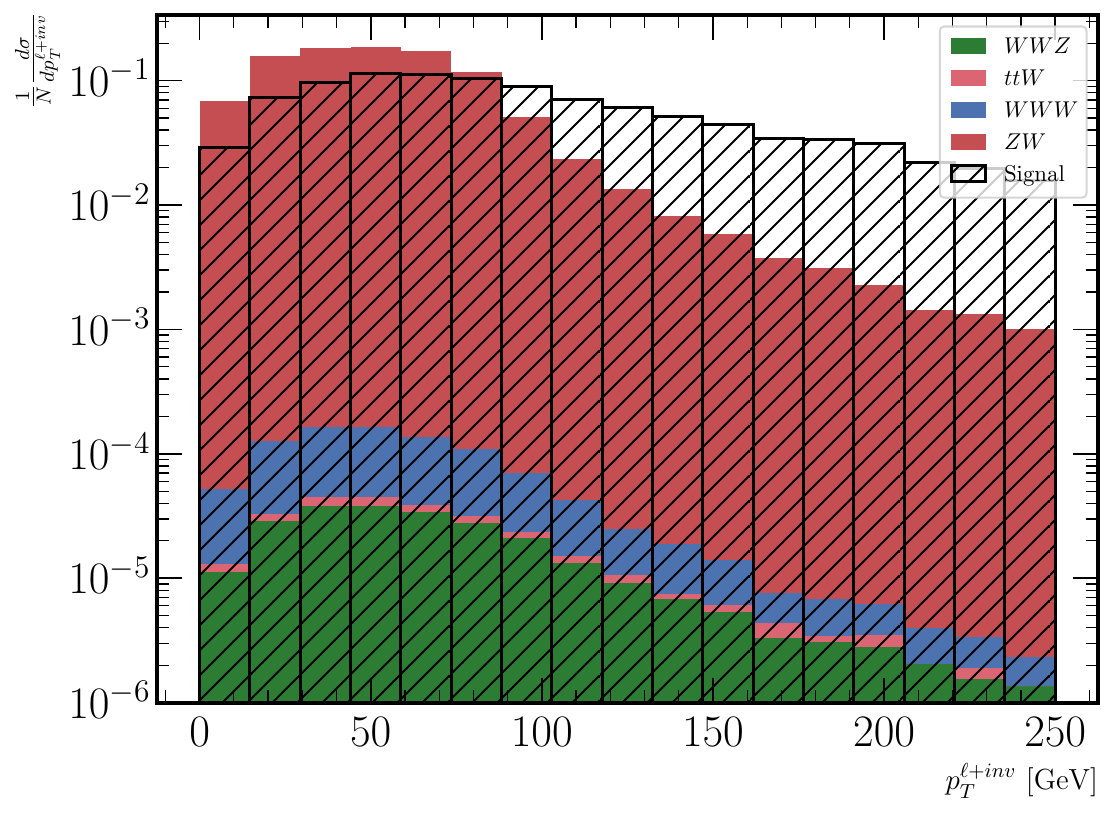}
    \caption{}
    \label{fig:mgg_azw}
  \end{subfigure}
%  \hfill 
  \begin{subfigure}{0.45\textwidth}
    \includegraphics[width=\textwidth]{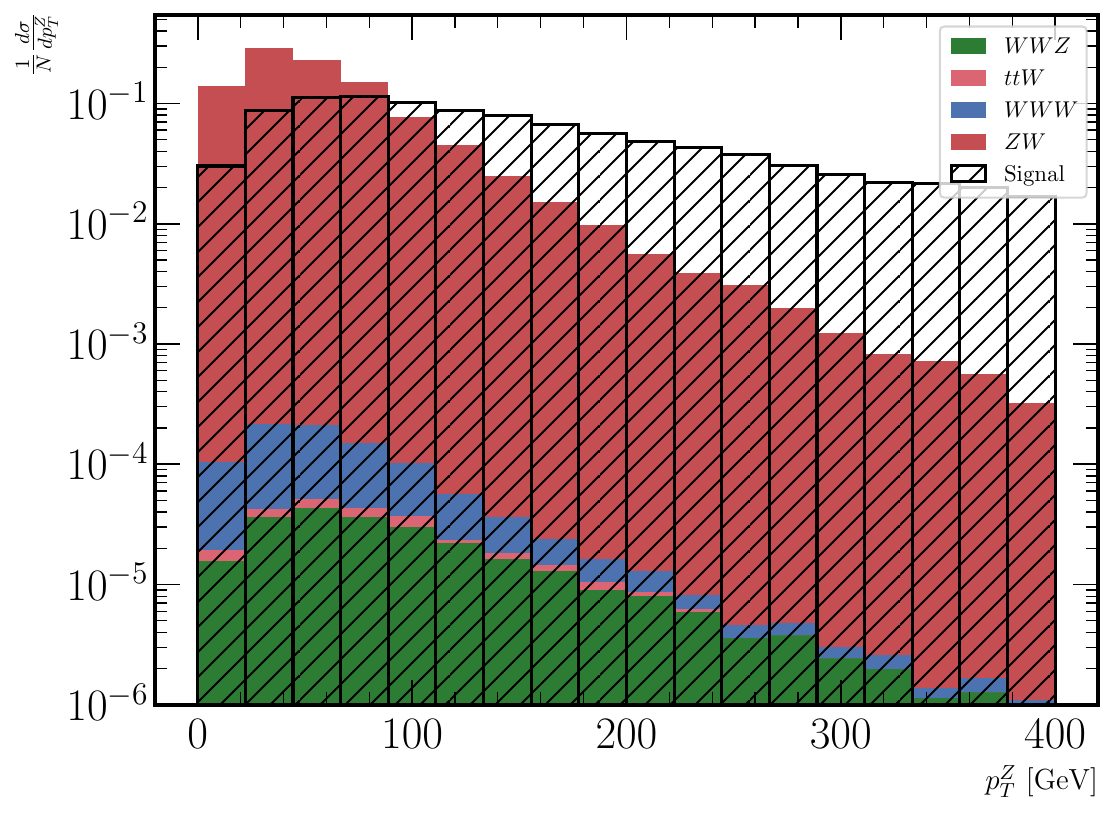}
    \caption{}
    \label{fig:pt_azw}
  \end{subfigure}
  \caption{Normalized distributions for the simulated signal and background contributions to the $\azw$ channel at the LHC with $\sqrt{s}=14$~TeV using the benchmark point BP2 of Eq.~\eqref{eq:BP2}. The hatched histograms represent the signal, whereas the filled histograms represent the $t\bar{t}W$~(green), $WWZ$~(blue), $WWW$~(pink), and $ZW$~(red) background contributions.}
\label{fig:agg-sig-back_azw}
\end{figure}

\begin{figure}[!tp]
  \centering
      \includegraphics[width=0.65\textwidth]{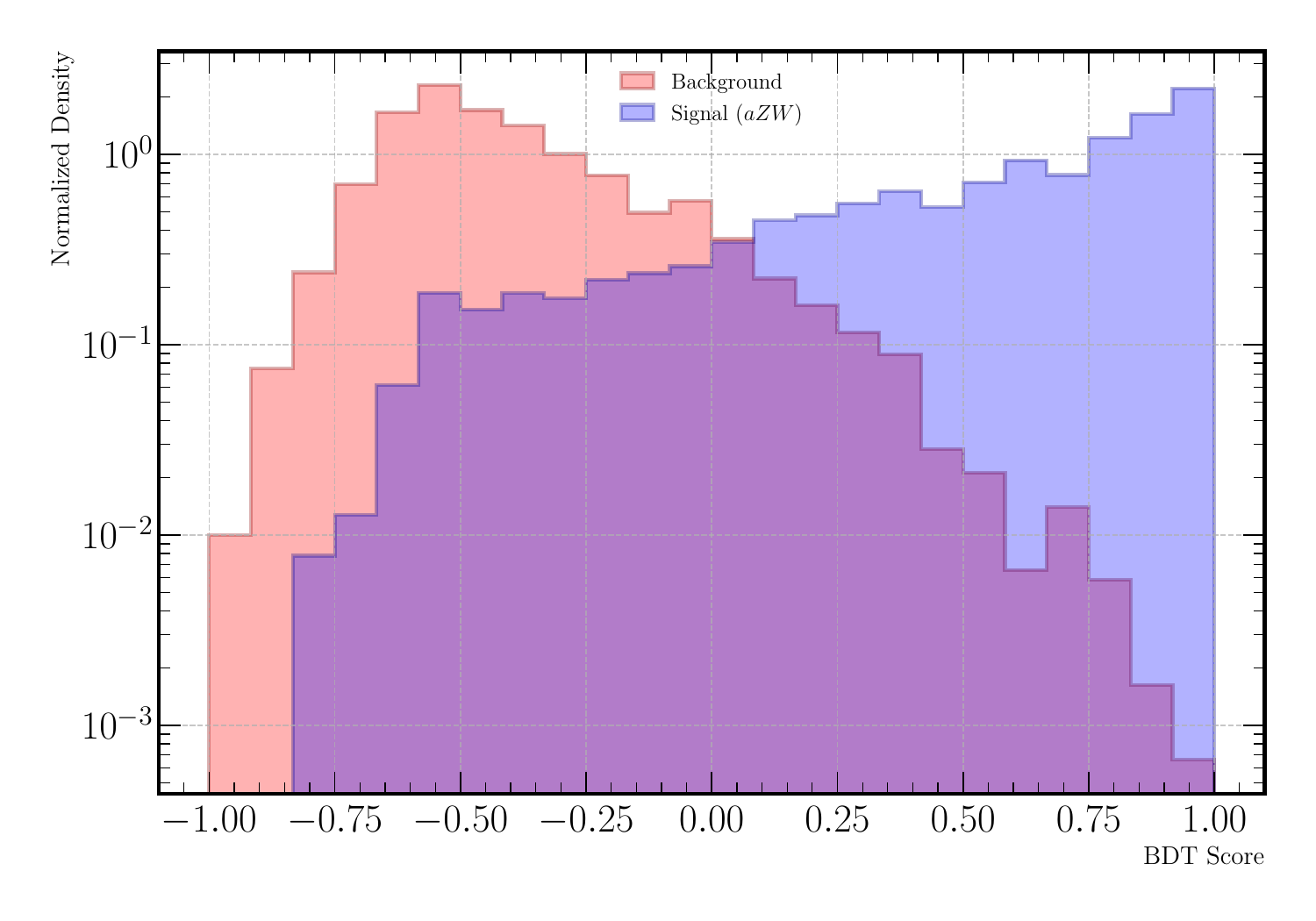}
    \caption{
      Normalized BDT score distributions of the $aZW$ signal versus the sum of the $t\bar{t}W$, $WWZ$, $WWW$ and $ZW$ background processes for the benchmark point BP2 of Eq.~\eqref{eq:BP2}.} 
\label{fig:azw_bdt}
\end{figure}

The results of our analysis are shown in Fig.~\ref{fig:agg-sig-back_azw}. 
%%
%%%%%%%%%%%%
%
%
%%%%%%%%
%
In Fig.~\ref{fig:dr_azw} we display the normalized distribution of the missing transverse energy. The signal contribution exhibits a significantly broader distribution extending to much higher values than all background contributions, which are concentrated at low $E_T^{miss}$ instead. The missing transverse energy observable provides the strongest discrimination between signal and background among all variables considered for this channel, surpassing the discriminating power observed in the $\agg$, $\awg$ and $\azg$ processes. 
Fig.~\ref{fig:met_azw} shows transverse mass of the $\linv$ system. The background distributions peak sharply at low values and rapidly decrease towards higher transverse masses, whereas the signal distribution is relatively flat across the entire range, providing good discriminating power, similar to the transverse momentum distribution of the $Z$ boson given in Fig.~\ref{fig:pt_azw}.
The transverse momentum of the $\linv$ system is shown in Fig.~\ref{fig:mgg_azw}. For this observable both signal and background distributions exhibit slightly more similar shapes, meaning this variable provides a somewhat weaker, but still valid, discriminating power between signal and background as a BDT candidate.
The ranking and separation power of the variables entering our BDT for the $\azw$ process, obtained after applying the preselection cuts, are summarized in Tab.~\ref{tab:bdt_var_separation_azw}.
\begin{table}[t!]
  \centering
  \begin{tabular}{c||c|c|c|c}
    variable & $E_T^{miss}$ & $\ptz$ & $m_T^{\linv}$ & $p_T^{\linv}$ \\
    \hline
    separation & 0.377 & 0.249 & 0.180 & 0.156 \\
  \end{tabular}
  \caption{BDT input variables and their separation power for the $aZW$ channel.}
  \label{tab:bdt_var_separation_azw}
\end{table}
The BDT score distribution corresponding to these variables is shown in Fig.~\ref{fig:azw_bdt}.

%
%%%%%%%%%%
%
%
%%%%%%%%%
%

\begin{figure}[!tp]
  \centering
  \begin{subfigure}{0.45\textwidth}
  \includegraphics[width=\textwidth]{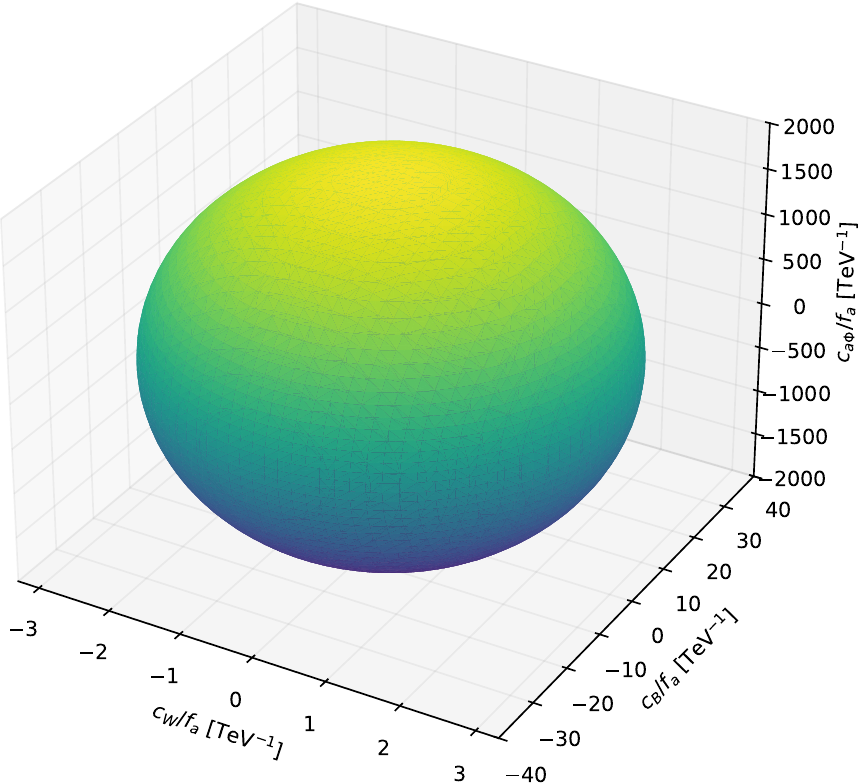}
  \caption{}
  \label{fig:azw_lep_a}
  \end{subfigure}
  \hfill
  \begin{subfigure}{0.45\textwidth}
  \includegraphics[width=\textwidth]{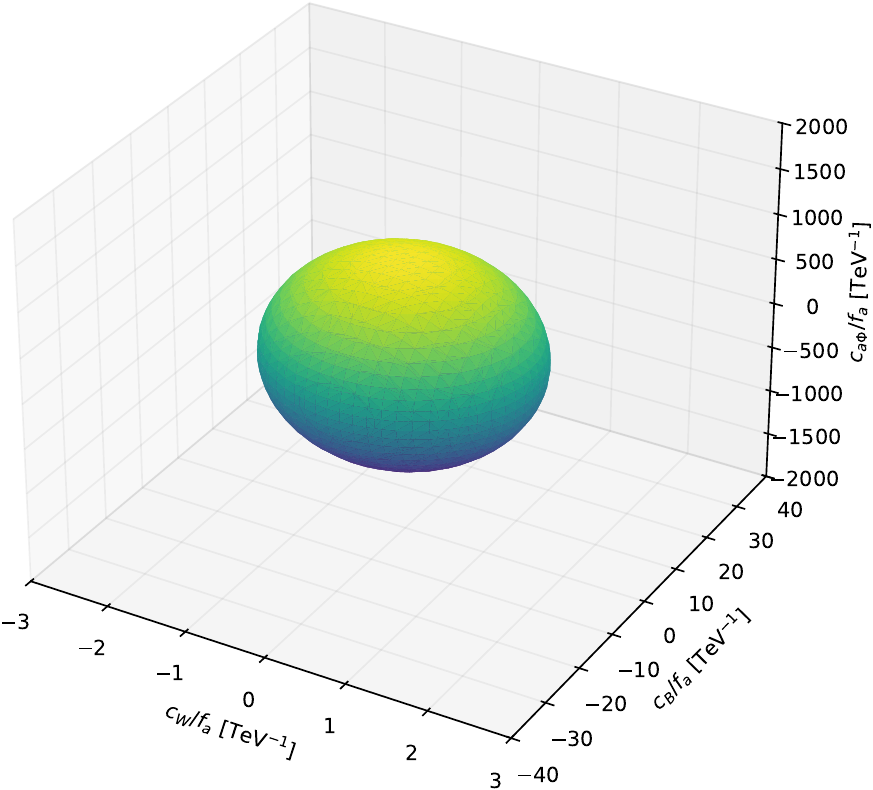}
  \caption{}
  \label{fig:azw_lep_b}
  \end{subfigure}
  \caption{Projected 95\cl~exclusion contours for the $\azw$ channel in the $(c_B/f_a, c_W/f_a, c_{a\Phi}/f_a)$ system. The assumed integrated luminosity is $\int L=450~\text{fb}^{-1}$ for~(a) and $\int L=3000~\text{fb}^{-1}$ for~(b).}
  \label{fig:azw_lep}
\end{figure}

The resulting exclusion limits are presented in Fig.~\ref{fig:azw_lep}. 
%
%%%%%%%%%
%
The constraints in the $(c_W/f_a, c_B/f_a)$ plane show a similar behavior to the $aW\gamma$ channel, reflecting the dependence on both electroweak couplings, but with different sensitivity due to the massive nature of the final state $Z$~bosons.

%%%%%%%%%%%%%%%%%%%%%%%%%%
%
\subsection{The $\aww$ final state}
We now turn to the associated production of an ALP with a pair of $W$~bosons, $pp \to \aww$. This process is sensitive to the $c_W$ coupling through diagrams involving a triple gauge-boson vertex including  $W$~bosons, as well as to the $c_G$ coupling via gluon-initiated diagrams,  similar to the neutral diboson channels considered above. Thus, similar to the $\agg$ and $\azg$ channels, the $\aww$ process involves topologies with $N_a=1$ and $N_a=2$. 

We focus on the fully leptonic decay channel where both $W$~bosons decay into a charged lepton and a neutrino. The final state then consists of two oppositely charged leptons, which can be of different flavors ($e^+\mu^-, \mu^+ e^-$) or same flavors ($e^+e^-, \mu^+\mu^-$), accompanied by significant missing transverse energy from the two neutrinos and the escaping ALP. We analyze the different-flavor and same-flavor channels separately, as they have different background compositions, and combine the results at the end. 
The fully leptonic decay mode of the $\aww$ channel provides a relatively clean environment compared to channels with semi-leptonic or hadronic decays of the $W$~bosons, although the presence of several invisible particles complicates the reconstruction of the individual $W^\pm$~bosons' kinematics.
The primary background for this signature is $W^+W^-$ production within the SM. Other significant backgrounds include $t\bar{t}$ and $t\bar{t}Z$ production (where $b$-jets are vetoed or lost), Drell-Yan processes (particularly for same-flavor lepton pairs), and $ZZ$ events where one of the $Z$~bosons decays leptonically, and the other one into a pair of neutrinos. 
For our $\aww$ analysis we require exactly two charged leptons and apply the following preselection cuts on the charged leptons: %
\beq
\label{eq:aww-cuts-leptons}
\ptl > 25~\text{GeV}\,, \quad |\etal| < 2.5\,.
\eeq
To suppress the top-quark background, we apply a $b$-jet veto requiring the absence of $b$-tagged jets with 
\beq
p_T^b > 20~\text{GeV}\,,\quad |\eta^b| < 2.5. 
\eeq
Additional selection criteria are imposed as follows:
\beq
\label{eq:aww-cuts}
E_T^{miss} > 30~\text{GeV}\,, \quad \Delta R_{\ell\ell} > 0.4\,, \quad m_{\ell\ell} > 20~\text{GeV}\,.
\eeq

Since the full kinematics of the $W^\pm$~bosons cannot be reconstructed due to the three invisible particles of the signal signature, we rely on variables reconstructed from the momenta of the visible charged leptons, i.e.\ the invariant mass $\mll$, transverse momentum $\ptll$, and separation in the pseudorapidity-azimuthal angle plane $\drll$, and the missing transverse mass system, $\etmiss$. Their distributions for the signal and background contributions, after applying the baseline selection cuts of Eqs.~\eqref{eq:aww-cuts-leptons} to \eqref{eq:aww-cuts}.
%
%%%%%%%%
%
\begin{figure}[!tp]
  \centering
  \begin{subfigure}{0.45\textwidth}
    \includegraphics[width=\textwidth]{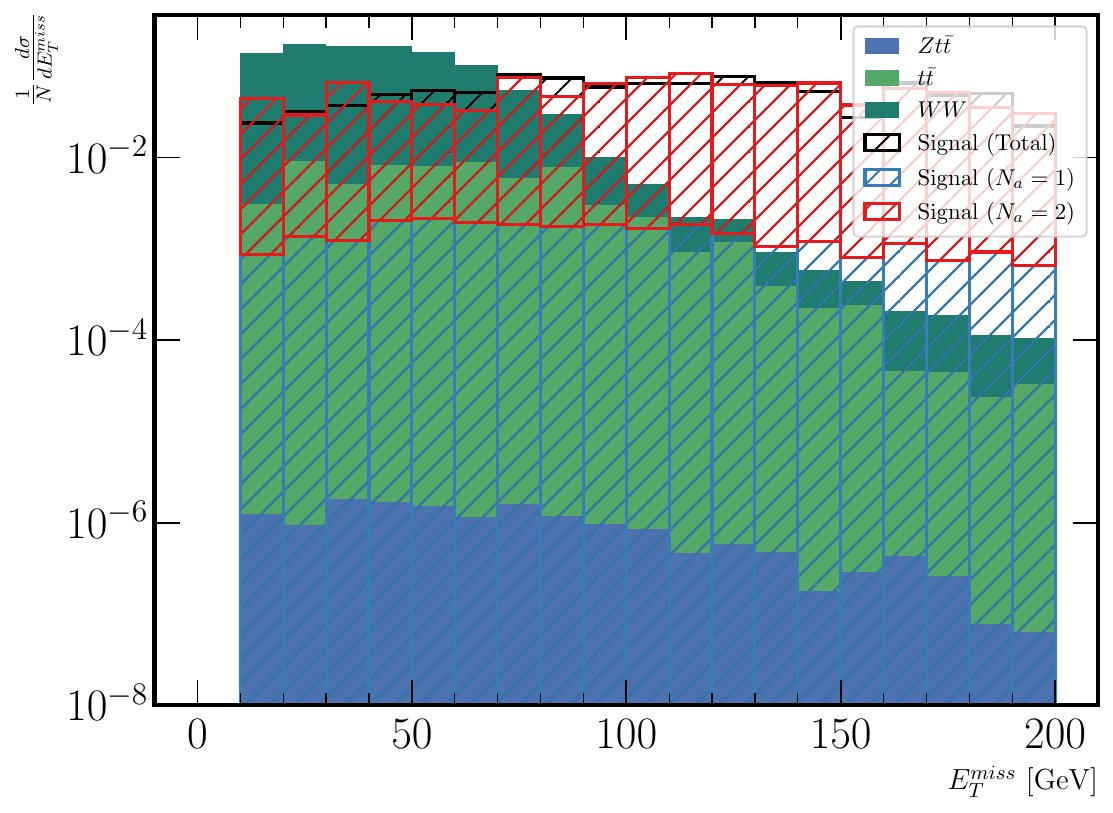}
    \caption{}
    \label{fig:aww_etmiss_nozz}
  \end{subfigure}
  \begin{subfigure}{0.45\textwidth}
    \includegraphics[width=\textwidth]{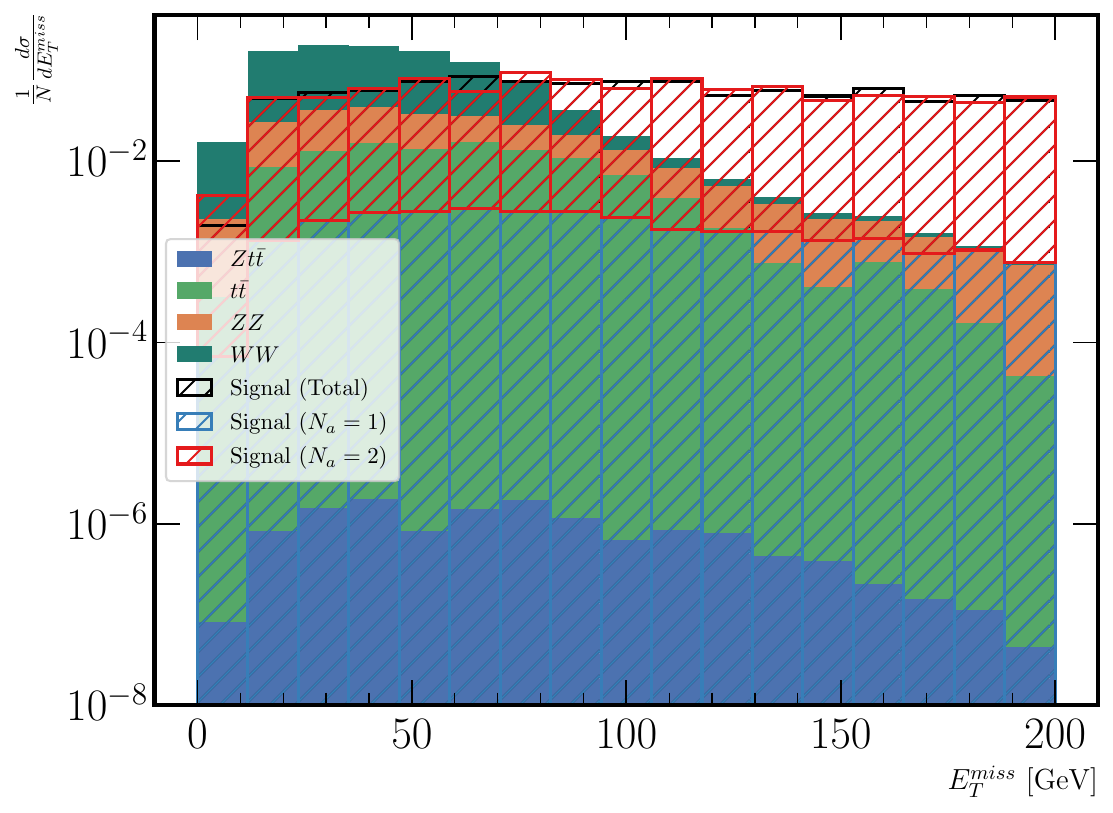}
    \caption{}
    \label{fig:aww_etmiss_zz}
  \end{subfigure}
  \\[5mm]
  \begin{subfigure}{0.45\textwidth}
    \includegraphics[width=\textwidth]{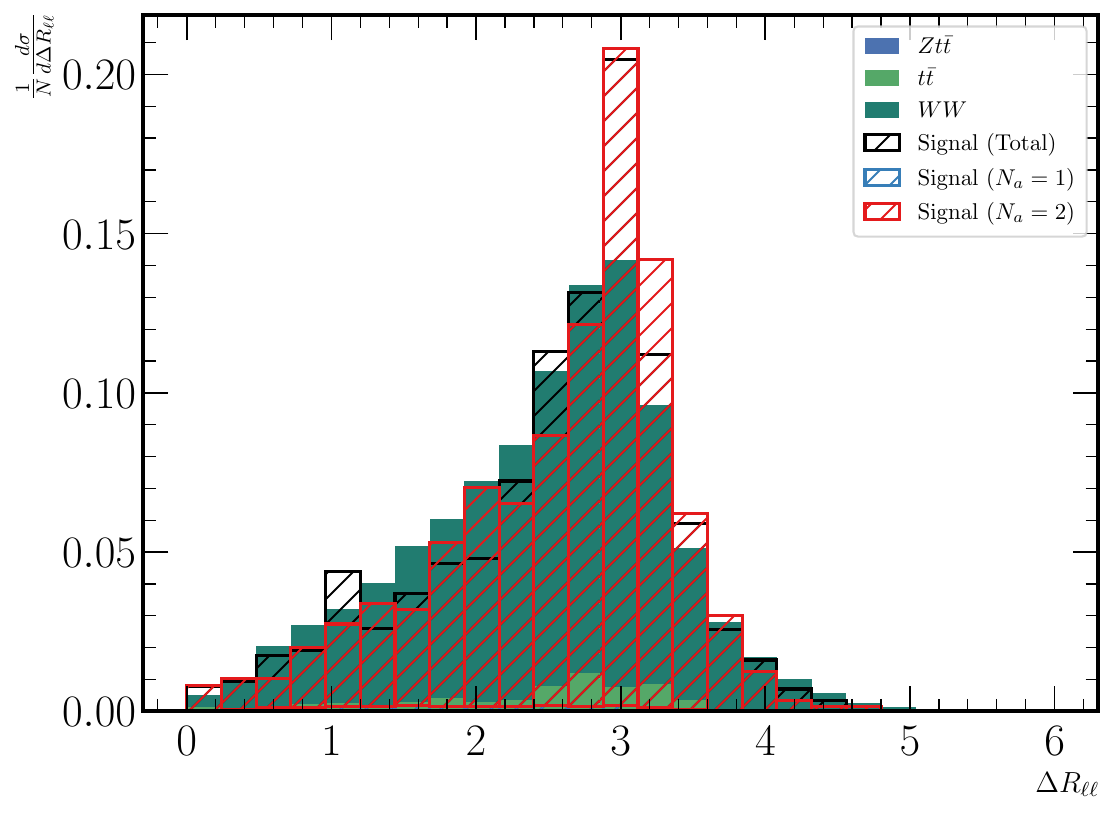}
    \caption{}
    \label{fig:aww_drll_nozz}
  \end{subfigure}
  \begin{subfigure}{0.45\textwidth}
    \includegraphics[width=\textwidth]{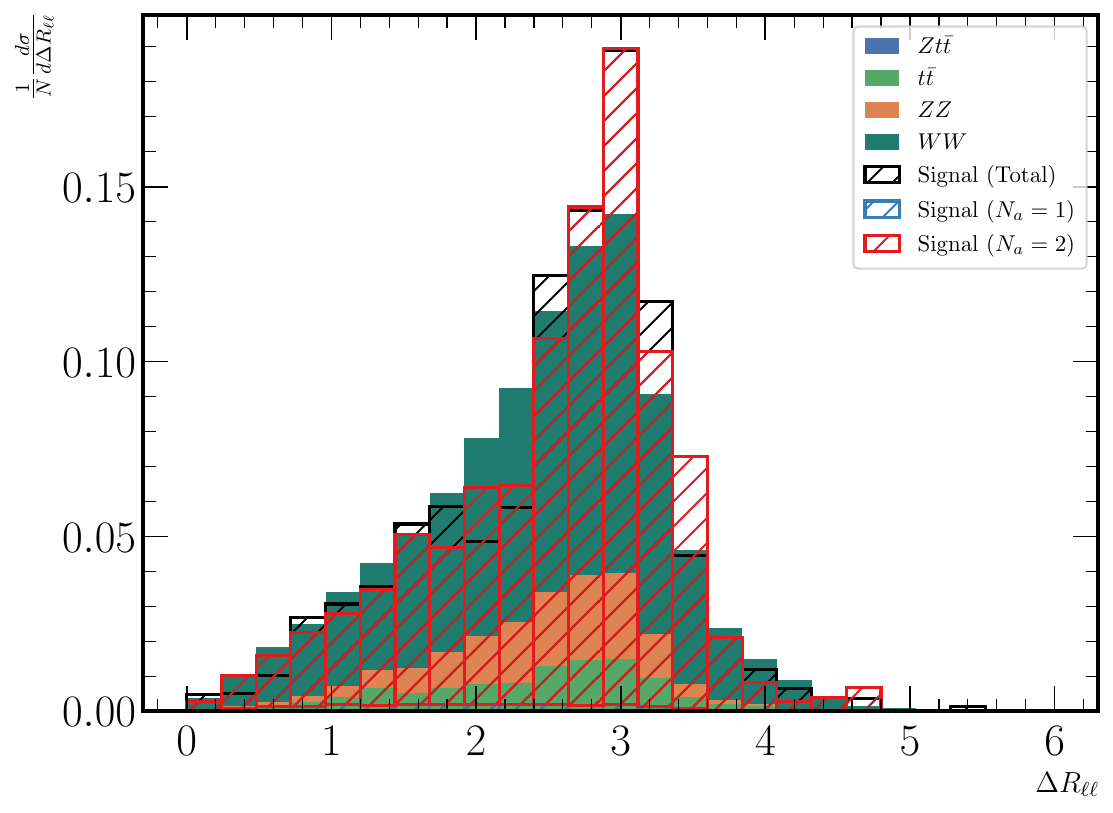}
    \caption{}
    \label{fig:aww_drll_zz}
  \end{subfigure}
  \\[5mm]
  \begin{subfigure}{0.45\textwidth}
    \includegraphics[width=\textwidth]{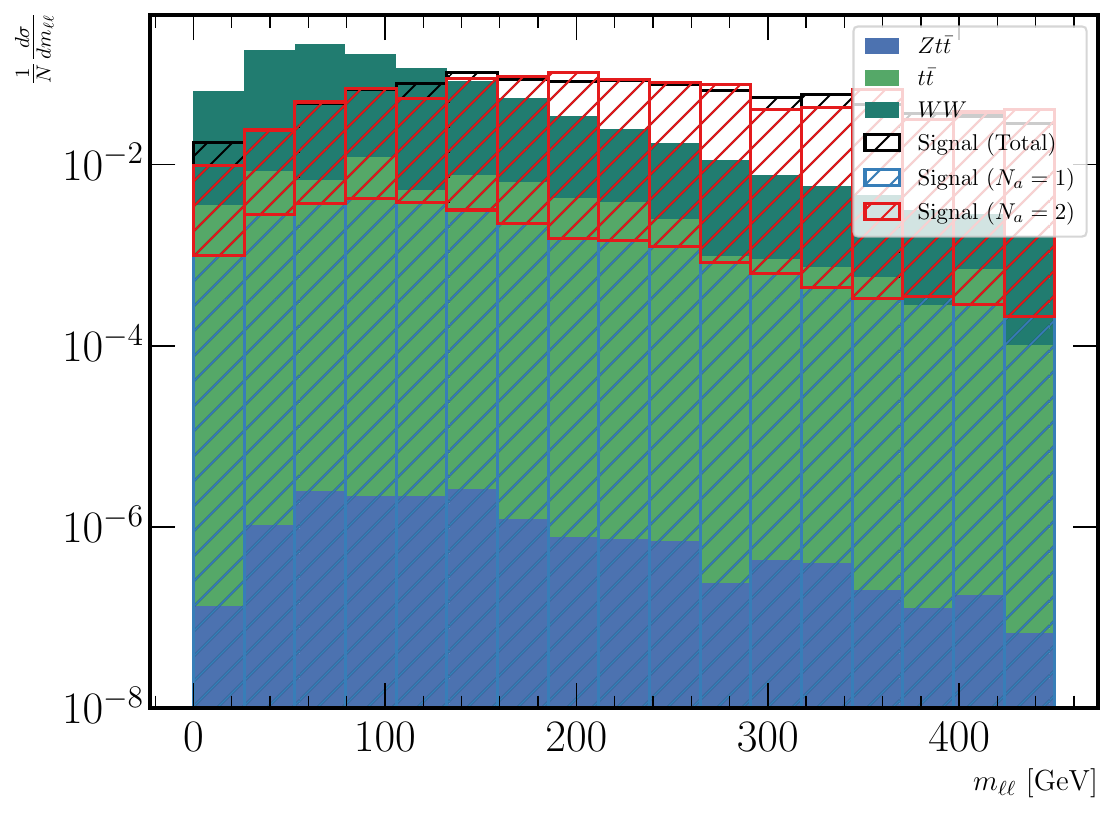}
    \caption{}
    \label{fig:aww_mll_nozz}
  \end{subfigure}
  \begin{subfigure}{0.45\textwidth}
    \includegraphics[width=\textwidth]{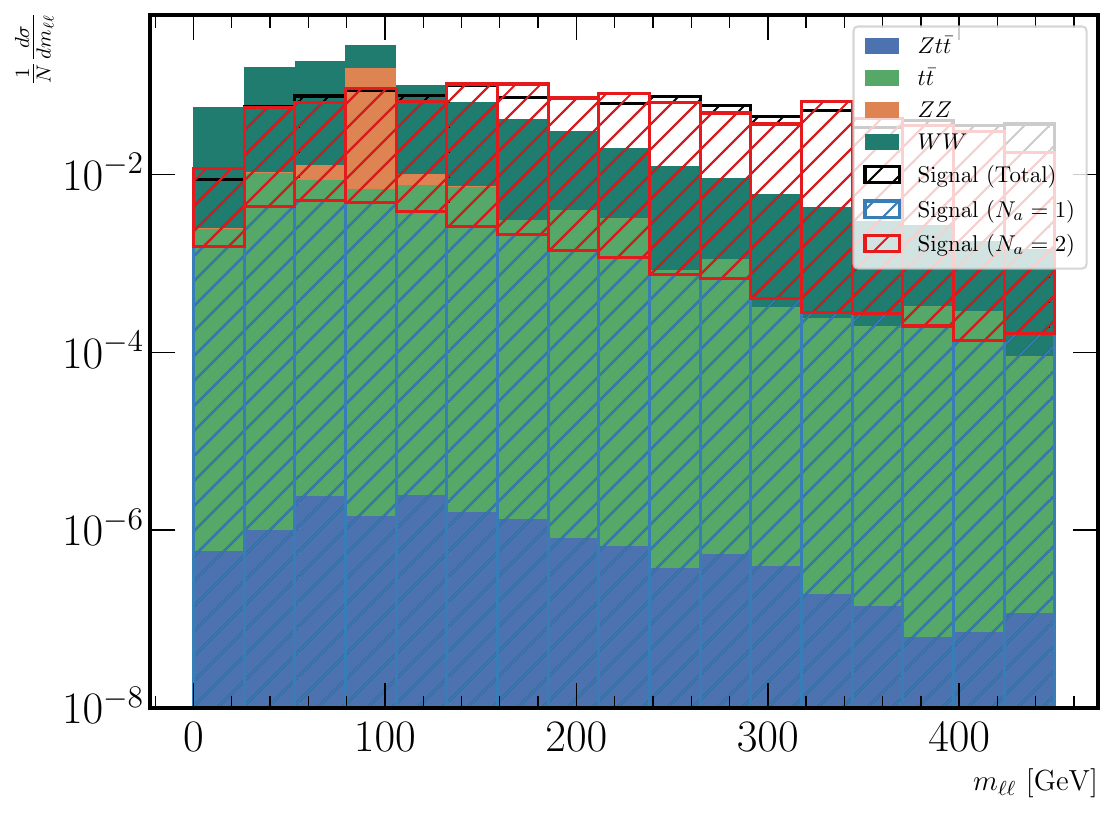}
    \caption{}
    \label{fig:aww_mll_zz}
  \end{subfigure}
  \caption{Normalized distributions for the simulated signal and background contributions to the $\aww$ channel at the LHC with $\sqrt{s}=14$~TeV using the benchmark point BP3 specified in Eq.~\eqref{eq:BP3}. In each case the hatched histograms represent the signal scenarios with $N_a=1$~(blue), $N_a=2$~(red) and their sum~(black), whereas the filled histograms represent the $t\bar{t}Z$~(blue), $ZZ$~(orange), $WW$~(dark green) and $t\bar{t}$~(light green) background contributions. The left (right) panels correspond to the different-flavor (same-flavor) lepton decay modes, respectively.} 
\label{fig:aww}
\end{figure}

The dilepton invariant-mass distributions in Fig.~\ref{fig:aww_mll_zz} show that, in the same-flavor channel, the $ZZ$ background exhibits a pronounced peak around the $Z$-boson mass, motivating an additional $Z$-mass veto. Therefore, for same-flavor contributions, we veto events with an invariant di-lepton mass close to the $Z$-boson mass by requiring
\beq
\label{eq:aww-cuts-sf}
|m_{\ell\ell} - m_Z| > 15~\text{GeV}\,.
\eeq
Subsequently,  we apply this additional cut on top of the baseline selection.
%
%%%%%%%
%
The missing transverse energy distributions of Figs.~\ref{fig:aww_etmiss_nozz} and \ref{fig:aww_etmiss_zz} exhibit a broad signal distribution, as expected from the presence of multiple invisible particles in the final state. However, the same-flavor channel (Fig.~\ref{fig:aww_etmiss_zz}) suffers from a larger background tail, primarily due to the $ZZ$ contribution, which reduces the discriminating power compared to the different-flavor case. The angular separation between the two leptons, $\drll$, is displayed in Figs.~\ref{fig:aww_drll_nozz} and \ref{fig:aww_drll_zz}. Both decay modes result in qualitatively similar distributions, providing very weak discrimination power. Finally, the dilepton invariant mass distributions are shown in Figs.~\ref{fig:aww_mll_nozz} and \ref{fig:aww_mll_zz}. In contrast to the $\etmiss$  distributions, the same-flavor channel  provides better discrimination, as the $ZZ$ background exhibits a pronounced peak around the $Z$~boson mass, $m_{\ell\ell} \approx m_Z$, which is effectively suppressed by the $Z$-mass veto of Eq.~\eqref{eq:aww-cuts-sf}. The distributions shown here are obtained before the application of this specific veto, in order to illustrate its necessity and the kinematic properties of the background components. Although not shown here, the transverse momentum of the dilepton system, $\ptll$, also provides good discriminating power, with similar behavior as the missing transverse energy.

\begin{figure}[!tp]
  \centering
      \begin{subfigure}{0.45\textwidth}
        \includegraphics[width=\textwidth]{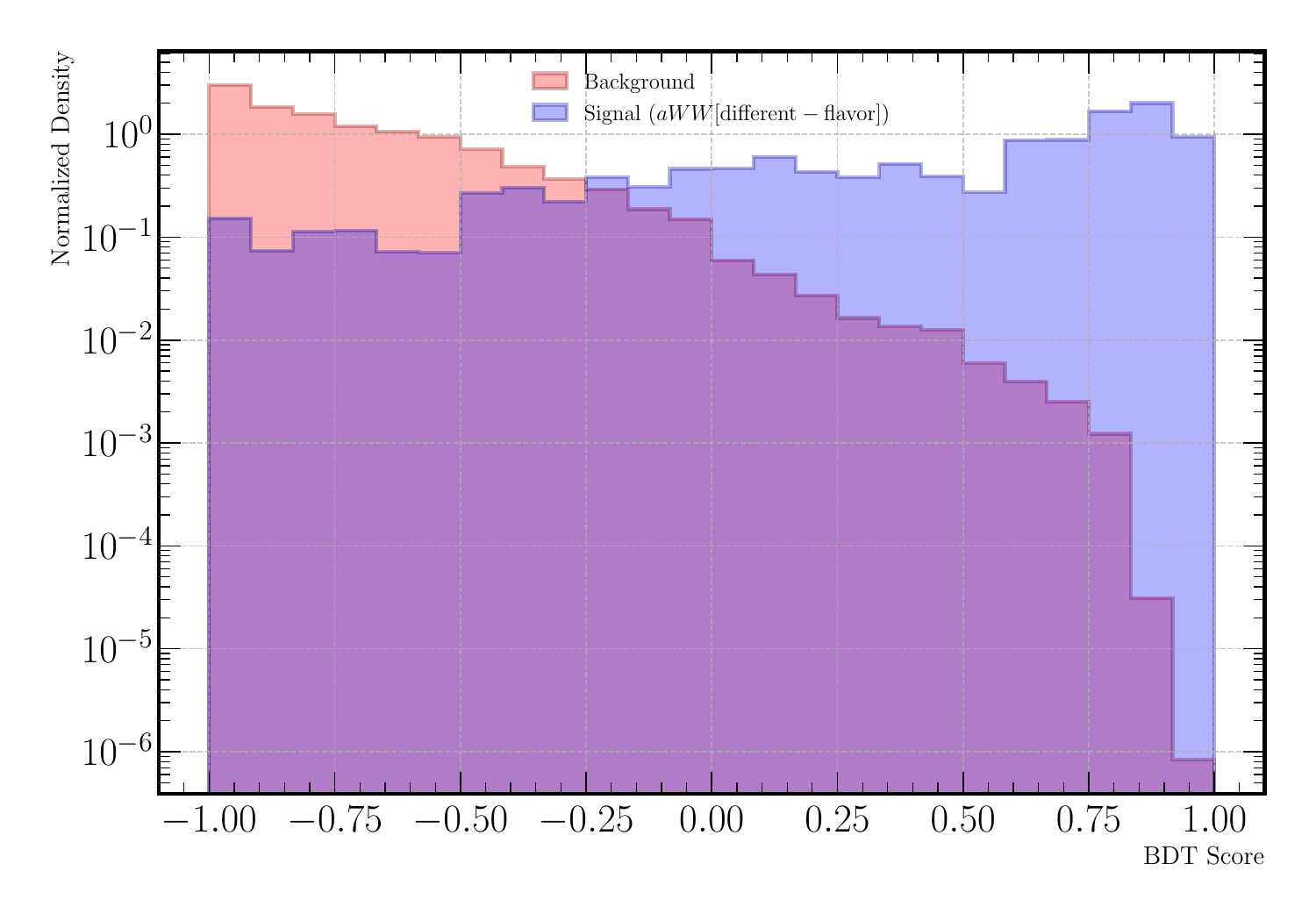} 
        \caption{}     
        \label{fig:aww_bdt_a}
      \end{subfigure}
      \begin{subfigure}{0.45\textwidth}
        \includegraphics[width=\textwidth]{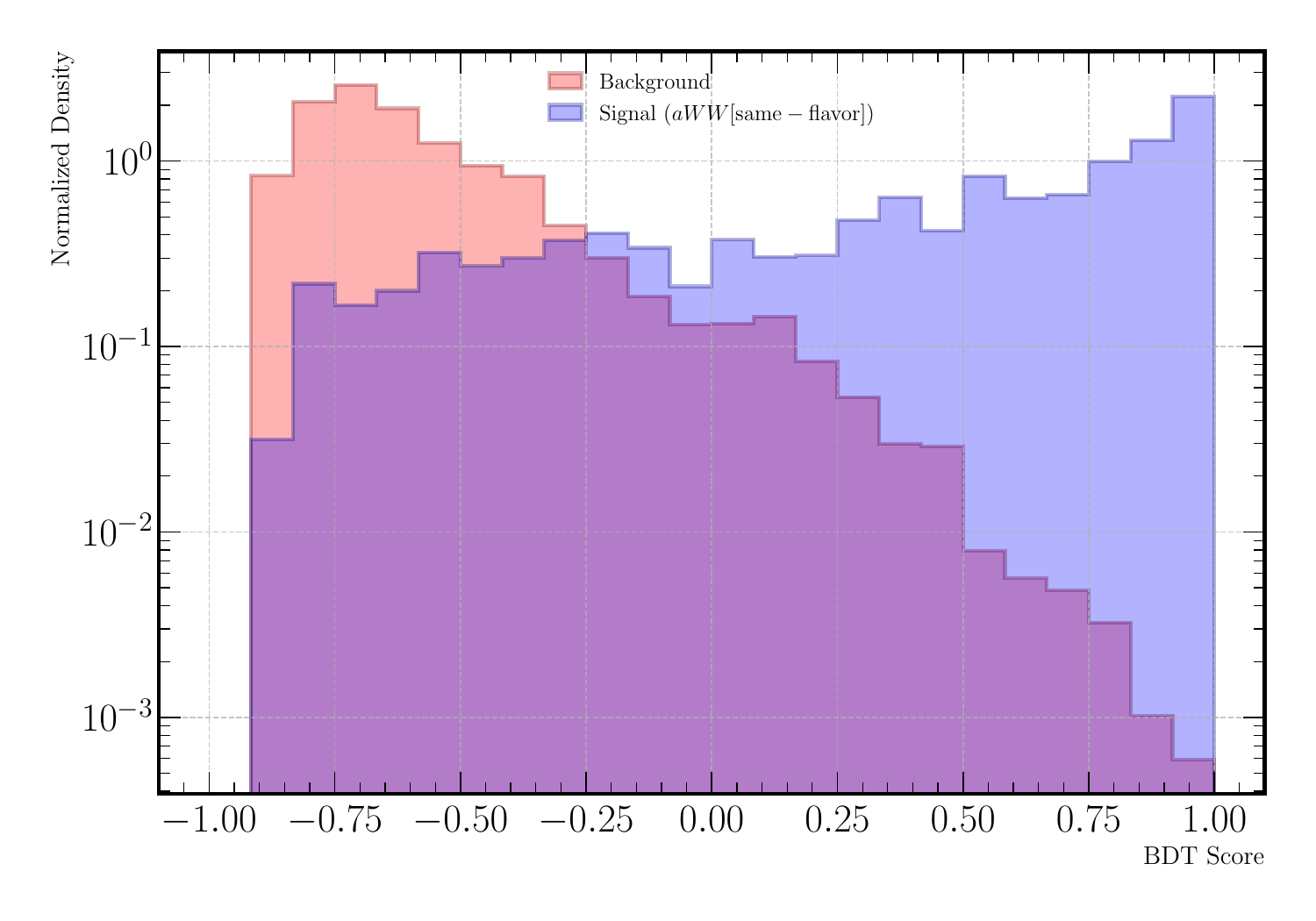}   
        \caption{}  
        \label{fig:aww_bdt_b}
      \end{subfigure}
      \caption{Normalized BDT score distributions for the $\aww$ signal versus the sum of the relevant background processes given in Tab.~\ref{tab:bdt_var_separation_aww} for the benchmark point BP3 of Eq.~\eqref{eq:BP3} for the different-flavor (a) and same-flavor (b) lepton channels.}
      \label{fig:aww_bdt}
\end{figure}

\begin{figure}[!tp]
  \centering
  \begin{subfigure}{0.45\textwidth}
  \includegraphics[width=\textwidth]{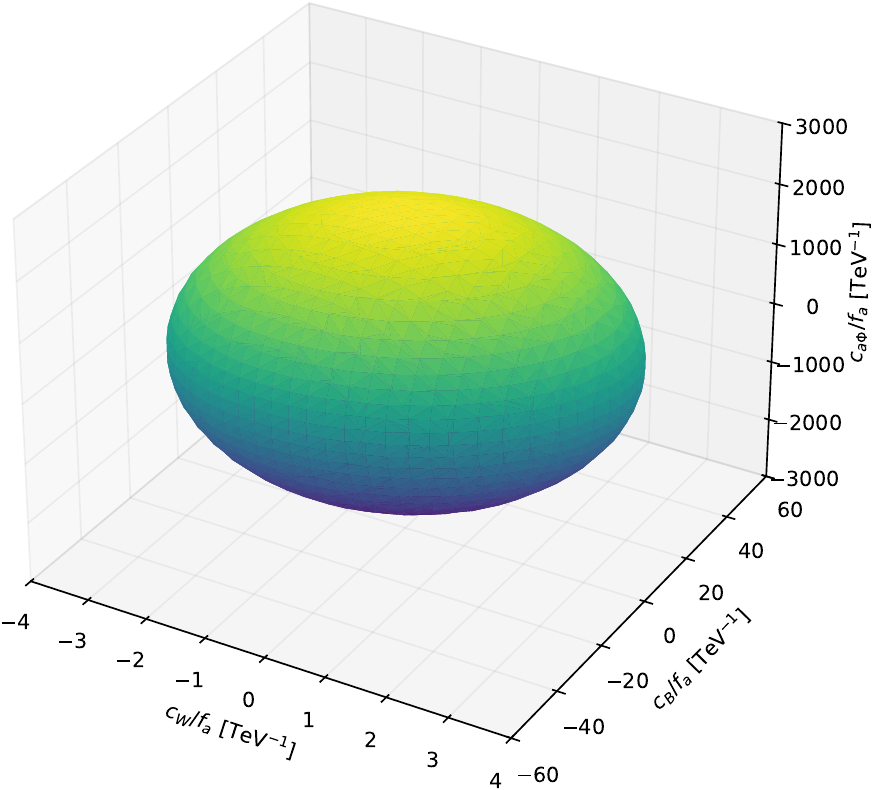}
  \caption{}
  \label{fig:aww_lep_a}
  \end{subfigure}
  \hfill
  \begin{subfigure}{0.45\textwidth}
  \includegraphics[width=\textwidth]{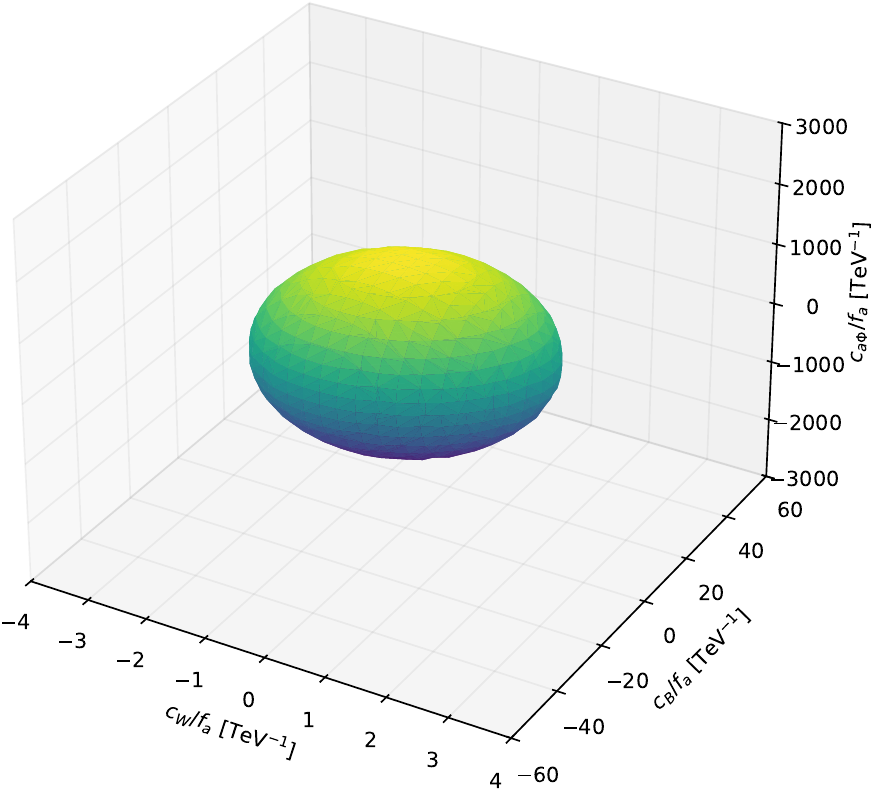}
  \caption{}
  \label{fig:aww_lep_b}
  \end{subfigure}
  \caption{Projected 95\cl~exclusion contours for the $\aww$ channel in the $(c_B/f_a, c_W/f_a, c_{a\Phi}/f_a)$ system with $c_G/f_a$ set to $0.1~\text{TeV}^{-1}$. The assumed integrated luminosity is $\int L=450~\text{fb}^{-1}$ for (a) and $\int L=3000~\text{fb}^{-1}$ for (b).}
  \label{fig:aww_lep}
\end{figure}

For the multivariate analysis, we train a BDT using the variables $\etmiss,\ptll,\drll$ and $\mll$. Their ranking and separation power, after the application of the pre-selection cuts, is summarized in Tab.~\ref{tab:bdt_var_separation_aww}.
%
%%%%%%%%
%
\begin{table}[t]
  \centering
  \begin{tabular}{c||c|c}
    &\textbf{different-flavor} & \textbf{same-flavor} \\
    \hline
    variable & \multicolumn{2}{c}{separation} \\
    \hline
    $\etmiss$ & 0.3093 & 0.3422 \\
    $\ptll$ & 0.3045 & 0.3081 \\
    $\drll$ & 0.2151 & 0.1767 \\
    $\mll$ & 0.1711 & 0.1731
  \end{tabular}
    \caption{BDT input variables and their separation power for the $\aww$ channel after applying the pre-selection cuts, where the first (second) column corresponds to the different-flavor (same-flavor) lepton channels.}
  \label{tab:bdt_var_separation_aww}
\end{table}
%
%%%%%%%
%
%
As expected from the distributions shown in Fig.~\ref{fig:aww}, the missing transverse energy provides the highest separation power in both lepton channels, followed by the dilepton transverse momentum and angular separation. The dilepton invariant mass has the lowest discriminating power. The normalized BDT score distributions for both lepton channels are shown in Fig.~\ref{fig:aww_bdt}. 
%
%
%%%%%%%%%
%
%
%%%%%%%%
%
The resulting exclusion limits for the $\aww$ channel, for the example of $c_G/f_a=0.1~\text{TeV}^{-1}$, are shown in Fig.~\ref{fig:aww_lep}. 
Although looking quite similar, the different-flavor channel of Fig.~\ref{fig:aww_bdt_a} exhibits a slightly better separation between signal and background compared to the same-flavor case of Fig.~\ref{fig:aww_bdt_b}, consistent with the observations from the individual kinematic distributions. 

In Fig.~\ref{fig:aww_lep} 
%
%%%%%%
%
%
%%%%%%%%
%
our results are presented in the $(c_B/f_a, c_W/f_a, c_{a\Phi}/f_a)$ parameter space. The exclusion contours exhibit a similar elliptical shape as in the $\azg$, $\awg$ and $\azw$ channels considered above, reflecting the dependence on both electroweak couplings. However, the sensitivity to $c_W$ is now notably enhanced due to the direct involvement of $W^\pm$~bosons in the final state. As a result, the constraints on $c_W$ are significantly tighter than those on $c_B$. We note that the bounds are mostly determined by the different-flavor lepton channel. 

%%%%%%%%%%%%%%%%%%%%%%%%%%
%
\subsection{The $\azz$ final state}

Finally, we investigate the associated production of an ALP with a pair of $Z$ bosons. This channel offers a unique probe of the $g_{aZZ}$ coupling, which depends on the linear combination $c_W \cos^2\theta_W + c_B \sin^2\theta_W$.  
The $\azz$ final state allows for a rich variety of decay topologies. While the fully hadronic decay modes suffer from overwhelming QCD backgrounds, and the fully invisible modes where both $Z$~bosons decay into neutrinos are experimentally challenging to trigger  without additional hard recoil objects, the semi-leptonic and fully leptonic channels provide cleaner signatures.
We focus on two primary decay channels of the $ZZ$~system: the 4-lepton ($4\ell$) channel, where both $Z$~bosons decay into charged lepton pairs, resulting in an $e^+e^-e^+e^-$ $(4e)$,  $e^+e^-\mu^+\mu^-$ $(2e2\mu)$, or $\mu^+\mu^-\mu^+\mu^-$ $(4\mu)$ system, and the 2-lepton+2-neutrino ($2\ell2\nu$) channel, where one $Z$~boson decays into an $e^+e^-$ or $\mu^+\mu^-$ pair and the other one into a pair of neutrinos.

The $4\ell$ channel is characterized by very low SM backgrounds but a small branching ratio. The signal consists of exactly four isolated leptons forming two distinct opposite-sign, same-flavor pairs, each with an invariant mass consistent with the $Z$~boson mass, accompanied by significant missing transverse energy from the ALP. We apply the following preselection cuts on these four charged leptons: 
\beq
\label{eq:azz-4l-cuts-leptons}
\ptl > 25~\text{GeV}\,, \quad |\etal| < 2.5\,.
\eeq
To reconstruct the two $Z$~boson candidates in events with exactly four charged leptons, we form two opposite-sign, same-flavor dilepton pairs. In the $2e2\mu$ case the assignment is unique ($Z_1\to e^+e^-$, $Z_2\to\mu^+\mu^-$), while for the $4e$ and $4\mu$ cases we consider all distinct opposite-sign, same-flavor pairings and choose the one minimizing $|m_{Z_1}-m_Z|+|m_{Z_2}-m_Z|$, where $m_{Z_{1,2}}$ denote the corresponding dilepton invariant masses. Both lepton pairs identified as the $Z$~boson decay candidates are required to satisfy 
\beq
\label{eq:azz-4l-cuts-mZ}
|\mll^{(i)} - m_Z| < 15~\text{GeV},
\eeq
and each of the leptons must be separated by 
\beq
\Delta R_{\ell\ell}>0.4
\eeq
from the other leptons.
The dominant SM backgrounds for this topology include $ZZ$, $t\bar{t}Z$, $t\bar{t}t\bar{t}$, and rare multiboson production processes such as $W^+W^-Z$ and $W^+W^-W^+W^-$ production. Reducible backgrounds from $t\bar{t}$ and $Z+\text{jets}$ production are effectively suppressed by our strict isolation requirements on the four leptons.

The $2\ell2\nu$ channel benefits from a larger branching ratio but also faces larger backgrounds. The signal  signature consists of a pair of exactly two same-flavor oppositely charged leptons with an invariant mass consistent with the mass of the $Z$~boson, and large missing transverse energy. We apply the same preselection cuts on the charged leptons as in Eq.~\eqref{eq:azz-4l-cuts-leptons}. The would-be $Z$ boson is reconstructed from the two opposite-sign, same-flavor leptons with an invariant mass satisfying the condition 
\beq
\label{eq:azz-2l2v-cuts-mZ}
|\mll - m_Z| < 15~\text{GeV},\quad \Delta R_{\ell^+\ell^-}>0.4.
\eeq
The missing transverse energy now receives contributions from both the invisible $Z$-boson decay products and the escaping ALP. 
We conservatively require  
\beq
\label{eq:azz-2l2n-cuts-etmiss}
E_T^{miss} > 100~\text{GeV}
\eeq
to suppress backgrounds with low genuine missing energy, and rely on the multivariate analysis to discriminate signal from background based on the full kinematic correlations. 
Major SM backgrounds to the $2\ell2\nu$ decay mode of the $\azz$ process are the $ZZ$, $Zt\bar{t}$, $WW$, and $t\bar{t}$ production processes. 

For the multivariate analysis, we train separate BDTs for the $4\ell$ and $2\ell2\nu$ channels.  
In the $4\ell$ case, we use the transverse momentum $p_T^{\ell\ell,1}$ of the leading dilepton pair, computed from the transverse momenta of the two relevant leptons; 
the invariant mass $m_{4\ell}$ of the four-lepton system, calculated from the four-momenta of all four leptons; the scalar sum of the transverse momenta of all four leptons, $P_T = \sum_{i=1}^{4} p_T^{\ell_i}$; the missing transverse energy $\etmiss$. 

For the $2\ell2\nu$ channel, we utilize the transverse momentum $p_T^{\ell\ell}$ of the reconstructed dilepton system, computed from the momenta of two leptons that are assumed to originate from the $Z$~boson decay;   
the pseudorapidity $\eta_{\ell\ell}$ of the reconstructed dilepton system; 
the transverse mass $\mtllinv$ of the system containing the reconstructed dilepton system and missing transverse energy, defined as 
\beq
  \mtllinv = \sqrt{2\, p_T^{\ell\ell}\, E_T^{miss}\, (1 - \cos\dpllinv)},
\eeq
where $\dpllinv$ is the azimuthal angle separation of the reconstructed dilepton system and the missing transverse momentum vector; 
the missing transverse energy; 
the separation $\Delta R_{\ell\ell}$ of  the two leptons in the pseudorapidity-azimuthal angle plane.

\begin{figure}[!tp]
  \centering
  \begin{subfigure}{0.45\textwidth}
    \includegraphics[width=\textwidth]{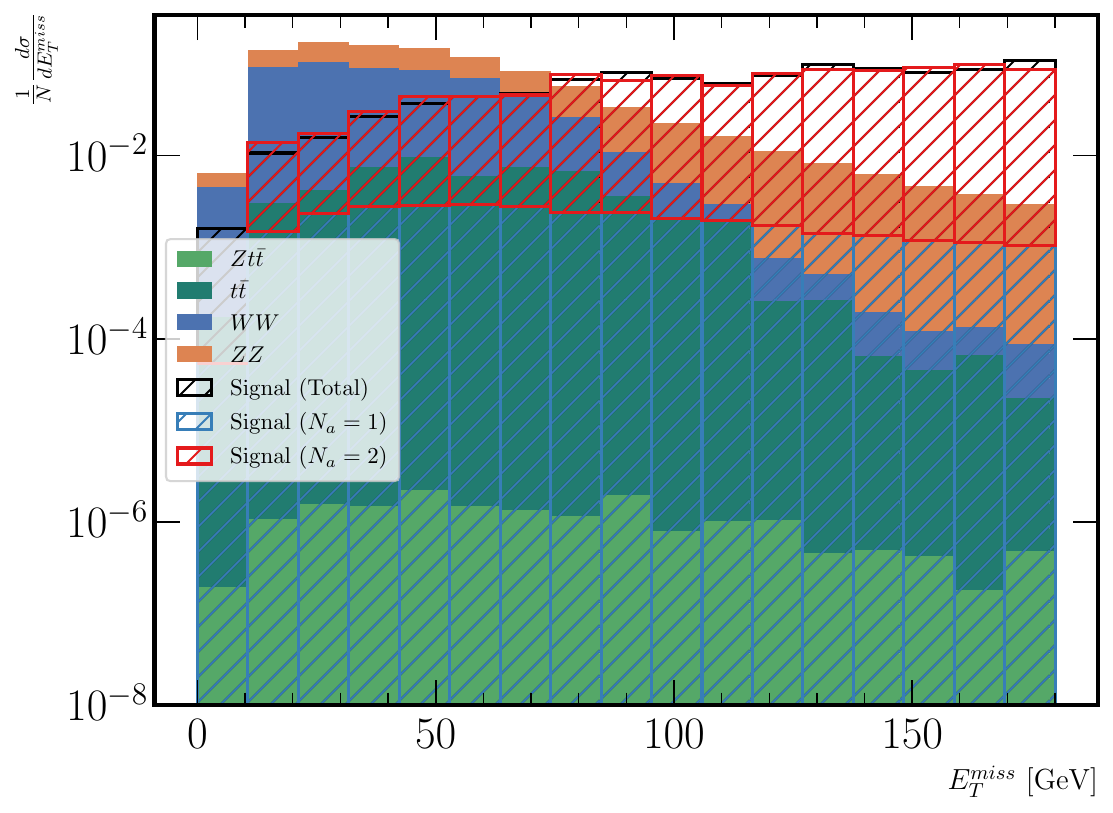}
    \caption{}
    \label{fig:azz_etmiss}
  \end{subfigure}
  \begin{subfigure}{0.45\textwidth}
    \includegraphics[width=\textwidth]{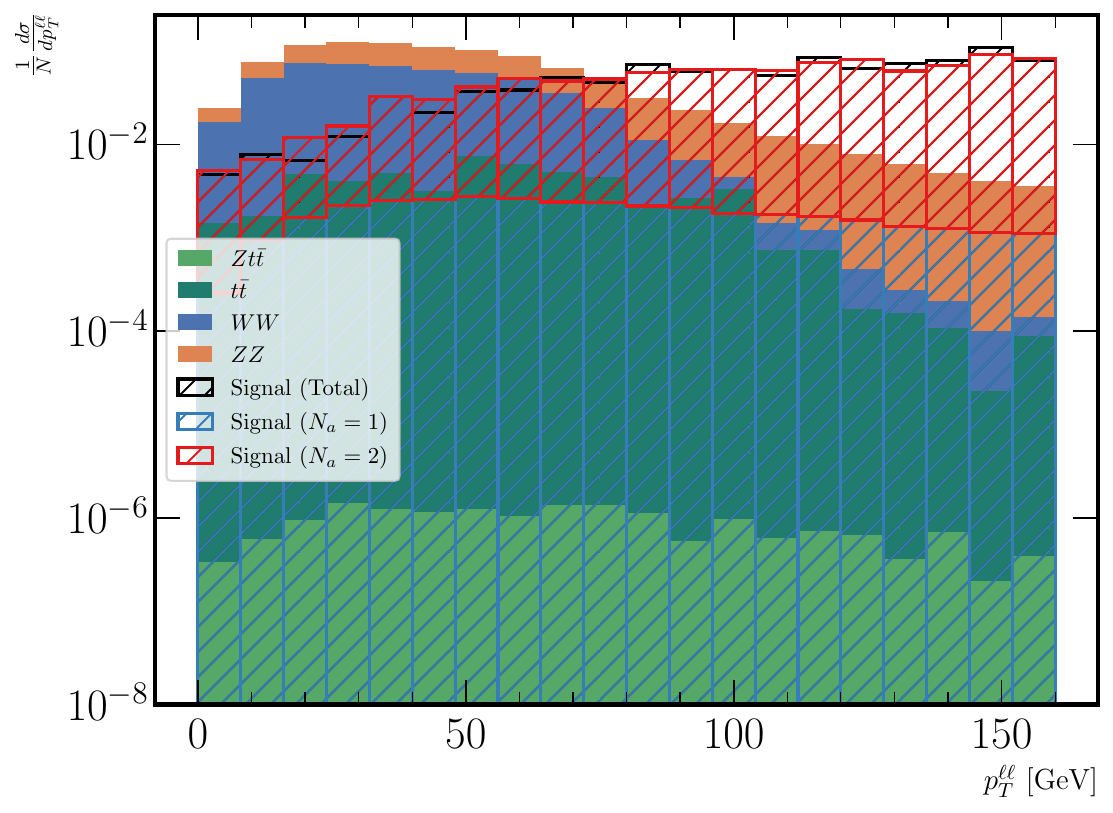}
    \caption{}
    \label{fig:azz_pT_Z}
  \end{subfigure}
  \\[5mm]
  \begin{subfigure}{0.45\textwidth}
    \includegraphics[width=\textwidth]{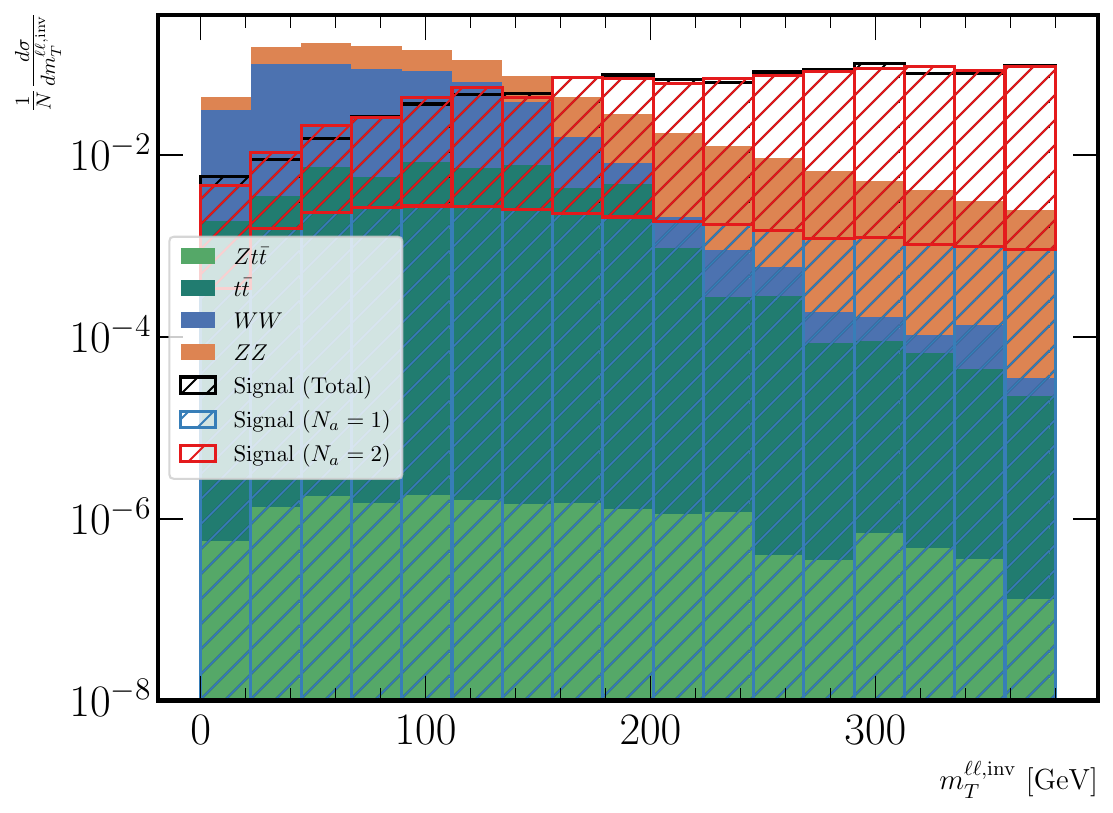}
    \caption{}
    \label{fig:azz_mt_Zmiss}
  \end{subfigure}
  \begin{subfigure}{0.45\textwidth}
    \includegraphics[width=\textwidth]{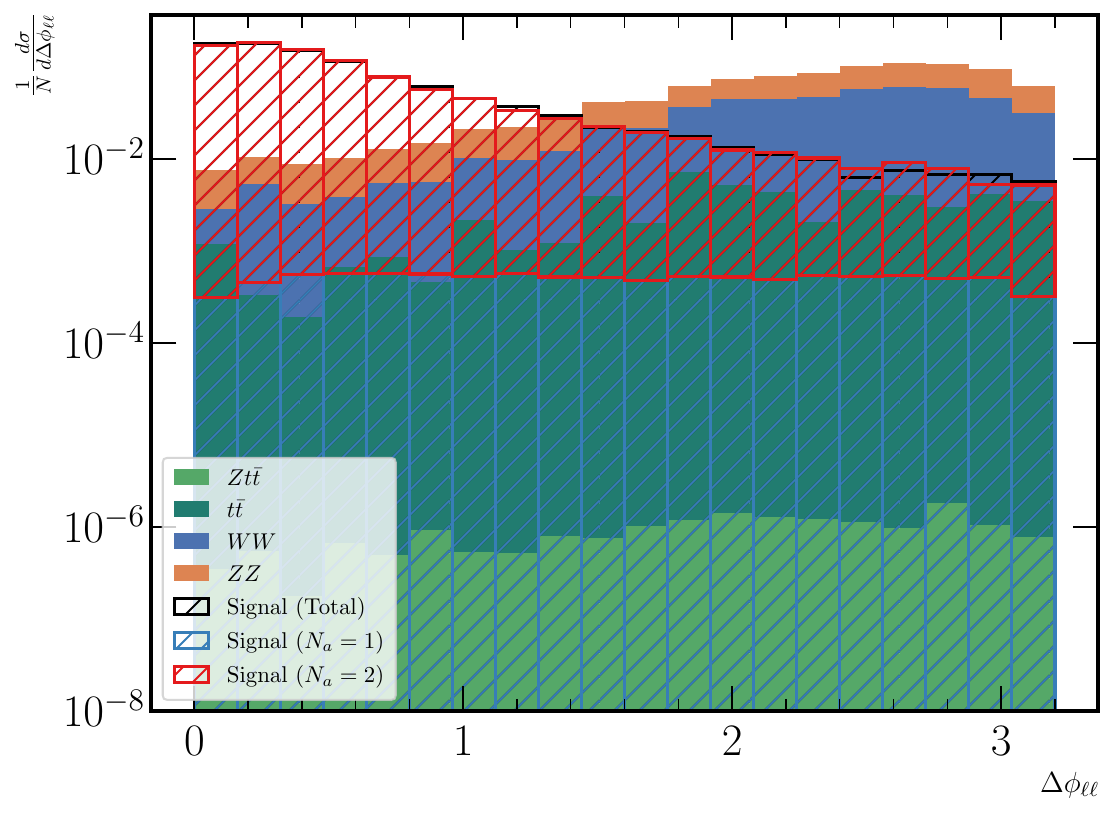}
    \caption{}
    \label{fig:azz_deltaPhi_ll}
  \end{subfigure}
  \caption{Normalized distributions for the simulated signal and background contributions to the $\azz$ process in the $2\ell2\nu$ decay mode at the LHC with $\sqrt{s}=14$~TeV using the benchmark point BP3 specified in Eq.~\eqref{eq:BP3}. The hatched histograms represent the signal scenarios with $N_a=1$~(blue), $N_a=2$~(red) and their sum~(black), whereas the filled histograms represent the $WW$~(blue), $ZZ$~(orange), $t\bar{t}$~(dark green) and $Zt\bar{t}$~(light green) background contributions.} 
\label{fig:azz_vars}
\end{figure}

As an example, the kinematic distributions of some of these variables for the $2\ell2\nu$ decay mode are shown in Fig.~\ref{fig:azz_vars}. 
%
%%%%%%%%
%
%
%%%%%%%
%
The missing transverse energy distribution in Fig.~\ref{fig:azz_etmiss} exhibits a broad shape for the signal, as expected from the invisible ALP and the neutrinos in the final state. The background distribution is also broadened, primarily due to the neutrinos from the SM $ZZ \to 2\ell2\nu$ process which contribute genuine missing energy; however, its high-$E_T^{\rm miss}$ tail is more strongly suppressed than for the signal.
Figure~\ref{fig:azz_pT_Z} displays the transverse momentum of the dilepton system. While the overall shape is similar to the missing energy distribution, the background shows a more pronounced peak at low transverse momentum values, reflecting the softer $p_T$ spectrum of $Z$~bosons in SM diboson production compared to the signal with an additional invisible particle. The $\mtllinv$ distribution is shown in Fig.~\ref{fig:azz_mt_Zmiss}. The behavior of this distribution is somewhat similar as in the case of the transverse momentum distribution. Finally, Fig.~\ref{fig:azz_deltaPhi_ll} shows the azimuthal angle separation between the two leptons. Here, the signal is somewhat more concentrated at smaller angular separations, while the background tends to populate larger values of $\dpll$. However, the two distributions still exhibit substantial overlap, so that this anti-correlation provides only mild additional discriminating power beyond the first three observables, which perform very similary among themselves. 

\begin{figure}[!tp]
  \centering
      \begin{subfigure}{0.45\textwidth}
        \includegraphics[width=\textwidth]{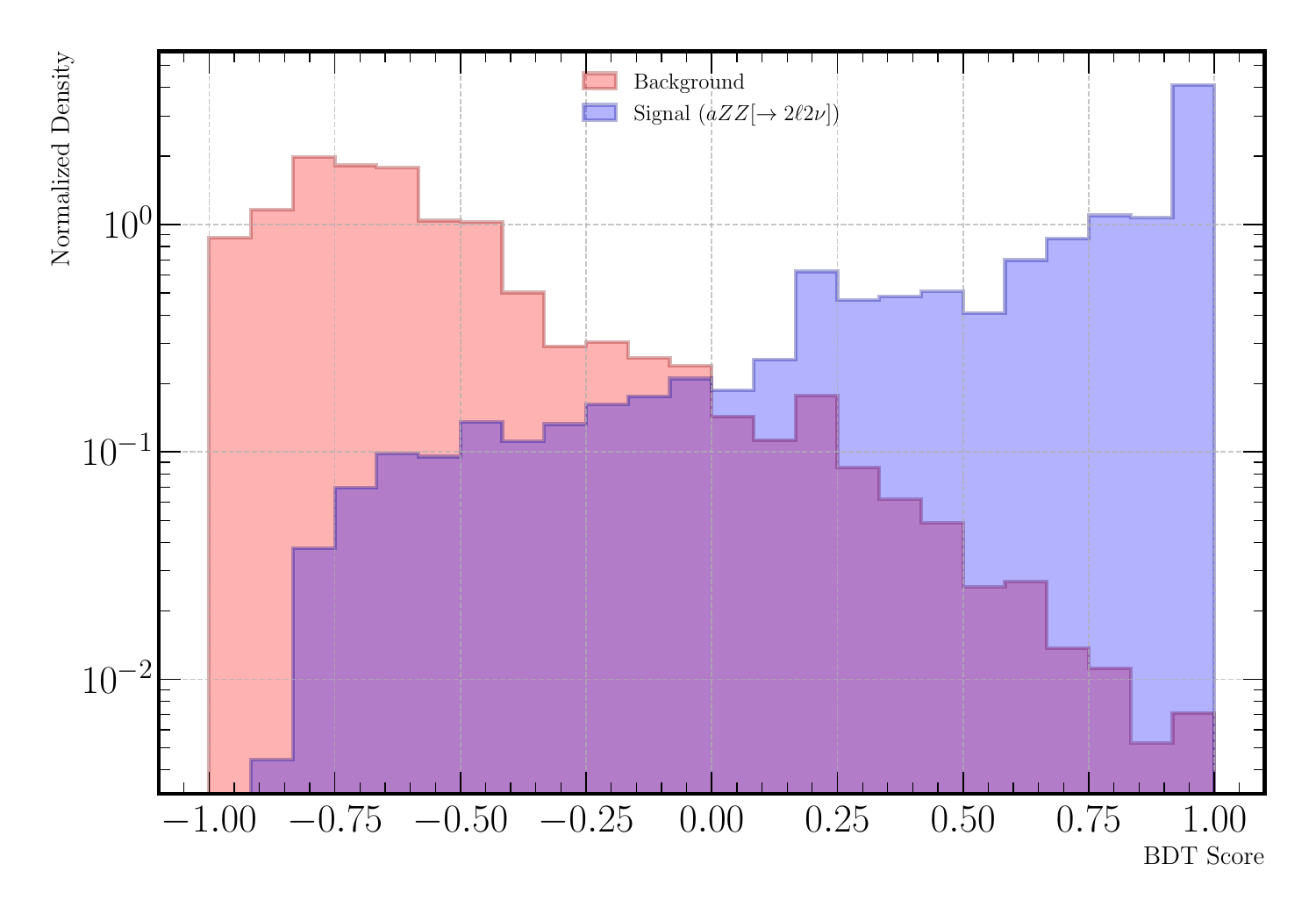}  
        \caption{}     
        \label{fig:azz_bdt_a}
      \end{subfigure}
      \begin{subfigure}{0.45\textwidth}
        \includegraphics[width=\textwidth]{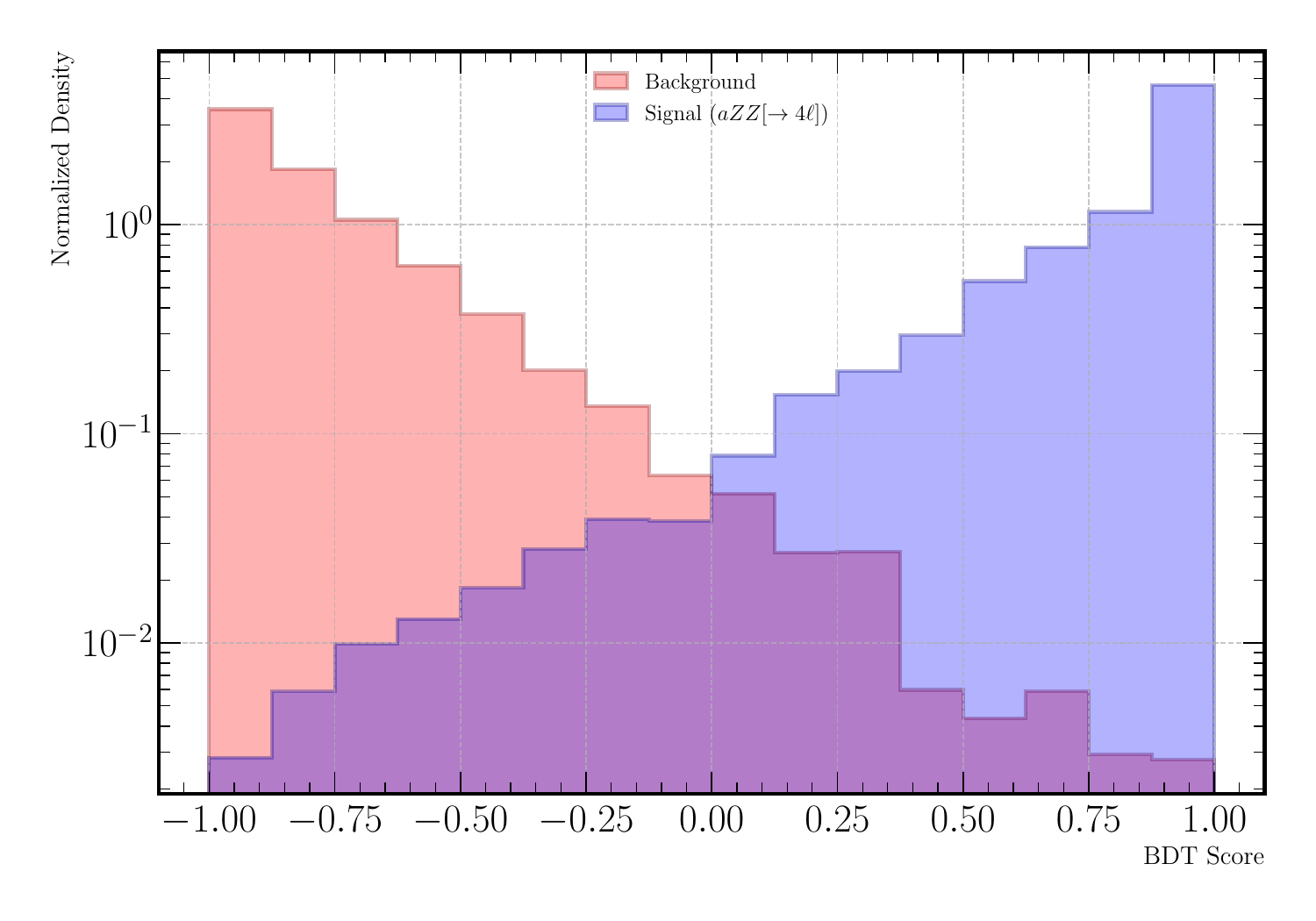}   
        \caption{}  
        \label{fig:azz_bdt_b}
      \end{subfigure}    
  \caption{Normalized BDT score distributions for the $\azz$ signal versus the sum of the relevant background processes given in Tab.~\ref{tab:bdt_var_separation_azz} for the benchmark point BP3 of Eq.~\eqref{eq:BP3} for the mixed (a) and purely leptonic (b) decay channels.}
  \label{fig:azz_bdt}
\end{figure}

The ranking and separation power of the variables entering our BDT for the $\azz$ process in the $2\ell2\nu$ and $4\ell$ channels, obtained after applying the above-mentioned preselection cuts, are summarized in Tab.~\ref{tab:bdt_var_separation_azz}.
%
%%%%%%
%
\begin{table}[t!]
  \centering
  \begin{tabular}{c|c||c|c}
    \multicolumn{2}{c||}{\textbf{$aZZ(\to2\ell2\nu)$}} & \multicolumn{2}{c}{\textbf{$aZZ(\to4\ell)$}} \\
    \hline
    variable & separation & variable & separation \\
    \hline
    $\etmiss$& 0.2622 & $\etmiss$ & 0.2727 \\
    $\ptll$ & 0.239 & $p_T^{\ell\ell,1}$ & 0.2609 \\
    $\mtllinv$ & 0.1727 & $m_{4\ell}$ & 0.2478 \\
    $\dpll$ & 0.1057 & $P_T$ & 0.1130 \\
    $\eta_{\ell\ell}$ & 0.0709 & - & - \\
    $\drll$ & 0.0680 & - & - \\
  \end{tabular}
  \caption{BDT input variables and their separation power for the $aZZ(\to2\ell2\nu)$ and $aZZ(\to4\ell)$ channels.}
  \label{tab:bdt_var_separation_azz}
\end{table}
%
%%%%%
%
As can be deduced from the histograms in Fig.~\ref{fig:azz_vars}, the missing energy, the (leading) dilepton transverse momentum, and the (transverse) invariant mass of the leptonic system provide similar discriminating power, with the missing energy being slightly more effective. For the $2\ell2\nu$ channel, the azimuthal angle separation of the two leptons also contributes somewhat to the signal-background separation, while the pseudorapidity and the $\Delta R$~separation of the two leptons have a comparably worse separation power. Overall, the BDT score distributions shown in Fig.~\ref{fig:azz_bdt} exhibit a slightly better separation between signal and background for the $4\ell$ than for the $2\ell2\nu$ channel. 
%
%%%%%%%
%
%
%%%%%%%
%

For the purely leptonic decay channel, the overall constraining performance is negatively affected by the smaller branching ratios, making both the $4\ell$ and $2\ell2\nu$ decay channels of the $\azz$ production mode poor probes of the ALP couplings.

\begin{figure}[!tp]
  \centering
      \begin{subfigure}{0.45\textwidth}
        \includegraphics[width=\textwidth]{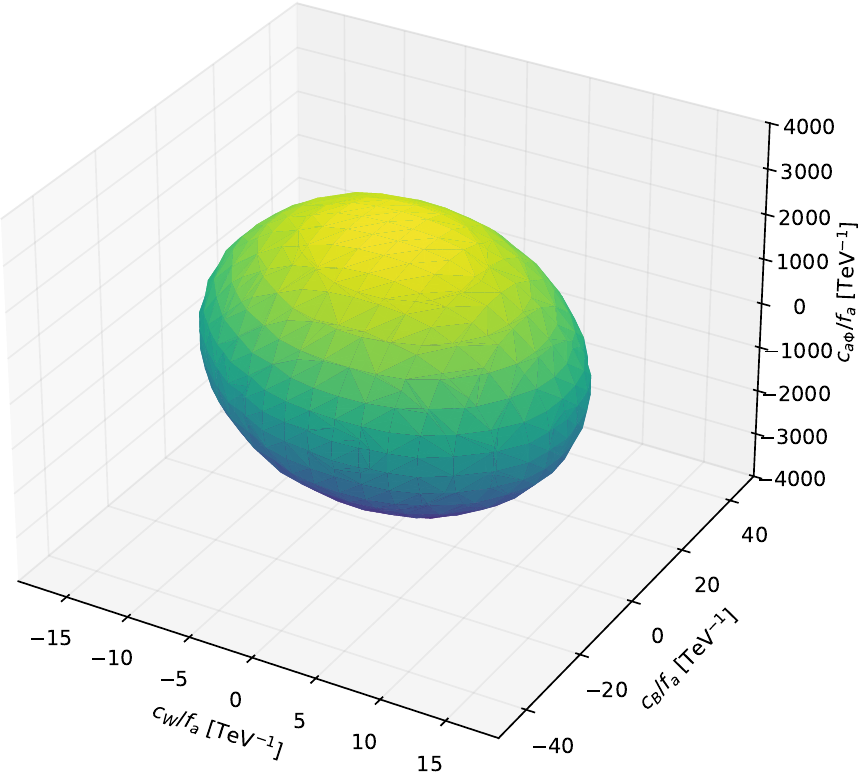}
        \caption{}
        \label{fig:azz_a}
        \end{subfigure}
        \hfill
        \begin{subfigure}{0.45\textwidth}
      \includegraphics[width=\textwidth]{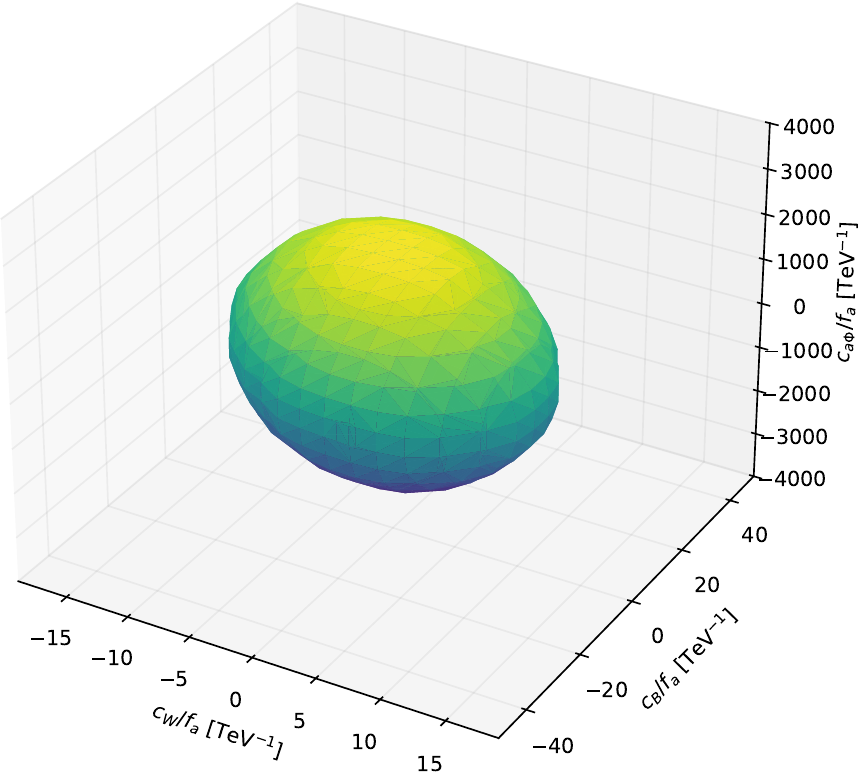}
      \caption{}
      \label{fig:azz_b}
        \end{subfigure}
        \par\bigskip
        \begin{subfigure}{0.45\textwidth}
      \includegraphics[width=\textwidth]{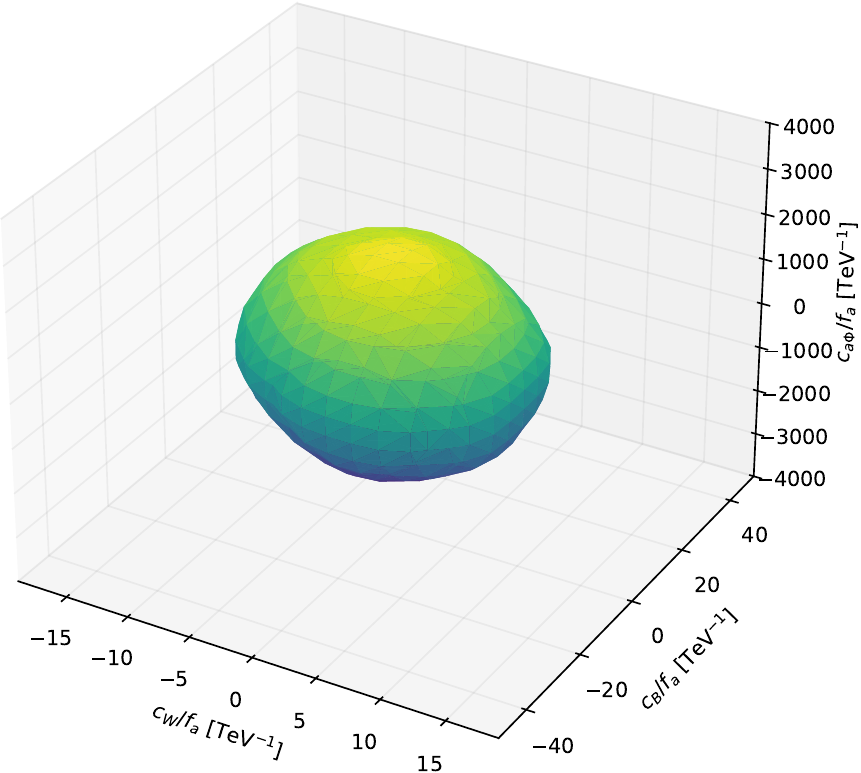}
      \caption{}
      \label{fig:azz_c}
        \end{subfigure}
        \hfill
        \begin{subfigure}{0.45\textwidth}
      \includegraphics[width=\textwidth]{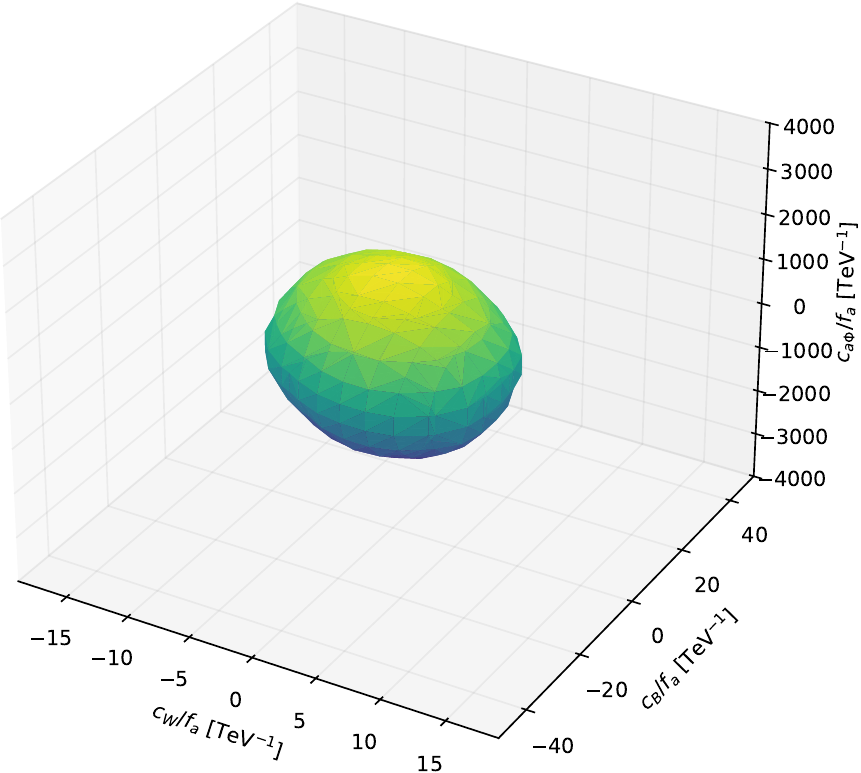}
      \caption{}
      \label{fig:azz_d}
        \end{subfigure}
          \caption{Projected 95\cl~exclusion contours for the $\azz$ channel in the $(c_B/f_a, c_W/f_a, c_{a\Phi}/f_a)$ system with $c_G/f_a$ set to $0.1~\text{TeV}^{-1}$. The top row (a,b) corresponds to the $2\ell2\nu$~decay channel, while the bottom row (c,d) corresponds to the $4\ell$~channel. The assumed integrated luminosity is $\int L=450~\text{fb}^{-1}$ for the left column (a,c) and $\int L=3000~\text{fb}^{-1}$ for the right column (b,d).} 
\label{fig:azz}
\end{figure}

The resulting constraints from the combination of the $4\ell$ and $2\ell2\nu$ channels are shown in Fig.~\ref{fig:azz}.
%
%%%%%%%
%
%
%%%%%%%%%
%
Presented in the $(c_B/f_a, c_W/f_a, c_{a\Phi}/f_a)$ parameter space with $c_G/f_a$ fixed to $0.1~\text{TeV}^{-1}$, the exclusion contours exhibit the expected elliptical shape, reflecting the dependence on both electroweak couplings. However, due to the limited discriminating power of the kinematic variables and the small branching ratios of the considered decay channels, the resulting constraints are relatively weak compared to those derived from other channels analyzed in this work. The $2\ell2\nu$ channel provides slightly weaker bounds than the $4\ell$ channel, primarily due to its larger background. 
%
%%%%%%%
%
\section{Summary and conclusions}\label{sec:results}
In this work, we have presented a comprehensive analysis of the sensitivity of the LHC and the future HL-LHC to ALPs in the sub-GeV mass range. By focusing on the linear EFT framework, we treated the ALP interactions with SM gauge bosons in a model-independent manner, allowing for a simultaneous determination of constraints on the Wilson coefficients $c_W$, $c_B$, $c_G$, and $c_{a\Phi}$. Unlike many previous studies that assume a hierarchical structure of couplings or the vanishing of specific interactions to simplify the parameter space, our approach embraces the codependencies inherent in the EFT. This is particularly crucial for processes involving neutral diboson final states, where the interplay between gluon-initiated production and electroweak decays significantly alters kinematic distributions and cross-sections, compared to purely non-gluon-initiated production topologies.

Our analysis relied on the associated production of ALPs with diboson pairs, a signature that becomes particularly relevant for light, long-lived ALPs that escape the detector. The resulting signature, large missing transverse energy associated with leptons or photons of large transverse momenta, requires robust discrimination against the formidable SM backgrounds, particularly those arising from misidentified QCD multijet events. By employing BDTs, we successfully exploited subtle kinematic correlations, such as the angular separation between identified objects and the missing energy vector, to maximize signal sensitivity.

\begin{figure}[!tp]
  \centering
  \begin{subfigure}{0.45\textwidth}
        \includegraphics[width=\textwidth]{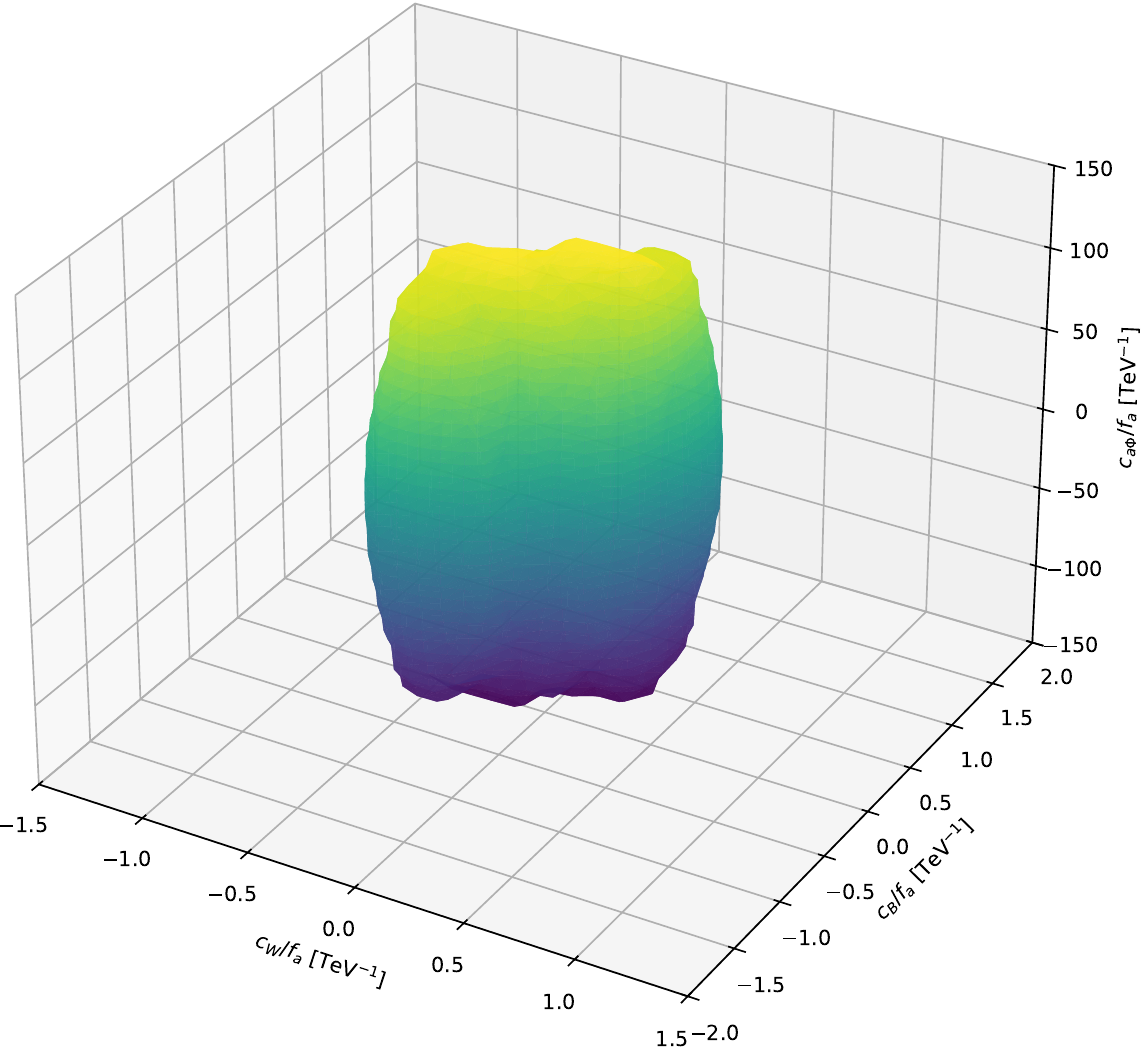}
        \caption{}
        \label{fig:fin_a}
        \end{subfigure}
        \hfill
        \begin{subfigure}{0.45\textwidth}
      \includegraphics[width=\textwidth]{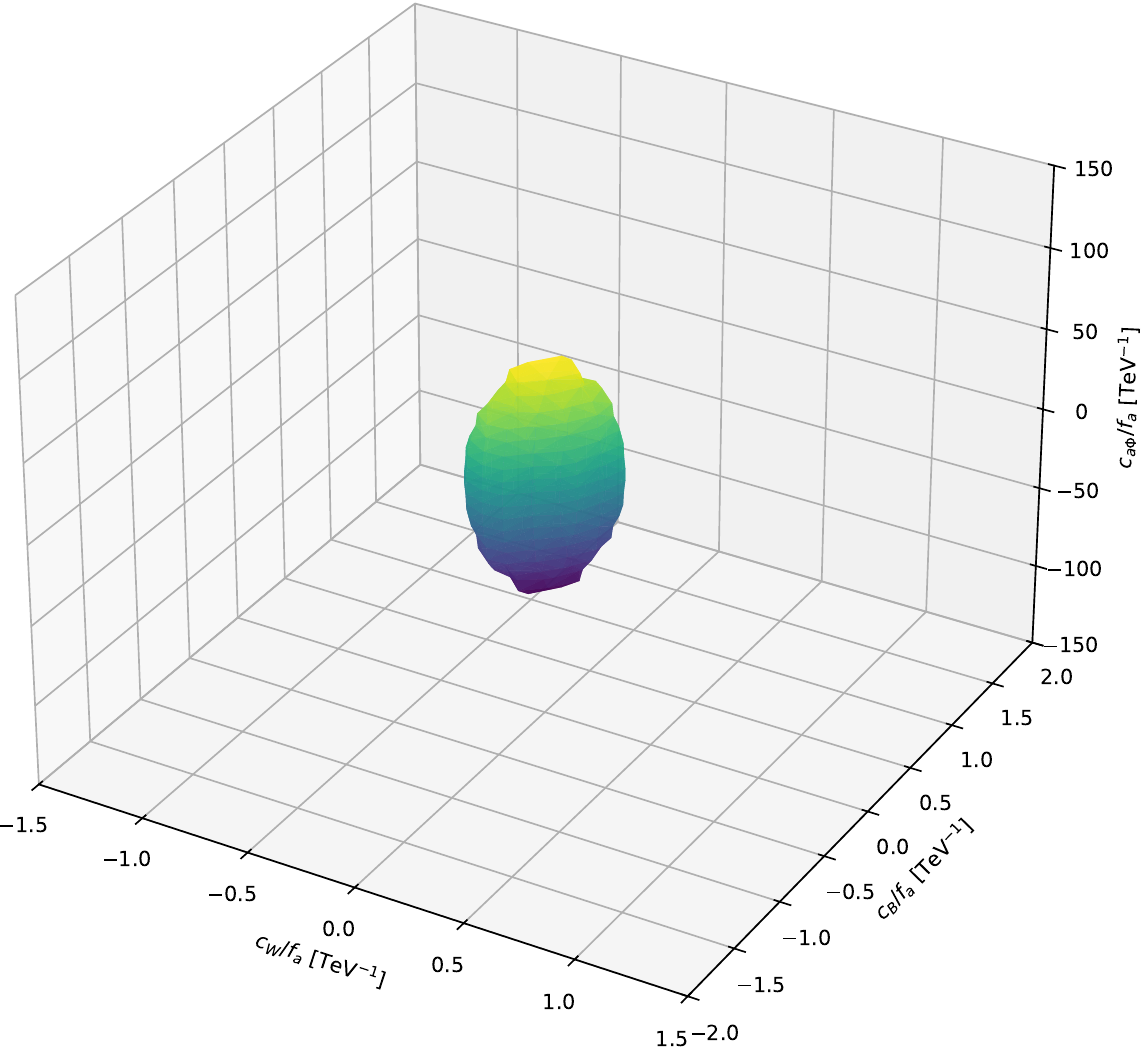}
      \caption{}
      \label{fig:fin_b}
        \end{subfigure}
        
        \par\bigskip
        \begin{subfigure}{0.45\textwidth}
      \includegraphics[width=\textwidth]{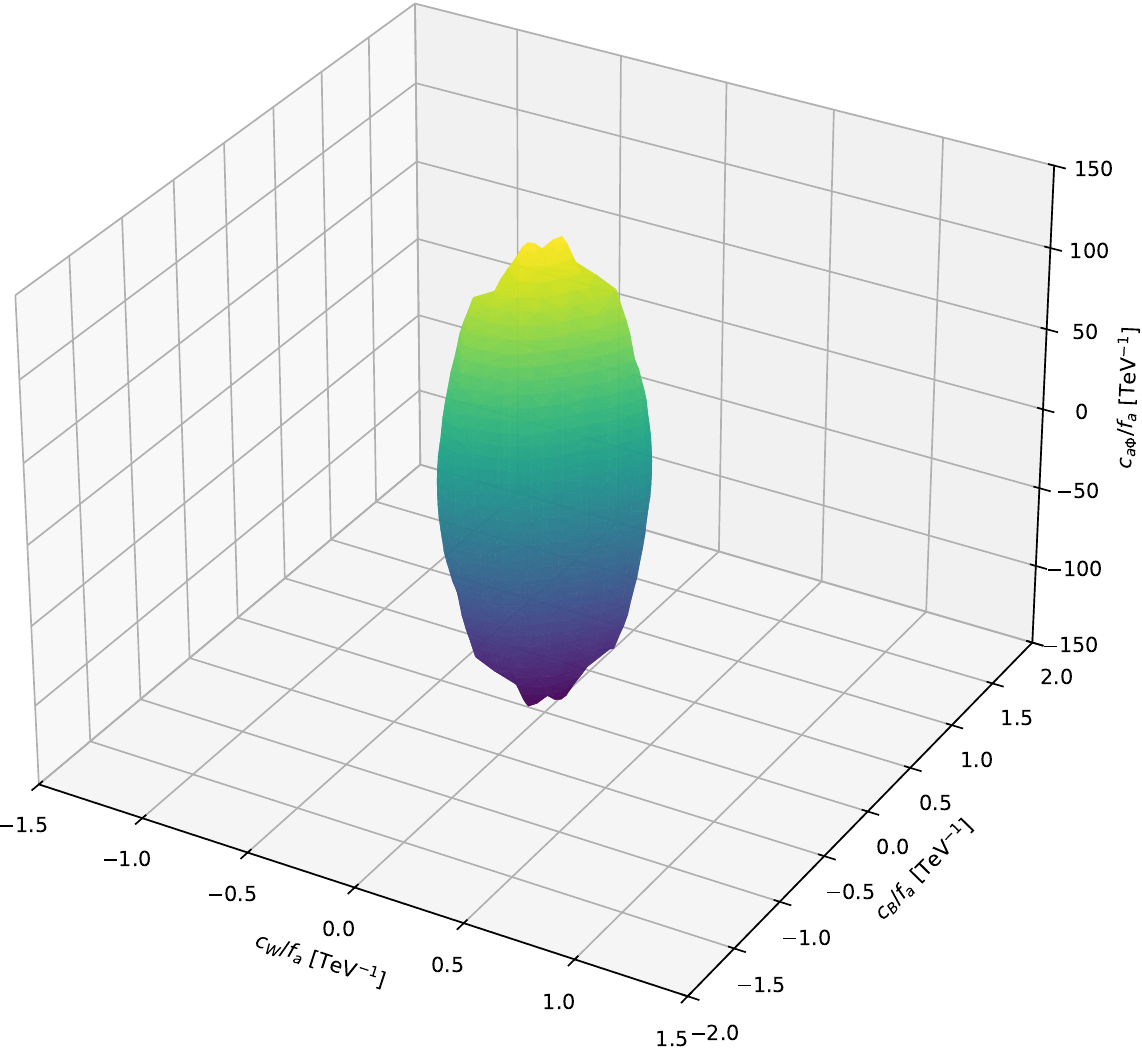}
      \caption{}
      \label{fig:fin_c}
        \end{subfigure}
        \hfill
        \begin{subfigure}{0.45\textwidth}
      \includegraphics[width=\textwidth]{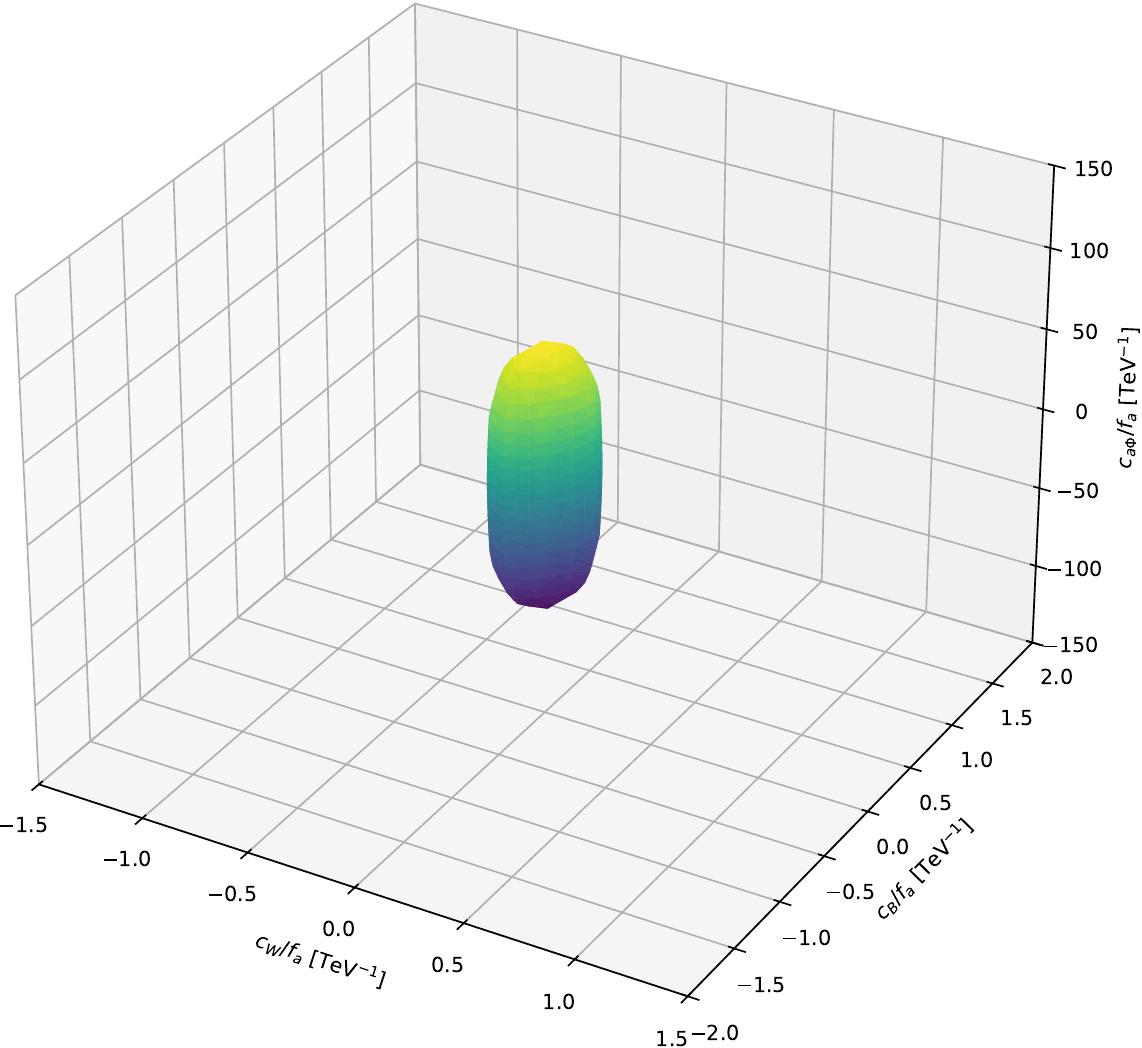}
      \caption{}
      \label{fig:fin_d}
        \end{subfigure}
  \caption{Combined 95\cl~exclusion contours for the bosonic ALP couplings derived from the combination  of the diboson channels $a\gamma\gamma$, $aW^\pm\gamma$, $aZ\gamma$, $aZW^\pm$, $aW^+W^-$, $aZZ$, considering the leptonic and invisible $Z$~boson decays as well as the leptonic $W^\pm$~boson decays. Panels (a) and (c) show the constraints for the LHC with $\int\mathcal{L} = 450~\text{fb}^{-1}$, and panels (b) and (d) the projections for the HL-LHC with $\int\mathcal{L} = 3000~\text{fb}^{-1}$. For (a) and (b) we used $c_G=0.1~\text{TeV}^{-1}$, and for (c) and (d)  $c_G=1~\text{TeV}^{-1}$. The regions outside the contours are excluded.}
  \label{fig:combined_results}
\end{figure}

The individual channels analyzed in this study exhibit distinct sensitivities to the various ALP couplings, highlighting the necessity of a combined analysis. 
The $a\gamma\gamma$ channel, driven largely by gluon fusion and ALP-photon interactions, provides the most stringent constraints on the effective ALP-gluon and ALP-photon couplings. However, due to the specific linear combination of $c_W$ and $c_B$ entering the $g_{a\gamma\gamma}$ vertex ($c_W \sin^2\theta_W + c_B \cos^2\theta_W$), this channel alone leaves a "blind direction" in the EW parameter space.
This degeneracy is not present in the the $aZ\gamma$ and $aZZ$ channels. The $aZ\gamma$ coupling depends on the orthogonal combination ($c_W - c_B$), while the $aZZ$ coupling introduces yet another combination, ($c_W \cos^2\theta_W + c_B \sin^2\theta_W$). Our results show that the $\azg$ channel with the $Z$~boson decaying into a neutrino pair, despite the lack of a reconstructible $Z$-mass peak, offers competitive sensitivity due to the large branching ratio and the enhanced $E_T^{miss}$ signature.
Furthermore, the charged diboson channels, $aW^\pm\gamma$, $aZW^\pm$, and $aW^+W^-$, play a critical role in isolating the ALP couplings to electroweak gauge bosons. Since these processes do not include gluon fusion topologies at tree level, they provide clean probes of $c_W$ and $c_B$ independent of the gluonic coupling $c_G$. The $aW^\pm\gamma$ channel, in particular, yields robust elliptical constraints centered at the origin of the $(c_W, c_B)$ plane, free of the asymptotic behavior seen in channels dominated by gluonic enhancements.
By statistically combining the likelihoods from all six analyzed channels, i.e.\  $a\gamma\gamma$, $aW^\pm\gamma$, $aZ\gamma$, $aZW^\pm$, $aW^+W^-$, and $aZZ$, considering the leptonic and invisible $Z$~boson decays as well as the leptonic $W^\pm$~boson decays as detailed in Sec.~\ref{sec:pheno}, we derive global exclusion limits for the bosonic ALP EFT. The intersection of these individual constraints defines the globally allowed region in the four-dimensional parameter space spanned by $\{\frac{c_W}{f_a},\frac{c_B}{f_a},\frac{c_G}{f_a},\frac{c_{a\Phi}}{f_a}\}$, see Fig.~\ref{fig:combined_results}. 
%
%
%%%%%%%%%%
%
%
%%%%%
%
Our results demonstrate a significant improvement in the reach of the LHC for the exclusion of sub-GeV ALPs. The higher luminosity of the HL-LHC tightens the bounds considerably, without imposing any assumptions on the coupling parameters. In the considered channels, the largest sensitivity arises from the $aZ\gamma$, $aW^\pm\gamma$ and $a\gamma\gamma$ final states. 

Our results underscore the potential of the LHC as a precision machine for ALP physics, even in the experimentally challenging regime of light, invisible particles. Our use of a conservative, flat jet-misidentification rate suggests that these limits are robust; however, they also point to areas where experimental improvements could yield significant gains. A better understanding and reduction of jet-faking-photon rates, particularly in the high-transverse momentum regime, would directly enhance the sensitivity of the $a\gamma\gamma$ and $aW\gamma$ channels, which are currently limited by systematic uncertainties in the background modeling.

Looking forward, the techniques developed here can be extended to other collider environments. While currently the LHC dominates in terms of energy reach, future colliders operating at yet higher energies (such as the FCC-hh) would offer complementary opportunities for probing ALP couplings to SM gauge bosons, potentially resolving the remaining degeneracies with even greater precision. Additionally, extending this EFT analysis to include non-linear realizations of the ALP symmetry could provide a more complete picture of the ALP sector.

\section*{Acknowledgments}
We gratefully acknowledge financial support from the Studienstiftung des deutschen Volkes.

\bibliography{bib}
\bibliographystyle{unsrt}
\end{document}